\documentclass[traditabstract]{aa}  
\usepackage{lscape,longtable}
\usepackage{graphicx}
\usepackage{amsmath}
\usepackage[varg]{txfonts}

\usepackage{natbib}
\bibpunct{(}{)}{;}{a}{}{,} 
\begin{document}
\hyphenation{data-points data-points velocities fxcor velocity sub-samples}
   \title{The globular cluster system of NGC\,1399 
\thanks{Based on observations made with ESO Telescopes at the  Paranal Observatories under programme ID 70.B-0174.}\fnmsep 
\thanks{Based on observations obtained at the Gemini Observatory, which
is operated by the Association of Universities for Research in
Astronomy, Inc., under a cooperative agreement with the NSF on behalf
of the Gemini partnership: the National Science Foundation (United
States), the Particle Physics and Astronomy Research Council (United
Kingdom), the National Research Council (Canada), CONICYT (Chile), the
Australian Research Council (Australia), CNPq (Brazil) and CONICET
 (Argentina).} 
}
   \subtitle{V.~dynamics of the cluster system out to 80\,kpc}

   \author{Y. Schuberth
     \inst{1,2}
     \and
     T. Richtler
     \inst{2}
     \and
     M. Hilker 
     \inst{3}
     \and
     B. Dirsch
     \inst{2}
     \and
    L.~P.~Bassino
     \inst{4}
     \and
     A.~J.~Romanowsky
     \inst{5,2}
     \and  
     L. Infante
     \inst{6}
          }

   \offprints{ylva@astro.uni-bonn.de}

   \institute{Argelander-Institut f\"ur Astronomie,
     Universit\"at Bonn, Auf dem H\"ugel 71, D-53121 Bonn, Germany 
     \and
     Universidad de Concepci\'on, Departamento de Astronomia, Casilla
     160-C, Concepci\'on, Chile 
     \and European Southern Observatory,
     Karl-Schwarzschild-Str.~2, D-85748 Garching, Germany 
     \and 
     Facultad de Ciencias Astron\'omicas y Geof\'isicas, 
     Universidad Nacional de La Plata, Paseo del Bosque S/N, 1900--La Plata, Argentina; and  Instituto de Astrof\'{\i}sica de La Plata (CCT La Plata --
CONICET -- UNLP)
\and UCO/Lick Observatory, University of California, Santa Cruz, CA 95064, USA
     \and 
     Departamento de Astronom\'ia y Astrof\'isica, Pontificia Universidad Cat\'olica de Chile, Casilla 306, Santiago 22, Chile
}

   \date{Received 14 May, 2009; accepted 16 October, 2009}


\abstract{
Globular clusters (GCs) are tracers of the gravitational
potential of their host galaxies. Moreover, their kinematic properties
may provide clues for understanding the formation of GC systems and
their host galaxies.  We use the largest set of GC velocities obtained
so far of any elliptical galaxy to revise and extend the previous
investigations (Richtler et al.~2004) of the dynamics of NGC\,1399,
the central dominant galaxy of the nearby Fornax cluster of galaxies.
The GC velocities are used to study the kinematics,
their relation with population properties, and the dark matter halo of
NGC 1399.
We have obtained 477 new medium--resolution spectra (of these, 292 are
 spectra from 265 individual GCs, 241 of which are not in the previous
 data set). with the VLT FORS\,2 and Gemini South GMOS multi--object
 spectrographs. We revise velocities for the old spectra and measure
 velocities for the new spectra, using the same templates to obtain an
 homogeneously treated data set.
Our entire sample now comprises velocities for almost 700 GCs with
projected galactocentric radii between 6 and 100\,kpc. In addition, we
use velocities of GCs at larger distances published by Bergond et
al.~(2007).  Combining the kinematic data with wide--field photometric
Washington data, we study the kinematics of the metal--poor and
metal--rich subpopulations.  We discuss in detail the velocity
dispersions of subsamples and perform spherical Jeans modelling.

The most important results are: The red GCs resemble the stellar field
population of NGC 1399 in the region of overlap. The blue GCs behave
kinematically more erratic.  Both subpopulations are kinematically
distinct and do not show a smooth transition. It is not possible to
find a common dark halo which reproduces simultaneously the properties
of both red and blue GCs.  Some velocities of blue GCs are only
to be explained by orbits with very large apogalactic distances, thus
indicating a contamination with GCs which belong to the entire Fornax
cluster rather than to NGC\,1399.  Also, stripped GCs from nearby
elliptical galaxies, particularly NGC\,1404, may contaminate the blue
sample.

We argue in favour of a
scenario in which the majority of the blue cluster population has been
accreted during the assembly of the Fornax cluster. The red cluster
population shares the dynamical history of the galaxy itself.
Therefore we recommend to use a dark halo based on the red GCs alone.

 The dark
halo which fits best is marginally less massive than the halo quoted by
Richtler et al.~(2004). The comparison with X-ray analyses is
satisfactory in the inner regions, but without showing evidence for a
transition from a galaxy to a cluster halo, as suggested by X-ray
work. 
}
\keywords{galaxies: elliptical and lenticular, cD --- galaxies:
kinematics and dynamics --- galaxies:individual: NGC\,1399}
\maketitle 
%

\section{Introduction}

\subsection{The globular cluster systems of central 
elliptical galaxies 
}

Shortly after M87 revealed its rich globular cluster system (GCS)
\citep{baum1955,racineGCLF} it became obvious
that bright ellipticals in general host globular clusters in much
larger numbers than  spiral galaxies 
\citep{harrisracine79}. Moreover, the richest GCSs are found for
elliptical galaxies in the centres of galaxy clusters, for which M87
in the Virgo cluster and NGC\,1399 in the Fornax cluster are the
nearest examples. For recent reviews of the field, see \cite{bs06}, and \cite{richtler06}.

The galaxy cluster environment may act in different ways to produce
these very populous GCSs. Firstly, there is the paradigm of giant
elliptical galaxy formation by the merging of disk galaxies
(e.g. \citealt{toomre77}, \citealt{renzini06}). That early--type
galaxies can form through mergers is evident by the identification of
merger remnants and many kinematical irregularities in elliptical
galaxies (counter-rotating cores, accreted dust and molecular rings).

 In fact, starting from the bimodal colour distribution of GCs in some
giant ellipticals, \cite{az92} predicted the efficient formation of
GCs in {{spiral--spiral}} mergers, before this was confirmed
observationally \citep{schweizer93}.  In their scenario, the blue
clusters are the metal--poor GCSs of the pre--merger components while
the red (metal--rich) GCs are formed in the material which has been
enriched in the starbursts accompanying the early merger {{
(gaseous merger model).  }}\par {{However, \cite{forbes97}
pointed out that the large number of metal--poor GCs found around
giant ellipticals cannot be explained by the gaseous merger model.
These authors proposed a `multi-phase collapse model' in which the blue
GCs are created in a pre--galactic phase along with a relatively low
number of metal--poor field stars. The majority of field stars,
i.e.~the galaxy itself and the red GCs, are then formed from the
enriched gas in a secondary star formation epoch.  }} {{ This
scenario is also supported by the findings of \cite{spitler08}
who studied an updated sample of 25 galaxies spanning a large range of
masses, morphological types and environments. They confirmed that the
spirals generally show a lower fraction of GCs normalised to host
galaxy stellar mass than massive ellipticals, thus ruling out the
possibility that the GCSs of massive ellipticals are formed through major wet
mergers.}}  {{Further, Spitler et al.~suggest that the number
of GCs per unit halo mass is constant -- thus extending the work by
\cite{blakeslee97} who found that in the case of central cluster
galaxies the number of GCs scales with the cluster mass -- pointing
towards an early formation of the GCs.}}

\par {{ In some scenarios 
(e.g.~\citealt{cote98,hilker99III,cote02,beasley02}), the accretion of
mostly metal--poor GCs is responsible for the richness of the GCSs of
central giant ellipticals.}}

\par But only during the last years it has become evident that
accretion may be important even for GCSs of galaxies in relatively low
density regions like the Milky Way (e.g.~\citealt{helmi08}), the most
convincing case being the GCs associated with the Sagittarius stream
{\citep{ibata01,bellazzini03}}. Therefore, GC accretion should
plausibly be an efficient process for the assembly of a GCS in the
central regions of galaxy clusters.

Given this scenario one expects huge dark matter halos around central
giant ellipticals, perhaps even the sum of a galaxy-size dark halo and
a cluster dark halo (e.g.~\citealt{ikebe96}).  However, dark matter
studies in elliptical galaxies using tracers other than X-rays were
long hampered by the lack of suitable dynamical tracers. Due to the
rapidly declining surface brightness profiles, measurements of stellar
kinematics are confined to the inner regions, just marginally probing
the radial distances at which dark matter becomes dominant. One
notable exception is the case of NGC\,6166, the cD galaxy in
Abell\,2199, for which \cite{kelson02} measured the velocity
dispersion profile out to a radius of 60\,kpc. Another one is the case
of NGC\,2974 where an H{\textsc{i}} disk traces the mass out to
20\,kpc \citep{weijmans08}. \par Only with the advent of 8m--class
telescopes and multi--object spectrographs it has become feasible to
study the dynamics of globular cluster systems (GCSs) of galaxies as
distant as 20\,Mpc. Early attempts
(\citealt{hb87,grillmair94,cohen97,minniti98,kisslerpatig98}) were
restricted to the very brightest GCs, and even today there are only a
handful of galaxies with more than 200 GC velocities measured.  Large
($N_{\rm{GC}}>200$) samples of GC radial velocities have been
published for M87 \citep{cote01} and NGC\,4472 \citep{cote03} in Virgo
and \object{NGC\,1399} in Fornax (\citealt{richtler04}, hereafter
Paper\,I).  {{Also the GCS dynamics of the nearby ($\sim4\,\rm{Mpc}$)
disturbed galaxy Cen\,A (NGC\,5128) has been studied extensively
\citep{peng04gcs,woodley07}, with 340 GC velocities  available to
date.}}

Investigating the kinematics and dynamics of GCSs of elliptical
galaxies  has a twofold objective.  Firstly, kinematical
information together with the population properties of GCs promise to
lead to a deeper insight into the formation history of GCSs with their
different GC subpopulations. Secondly, GCs can be used as dynamical
tracers for the total mass of a galaxy and thus  allow the
determination of the dark matter profile, out to large galactocentric
distances which are normally inaccessible to studies using the
integrated light. These results then can be compared to X--ray
studies.  A large number of probes is a prerequisite for the analysis
of these dynamically hot systems. Therefore, giant ellipticals, known
to possess extremely populous and extended GCSs with thousands of
clusters, have been the preferred targets of these studies.

Regarding the formation history of GCSs, a clear picture has not yet
emerged.  Adopting the usual bimodal description of a GCS by the
distinction between metal--poor and metal--rich GCs, the kinematical
properties seem to differ from galaxy to galaxy. For example, in M87
the blue and red GCs do not exhibit a significant difference in
their velocity dispersion (\citealt{cote01}, with a sample size of 280
GCs).  Together with their different surface density profiles,
C{\^o}t{\'e} et al. concluded that their orbital properties should be
different: the metal--poor GCs have preferentially tangential
orbits while the metal--rich GCs prefer more radial orbits.

In NGC\,4472, on the other hand, the metal--rich GCs have a
significantly lower velocity dispersion than their blue counterparts
(found for a sample 250 GCs, \citealt{cote03}), and
these authors conclude that the cluster system as a whole has an
isotropic orbital distribution.

NGC\,1399, the object
of our present study, is a galaxy, which, as a central cluster galaxy,
is similar to M87 in many respects (see \citealt{dirsch03}). Here, the
metal--poor GCs show a distinctly higher velocity dispersion than
the metal--rich GCs, but more or less in agreement with their
different density profiles (with a sample size of $\sim\!470$ GCs). A
similar behaviour has been found for NGC\,4636 \citep{schuberth06}.

In most other studies, the sample sizes are still too small to permit
stringent conclusions or the separate treatment of red and blue GCs,
but dark matter halos have been found in almost all cases.

\subsection{The case of NGC 1399}

NGC\,1399 has long been known to host a very populous globular cluster
system (e.g.~\citealt{dirsch03} and  references therein).
With the photometric study by \cite{bassino06} it
became clear that the GCS of NGC\,1399 extends to about 250\,kpc,
which is comparable to the core radius of the  cluster \citep{ferguson89}.
Accordingly, it has always been an attractive target for studying the
dynamics of its GCS.
One finds there the largest sample of GC velocities (469) available so far
(\citealt{richtler04} (Paper\,I), \citealt{dirsch04}).  It was shown in
 Paper\,I  that blue and red GCs are kinematically different, as was
expected from their different number density profiles: the red
GCs exhibit a smaller velocity dispersion than the blue GCs
in accordance with their respective density profiles. Evidence for
strong anisotropies has not been found. The radial velocity dispersion
profile was found to be constant for red and blue GCs.  However,
as we think now, this could have been a consequence of a velocity cut
introduced to avoid outliers.  A dark halo of the NFW type under
isotropy reproduced the observations satisfactorily.  No rotation was
detected apart from a slight signal for the outer blue GCs.  It
was shown that some of the extreme radial velocities in conjunction
with the derived dark halo were only understandable if they were being caused by
orbits with very large apogalactic distances.  In this paper, we
extend our investigation of the NGC\,1399 GCS to larger radii (80
kpc). We simultaneously revise the old velocities/spectra in order to
have an homogeneously treated sample. The case of Modified Newtonian
Dynamics (MOND) has already been  discussed in \cite{richtler08}, where
it has been shown that MOND still needs additional dark matter of the
order of the stellar mass. We do not come back to this issue in the
present contribution.  \par

The GCS of NGC 1399 is very extended. One can trace the blue GCs
out to about 250\,kpc, the red GCs only to 140\,kpc \citep{bassino06}.
Regarding total numbers, there are only half as many red GCs as there
are blue, suggesting that the formation of GCs in mergers is not the 
dominant mechanism producing a high
specific frequency, even if in the central regions of a proto--cluster
the merger rate is supposed to be particularly high.

\par Following Paper\,I, we assume a distance modulus of $31.40$.  At
the distance of 19\,Mpc, $1\arcsec{}$ corresponds to 92\,pc, and
$1\arcmin{}$ corresponds to 5.5\,kpc. \par

This paper is organised as follows: In Sect.~\ref{sect:obs}, we
describe the observations and the data reduction.  The velocity data
base is presented in Sect.~\ref{sect:database}. In
Sect.~\ref{sect:sample}, we present the photometric properties and the
spatial distribution of our GC velocity sample. The contamination by
interlopers is discussed in Sect.~\ref{sect:interlopers}. The
properties of the line--of--sight velocity distribution are studied in
Sect.~\ref{sect:losvd}. In Sect.~\ref{sect:rot} we test our GC
sample for rotation. The line--of--sight velocity dispersion and the
higher order moments of the velocity distributions are calculated in
Sect.~\ref{sect:dispersion}. 
The Jeans modelling and the derived mass models are described 
in Sects.~\,\ref{sect:jeans} and \ref{sect:massmodels}.
The results
are discussed and summarised in Sections\,\ref{sect:discussion} and
\ref{sect:conclusions}.

\section{Observations and  data reduction}
\label{sect:obs}

\begin{figure}
\centering
\resizebox{\hsize}{!}
{\includegraphics[width=0.48\textwidth]{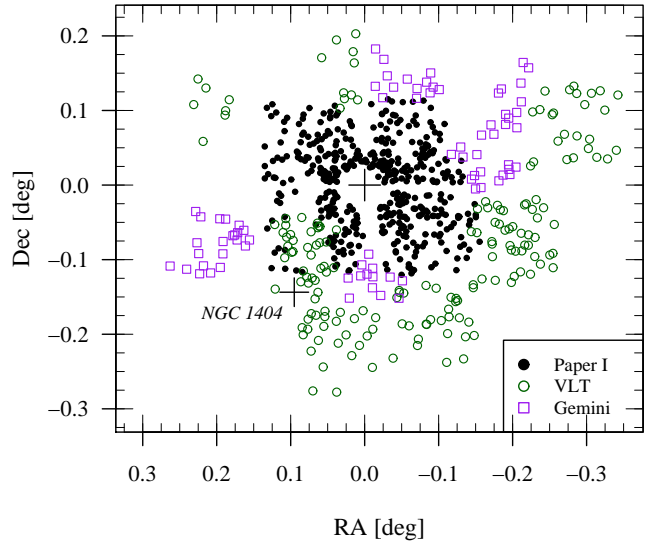}}
\caption{Spatial distribution of {{spectroscopically confirmed
GCs}} with respect to NGC\,1399 $(0,0)$. Dots represent GCs from
Paper\,I. Circles and squares are the new GCs, measured using
VLT/FORS\,2 and Gemini/GMOS, respectively. North is to the top and
East is to the left. The positions of NGC\,1399 and NGC\,1404 are
marked by crosses.}
\label{fig:map}
\end{figure}
To study the dynamics of the GCS of NGC\,1399 out to large radii, we use data
obtained with the multi--object--spectrographs FORS2/MXU and GMOS-S,
at the VLT and Gemini--South telescopes, respectively.  The spectral
resolution and wavelength coverage is similar for both datasets.
Details on the mask preparation and data reduction are given
below.\par The location of the {{GCs}} on the plane of the sky
is shown in Fig.~\ref{fig:map}. The {{GCs}} from Paper\,I
(shown as dots) occupy the inner region, and open symbols represent
the GCs added in the present study, raising the number of GC
velocities to almost 700 and extending the radial range by almost a
factor of two.
\subsection{Photometric data}
\label{sect:MOSAIC}
Wide--field imaging in the metallicity--sensitive Washington system
obtained for several fields in the Fornax cluster using the CTIO
MOSAIC camera\footnote{The camera is mounted in the prime focus of the
4--m Blanco telescope and the field--of--view is
$36\arcmin\times36\arcmin$.} forms the basis for the photometry used
in this work. The central field \citep{dirsch03} encompasses the area
covered by our spectroscopic study.  As in Paper\,I, these data were
used for target selection, and the photometric properties of our
velocity--confirmed GCs are presented in Sect.~\ref{sect:colour}.
\par Further, the analysis of the outer fields presented by
\cite{bassino06} provides the number density profiles of the GC
subpopulations out to large radii (cf.~Sect.~\ref{sect:powerlaws})
which are required for the dynamical modelling.

\subsection{VLT--FORS2/MXU spectroscopy}

\begin{table*}
\centering
\caption[Summary of observations]{{Summary of VLT FORS\,2/MXU
 observations (programme ID 70.B-0174 ). The six--digit number given
 in the second column is the mask identifier as given in the image
 header (keyword: \texttt{HIERARCH ESO INS MASK ID}). The seeing
 values (Col.~6) were determined from the acquisition images taken just before
 the MXU exposures. {{The numbers given in Cols.~7 and 8 refer to the total number of slits and the number of GCs found on a given mask, respectively.}}}}

\begin{tabular}{llllcccrl@{-}l@{-}lr@{:}l}\hline
\hline
{Mask} & \multicolumn{1}{c}{ID} &
\multicolumn{2}{c}{Centre Position} &
{Exp.~Time}&{Seeing}&\multicolumn{1}{c}{\#\,Slits}&
\multicolumn{1}{c}{\#\,GCs}&
\multicolumn{3}{c}{Night}\\
{} & {} &\multicolumn{2}{c}{{({J\,2000})}} &
{{({sec})}}&{{{}}}&
\multicolumn{1}{c}{{}}&
\multicolumn{1}{c}{\,}&
\multicolumn{3}{c}{\,}\\
\hline
Mask\,1&965\ 705& 03:37:15.9 &$-35$:22:18.2 & 2600  & 0 \farcs 90& 130 &9  &2002 &12&01 \\
Mask\,2&965\ 813& 03:37:22.1 &$-35$:22:18.5 & 2600  & 0 \farcs 90&    132 & 16&\multicolumn{3}{c}{\ldots} \rule[-1.5ex]{0pt}{0pt}\\
Mask\,3&965\ 930& 03:37:34.7 &$-35$:31:40.9 & 2600  & 0 \farcs 65&    130 & 24&\multicolumn{3}{c}{\ldots} \\
Mask\,4&970\ 048 & 03:37:40.9 &$-35$:31:41.9 & 2600  & 0 \farcs 75&    132 & 27&\multicolumn{3}{c}{\ldots} \rule[-1.5ex]{0pt}{0pt}\\
Mask\,5&970\ 253& 03:38:02.8 &$-35$:38:15.7 &  2600 & 0 \farcs 85&    131 & 21 &\multicolumn{3}{c}{\ldots} \\

Mask\,7&971\ 436& 03:38:48.4 &$-35$:32:53.6 & 2600  & 0 \farcs 80&    128 & 36&2002 &12&02 \\
Mask\,8&973\ 401& 03:39:21.8 &$-35$:21:27.2 & 2600  & 0 \farcs 65&    127 &7&\multicolumn{3}{c}{\ldots} \rule[-1.5ex]{0pt}{0pt}\\
Mask\,9&970\ 843&  03:38:33.4  &$-35$:40:34.1  & 2600 & 0 \farcs 85&    122 &9&\multicolumn{3}{c}{\ldots} \\
Mask\,10&971\ 105& 03:38:39.6   &$-35$:40:35.3  & 2600  & 0 \farcs 80&    124 &18&\multicolumn{3}{c}{\ldots} \rule[-1.5ex]{0pt}{0pt}\\

Mask\,12&970\ 540& 03:38:09.1  &$-35$:38:14.6 & 2600  & 1 \farcs 35&  129   &12&\multicolumn{3}{c}{\ldots} \rule[-1.5ex]{0pt}{0pt}\\
Mask\,13&972\ 414& 03:38:40.5  &$-35$:17:24.9  & 2600   & 1 \farcs 45&  127   &10& 2002&12&03 \\
\hline
\hline
\end{tabular}
\label{tab:vlt}
\end{table*}

The spectroscopic observations of 11 masks
in seven fields (see Table\,\ref{tab:vlt} for details) were carried
out in visitor mode during three nights (2002 December 1--3) at the
European Southern Observatory Very Large Telescope (VLT) facility on
Cerro Paranal, Chile. We used the FORS\,2 (Focal Reducer/low
dispersion Spectrograph) instrument equipped with the Mask Exchange
Unit (MXU).  \par Our spectroscopic targets were selected from the
wide--field photometry by Dirsch et al.~(\citeyear{dirsch03},
hereafter D+03).  The masks were designed using the FIMS software. To
maximise the number of objects per mask, we observed objects and sky
positions through separate slits of $2\arcsec$ length. This strategy
has previously been used for the study of NGC\,1399
(\citealt{dirsch04}, D+04 hereafter) and NGC\,4636
\citep{schuberth06}. \par All observations were performed using the
grism 600\,B which provides a spectral resolution of
$\sim\!2.5\,\rm{\AA} {\simeq150\,\rm{km\,s}^{-1}}$ (as measured from the line--widths of the
wavelength calibration exposures).

\par The reduction of the FORS2/MXU spectra was performed in the same
way as described in \cite{dirsch04}, so we just give a brief
description here. After the basic reduction steps (bias subtraction,
flat fielding, and trimming), the science and calibration frames were
processed using the \texttt{apextract} package in IRAF. The wavelength
calibration was performed using \texttt{identify}. Typically 18 lines
of the Hg--Cd--He arc lamp were used to fit the dispersion relation,
and the residuals were of the order $0.04\,\rm{\AA}$. To perform the
sky subtraction for a given GC--spectrum, the spectra of two or three
nearby sky--slits were averaged and subsequently subtracted using the
\texttt{skytweak} task.

\subsection{Gemini GMOS spectroscopy}

\begin{table*}
\centering
\caption[Summary of observations]{{Summary of Gemini South GMOS
 observations (program IDs GS\,2003B\,Q031 and GS\,2004B\,Q029). The
 seeing (Col.~5) as estimated from the 2D-spectra of bright
 stars. {{The number of slits and the number of GCs detected on
 a given mask are given in Cols.~7 and 8, respectively.}}}}

\begin{tabular}{cllcccrl@{-}l@{-}l}\hline
\hline
{Mask} & \multicolumn{2}{c}{Centre Position} &
{Exp.~Time}&{Seeing}&
\multicolumn{1}{c}{\#\,Slits}&
\multicolumn{1}{c}{\#\,GCs} &
\multicolumn{3}{c}{Night}\\
{} & \multicolumn{2}{c}{{({J\,2000})}} &
{{({sec})}}&{{{}}}&
\multicolumn{1}{c}{{}}&
\multicolumn{1}{c}{\,}&
\multicolumn{3}{c}{\,}\\
\hline
m2003-01& 03:39:20.2& $-35$:31:58.6 & 2600  & 0 \farcs 58& 31 &14 &2003 &11&19\\
m2003-02& 03:39:20.2& $-35$:31:58.6 & 3900  & 0 \farcs 80& 21 &8   &2003 &11&20 \rule[-1.5ex]{0pt}{0pt}\\
m2003-03& 03:38:13.4& $-35$:18:03.0 & 3900  & 0 \farcs 73& 27 &7  &2003 &11&23 $\quad^\star$ \\
m2003-04& 03:38:13.4& $-35$:18:03.0 & 3900  & 0 \farcs 55& 24 &3  &2003 
&11&23 $\quad^\dagger$ \rule[-1.5ex]{0pt}{0pt}\\
m2004-01& 03:38:13.4& $-35$:18:03.0 & 2600  & 0 \farcs 66& 27 &8  &2004 &12&07 $\quad^\star$ \\
m2004-02& 03:38:13.4& $-35$:18:03.0 & 3900  & 0 \farcs 70& 24 &4  &2004 &12&07 $\quad^\dagger$ \rule[-1.5ex]{0pt}{0pt}\\
m2004-03& 03:37:44.9& $-35$:19:06.2 & 3900  & 0 \farcs 57& 30 &10  &2004 &12&10 \\
m2004-04& 03:37:44.9& $-35$:19:06.2 & 3900  & 0 \farcs 73& 21 &1  &2004 &12&10 \rule[-1.5ex]{0pt}{0pt}\\
m2004-05& 03:37:46.0& $-35$:24:41.1 & 3900  & 0 \farcs 50& 31 &14  &2004 &12&11 \\
m2004-06& 03:37:46.0& $-35$:24:41.1 & 3900  & 0 \farcs 58& 25 &5  &2004 &12&11 \rule[-1.5ex]{0pt}{0pt}\\
m2004-07& 03:38:22.8& $-35$:34:59.2 & 3900  & 0 \farcs 58& 30 &18  &2004 &12&12 \\
m2004-08& 03:38:22.8& $-35$:34:59.2 & 3900  & 0 \farcs 70& 25 &5  &2004 &12&12 \\
\hline
\hline
\multicolumn{10}{l} {$(\star)$  \footnotesize{The masks m2003-03 and m2004-01 are identical; $(\dagger)$ m2003-04 and m2004-02 are identical. 
}} 
\\
\multicolumn{10}{l} {\footnotesize{See Fig.~\ref{fig:velcomp} for a comparison  of the velocity   measurements. }} \\
\end{tabular}

\label{tab:gmos}
\end{table*}

We used the Gemini Multi--Object Spectrograph (GMOS) on Gemini--South,
and the observations were carried out in queue mode in November 2003
and December 2004. A total of ten spectroscopic masks in five fields
were observed. Table\,\ref{tab:gmos} summarises the observations.  The
mask layout was defined using the GMOS Mask Design software.  Again,
we selected the GC candidates from the D+03 photometry. We chose slits
of $1\arcsec$ width and $5\arcsec$ length. We used the B600+\_G5323
grating, centred on $5500\,\rm{\AA}$, giving a resolution of
$0.9\,\rm{\AA}$ per (binned) pixel. The spectral resolution is
$\sim4.5\,\rm{\AA}$.

The GMOS field of view is $5.\!\arcmin5 \times 5.\!\arcmin5$, and the
detector array consists of three $2048\times4608$ CCDs arranged in a
row ($2\times2$ binning results in a pixel scale of
$0\,\farcs146\,\rm{pixel}^{-1}$).  Thus, with the chip gaps being
perpendicular to the dispersion direction, two gaps show up in the
spectra.  For each mask, the observations typically consisted of three
consecutive 1300\,sec exposures, which were bracketed by exposures of
a CuAr arc lamp and screen flat exposures.\par The data were reduced
using version~1.6 of the \texttt{gemini.gmos} {IRAF}--package in
conjunction with a number of customised scripts.  The two prominent
`bad columns' on the CCD--mosaic were corrected for using
\texttt{fixpix}, with the interpolation restricted to the dispersion
direction.  Cosmic ray (CR) rejection was done by combining the
science exposures using \texttt{gemcombine} with the CR rejection
option.  The wavelength calibration was performed using
\texttt{gswavelength}: Chebyshev polynomials of the $4^{\rm{th}}$ order
were used to fit the dispersion relation. The number of lines used in
these fits varied depending on the location of the slit, but typically
$\sim\!70$ lines were identified, and the residuals (r.m.s.)  were of
the order $0.15\rm{\AA}$.  We carefully inspected all calibration
spectra in order to exclude blended lines and lines in the proximity
of the chip gaps.  In the next step, the wavelength calibration was
applied to the science spectra using \texttt{gstransform}.  The sky
subtraction was done using the source--free regions of the slit, by
using the \texttt{gsskysub} task in interactive mode. The final
one--dimensional spectra were extracted using \texttt{gsextract}: The
apertures were typically 1\arcsec{} wide, and the tracing was done
using Chebyshev polynomials of the $4^{\rm{th}}$-- $8^{\rm{th}}$ order.

\section{The velocity data base}
\label{sect:database}
In this section, we detail how we build our velocity data base.  The
radial velocities are measured using Fourier--cross-correlation.
Coordinates, colours and magnitudes are taken from the D+03 Washington
photometry.  We anticipate here that we adopt $C\!-\!R=1.55$ to divide
blue (metal--poor) from red (metal--rich) GCs
(cf.~Sect.~\ref{sect:colour}), and the division between foreground
stars and bona--fide GCs is made at $\varv=450\,\rm{km\,s}^{-1}$
(cf.~Sect~\ref{sect:gcstars}).  \par To obtain a homogeneous data set,
we also re--measure the velocities for the spectra used in Paper\,I
(the velocities are tabulated in D+04).

\subsection{Radial velocity measurements}
The radial velocities are obtained using the IRAF--\texttt{fxcor}
task, which implements the Fourier cross--correlation technique by
\cite{TD79}. The templates (i.e.~reference spectra) are the
FORS\,2/MXU spectrum of NGC\,1396 $(\varv=815\pm8\,\textrm{km\,s}^{-1})$,
which was already used   by D+04,
and the spectrum of a bright GC in the NGC\,4636 GCS\footnote{This
spectrum is part of the dataset analysed in \cite{schuberth4636}.}.
The latter has a heliocentric velocity of $980\pm15\,\rm{km\,s}^{-1}$, its
colour is $C\!-\!R=1.62$ and the $R$--magnitude is $19.9$.  Since the
spectral resolution for both datasets is similar, we use these templates
for the FORS\,2/MXU as well as the GMOS data. \par 

The cross--correlation is performed on the 
wavelength interval  $4200 \lesssim \lambda \lesssim 5500\,\rm{\AA}$.
The upper
bound excludes sky--subtraction residuals from the most prominent
telluric emission line at $5577\,\rm{\AA}$, and the lower bound
ensures that we are well within the region for which the FORS\,2
wavelength calibration is reliable.  For the GMOS data, we did not
interpolate over the chip gaps. Hence these features are easily
identified and excluded from the spectral regions used for the
cross--correlation. \par Our spectral database contains velocities for
{1036} spectra where we could identify a clear peak in the
cross--correlation function (CCF). This number does not include obviously
redshifted background galaxies. \par 
For each spectrum, we adopt as velocity the \texttt{fxcor}
measurement with the highest value of the quality parameter
$\cal{R}_{\rm{TD}}$ (which is inversely proportional to the velocity
uncertainty $\Delta\varv$, see \citealt{TD79} for details).
For {{973}}
spectra, both templates yielded a velocity measurement. For the
remaining {63} spectra, only one of the templates returned a robust
result. 
\subsection{Velocity uncertainties of the GCs}
\label{sect:uncertainties}

\begin{figure}
\centering
\resizebox{\hsize}{!}
{\includegraphics[width=0.48\textwidth]{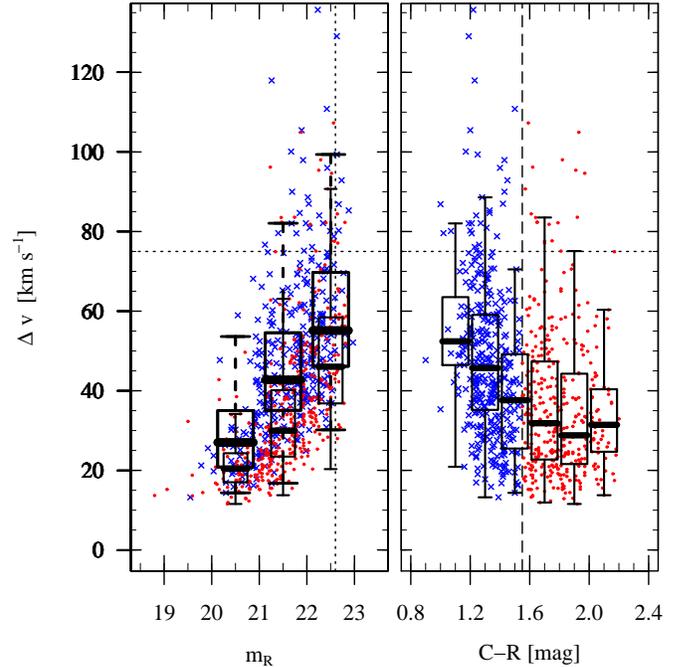}}
\caption{ Velocity uncertainties of the GC spectra as computed by
\texttt{fxcor}.  In both panels, crosses and dots represent blue and
red GCs, respectively. \emph{Left panel:}
\texttt{fxcor}--uncertainties versus \mbox{$R$--magnitude}.  Narrow
and wide box--plots show the data for red and blue GCs, respectively.
\emph{Right panel:} \texttt{fxcor}--uncertainties vs.~$C\!-\!R$ colour
overlaid with box--plots. The long--dashed line at $C\!-\!R=1.55$
shows the division between blue and red GCs. In both panels, the
short--dashed lines show the cuts used for assigning the quality flags
(cf.~Sect.~\ref{sect:qflag}). The boxes show the interquartile range
(IQR), with the band marking the median. The whiskers extend to 1.5
times the IQR or the outermost data point, if closer.}
\label{fig:uncertainties}
\end{figure}

The left panel of Fig.~\ref{fig:uncertainties} shows the velocity
uncertainties (as computed by \texttt{fxcor}) for the GCs as a function
of $R$--magnitude. As expected, the fainter GCs have larger velocity
uncertainties.  Red and blue GCs show the same trend, yet the offset
between the median values shows that the blue GCs, on average, have
larger velocity uncertainties.\par The right panel shows the
uncertainties versus $C\!-\!R$ colour. One indeed finds that the
uncertainties increase as the GCs become bluer.  While this might
partly be due to template mismatching, the paucity of absorption
features in the spectra of the metal--poor GCs by itself leads to
larger uncertainties. We compared the velocity measurements of some of
the bluest objects using the spectrum of a bright blue GC as template
and did not find any significant difference in the derived velocities
or uncertainties compared to the results obtained with the other
templates.

\subsection{Quality flags}
\label{sect:qflag}
Given that our spectroscopic targets have $R$--magnitudes in the range
$18.80 \leq m_R\leq 22.97 $, the accuracy of the velocity
determinations varies strongly, as illustrated in the left panel of
Fig.~\ref{fig:uncertainties}. With decreasing $S/N$, the risk of
confusing a prominent random peak of the CCF with the `true' peak of
the function increases. With the goal of weeding out spurious and
probably inaccurate velocity determinations, we therefore assign
quality flags (Class\,A or B) to the spectra: If at least one of the
criteria listed below is fulfilled, the velocity measurement is
regarded as `uncertain' and the corresponding spectrum is flagged as `Class\,B':
\begin{itemize}
\item{Only one template yields a velocity measurement.}
\item{Velocities measured with the two templates deviate by more than
$50\,\rm{km\,s}^{-1}$}
\item{Velocity uncertainty $\Delta\varv\geq 75\,\rm{km\,s}^{-1}$}
\item{Quality parameter ${\cal{R}}_{\rm{TD}} \le 4.5$}
\item{Relative height of the CCF peak $\rm{HGHT_{corr}} < 0.15$}
\item{Width of the CCF peak $\rm{FWHM_{corr}} \geq650\,\rm{km\,s}^{-1}$} 
\item{$R$--magnitude limit: $m_R \geq22.6$}  
\end{itemize}

\begin{figure}
\centering \resizebox{\hsize}{!}
{\includegraphics[width=0.48\textwidth]{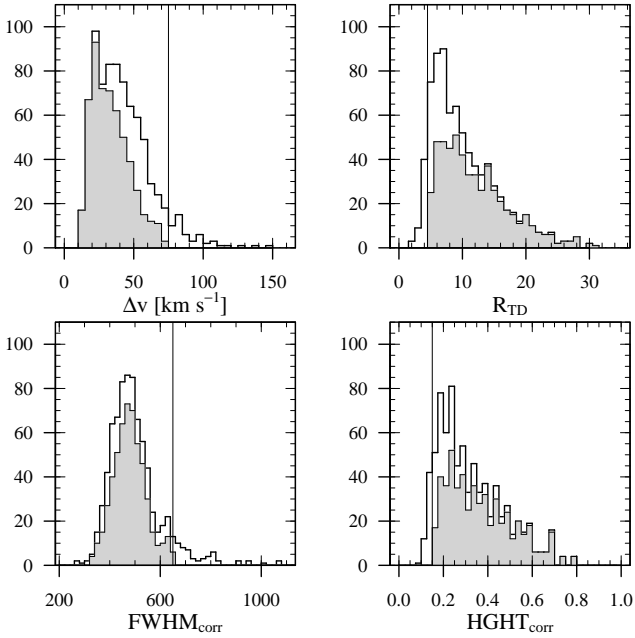}}
\caption{\texttt{fxcor} parameters and data classification. The panels
show the histograms for all (unfilled histograms) and the `Class\,A'
(grey histogram bars) GC spectra. \emph{Top left:} velocity
uncertainty, cut at $75\,\rm{km\,s}^{-1}$. \emph{Top right:}
$\cal{R}_{\rm{TD}}$, the cut is at $4.5$. \emph{Bottom left:} FWHM of
the fitted cross--correlation peak, the cut is at
$650\,\rm{km\,s}^{-1}$. \emph{Bottom right:} Relative height of the
cross--correlation peak, the cut is at $0.15$. In all panels, the
vertical line shows the corresponding cut.}
\label{fig:goodbad}
\end{figure}

Figure\,\ref{fig:goodbad} shows the distribution of the \texttt{fxcor}
 parameters for all GC spectra as
 unfilled histograms. In each sub--panel, the vertical line indicates the
 respective value defining the `Class\,B' (uncertain)
 measurements. The `Class\,A' spectra are shown as grey
 histograms. \par Assigning these quality flags to the spectra yields
 {723} Class\,A and {313} Class\,B measurements.

\subsection{The new spectra}
For the new data set, velocities were determined for {477}
spectra, {179 (139)} of which were obtained with GMOS, and
{298 (200)} with FORS\,2, where the numbers in brackets refer
to the Class\,A measurements. The slightly higher fraction of Class\,A
spectra found for the GMOS data is probably due to the different
treatment of the sky which, for the GMOS data, was subtracted prior to
the extraction.
\subsection{Re--measuring the spectra from Paper\,I}

\begin{figure}
\centering
\resizebox{\hsize}{!}
{\includegraphics[width=0.48\textwidth]{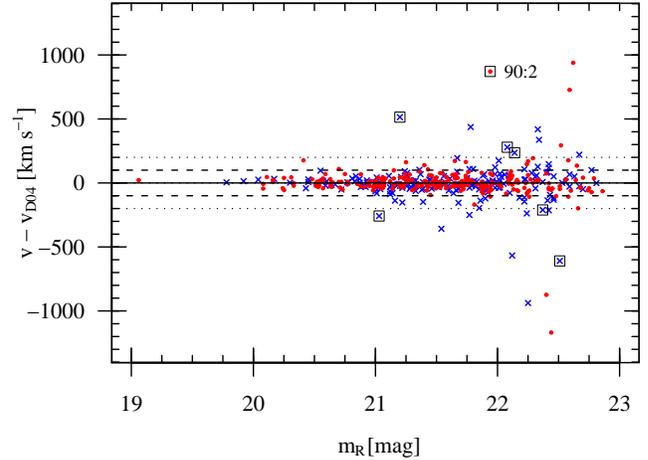}}
\caption{Comparison to the velocity measurements by \cite{dirsch04}.
Velocity difference vs.~$R$--magnitude. Dots and crosses represent red
and blue GCs, respectively. The dashed (dotted) lines are drawn at
$\pm100\,(\pm200)\,\rm{km\,s}^{-1}$.  The squares indicate the spectra
with a large ($>200\,\rm{km\,s}^{-1}$) deviations {but} a `Class\,A'
velocity determination according to the criteria given in
Sect.~\ref{sect:qflag}, {{and the label refers to the object 90:2 for which
a second spectrum exists}} (see text for details).}
\label{fig:R+04-TDR}

\end{figure}

Our re--analysis of the spectral data set from Paper\,I yielded 559
\texttt{fxcor} velocities (GCs and foreground stars).  Our database and
the D+04 catalogues have 503 (GCs and foreground stars) spectra in
common, and the results are compared below.

The GC catalogue by D+04 (their Table\,3) lists the
velocities\footnote{These authors give two velocities per spectrum,
$\varv_c$ is the value derived with \texttt{fxcor}, and $\varv_\ell$
refers to the direct line measurement (\texttt{rvidlines}).} for
{502} GC spectra, and the authors quote {468} as the
number of individual GCs. This number drops to {452} after
accounting for doubles overlooked in the published list.

We determine \texttt{fxcor} velocities for 455 (of the 502) spectra,
which belong to 415 individual GCs.  With a median $R$--magnitude of
$22.5$, the 47 spectra for which no unambiguous \texttt{fxcor}
velocity measurement could be achieved belong to the fainter objects
in our data set.

\par The
difference between our measurements and the values presented in
Table\,3 of D+04 are plotted against the $R$--magnitude in the upper
panel of Fig.~\ref{fig:R+04-TDR}. For most spectra, the agreement is
very good, but a couple of objects show disturbingly large
discrepancies.  For faint objects, deviations of the order several
hundred $\textrm{km\,s}^{-1}$ are possibly due to multiple peaks in
the CCF which occur in low $S/N$--spectra. 

\par However, we also find very large ($>200,\textrm{km\,s}^{-1}$)
differences for seven `Class\,A' spectra (marked with squares in
Fig.~\ref{fig:R+04-TDR}).  Of these GCs, one is also present on a
second mask: object 90:2 ({{the labelled object at}}  $\varv=1688\pm31$,
$\varv_{D04}=817\pm27\,\textrm{km\,s}^{-1}$) was observed with GMOS
(spectrum GS04-M07:171, $\varv=1710\pm34\,\textrm{km\,s}^{-1}$), thus
confirming our new measurement. The remaining six spectra (in order of
decreasing brightness: 86:19, 75:9, 90:2, 77:84, 78:102, 81:5, 75:24,
and 86:114), and the spectrum 81:55 (for which no photometry is
available) are re--classified as `Class\,B'.  \par Since, for three of
the very discrepant spectra (75:9, 78:102, and 90:2), the
line--measurements ($\varv_\ell$) by D+04 lie within just
$75\,\textrm{km\,s}^{-1}$ of our new  values, we suspect that, in some
cases, typographical errors in the published catalogue might be the
cause of the deviations.

\par Table\,4 in D+04 lists the velocities for {{72}}
spectra\footnote{the entry for 80:99 is a duplicate of 80:100. The
object in slit 80:99 is a GC listed in Table\,3 of D+04.}  of
foreground stars with velocities in the range $-350 \leq \varv_c
\leq320\, \textrm{km\,s}^{-1}$. Our data base contains measurements
for {48} of these spectra ({44} objects) .  The overall
agreement is good, but objects 77:6 and 91:82 show large
$>1000\,\textrm{km\,s}^{-1}$ deviations and are re-classified as GCs
(spectra of Class\,B). For the remaining spectra, the velocity
differences are of the order of the uncertainties. \par Finally, our
database contains velocities for {56 (55 objects) } spectra
({eleven} are foreground stars and {45 (44)} GCs) that
do not appear in the lists of D+04.\par
The reason for this discrepancy is unknown.
\subsection{Duplicate measurements}

In this section, we use the duplicate measurements to assess the
quality and robustness of our velocity determinations.  First, we
compare the velocities obtained when exposing the same mask on two
different occasions. Secondly, we have objects which are present on
more than one spectroscopic mask (but the instrument is the same).
Then, we compare the results for objects observed with both FORS\,2
and GMOS.  Finally, we compare common objects to values found in the
literature where a different instrument (FLAMES) was used.
\subsubsection{Double exposures of GMOS masks:}
\begin{figure*}
\centering

{\includegraphics[width=0.48\textwidth]{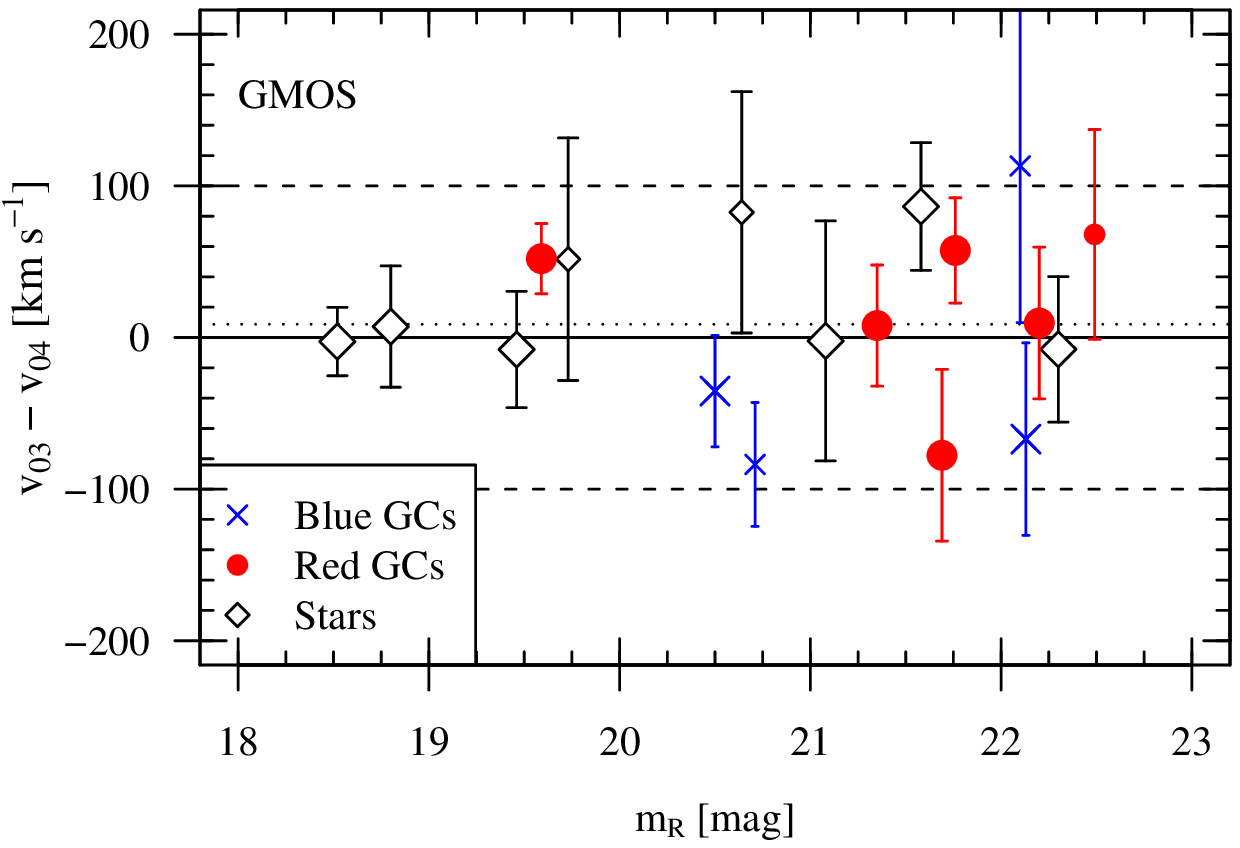}}
{\includegraphics[width=0.48\textwidth]{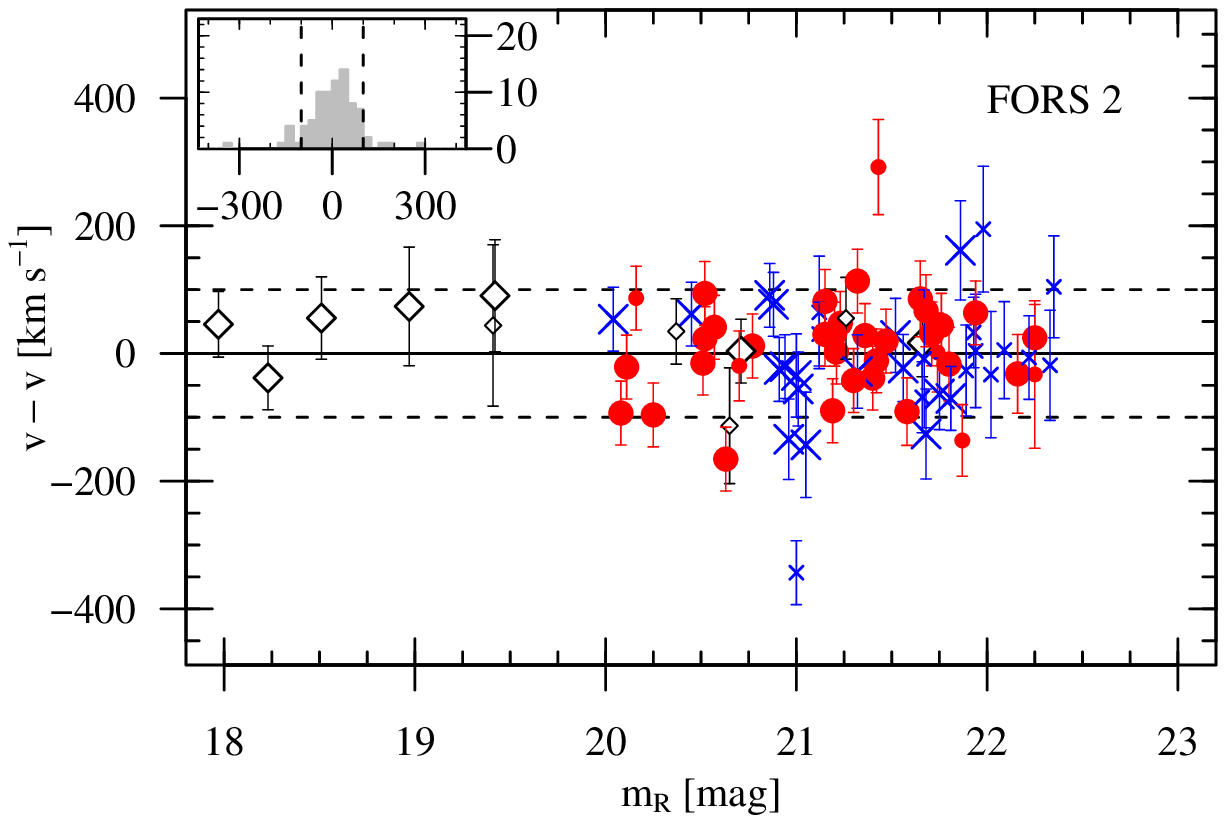}}\\
{\includegraphics[width=0.48\textwidth]{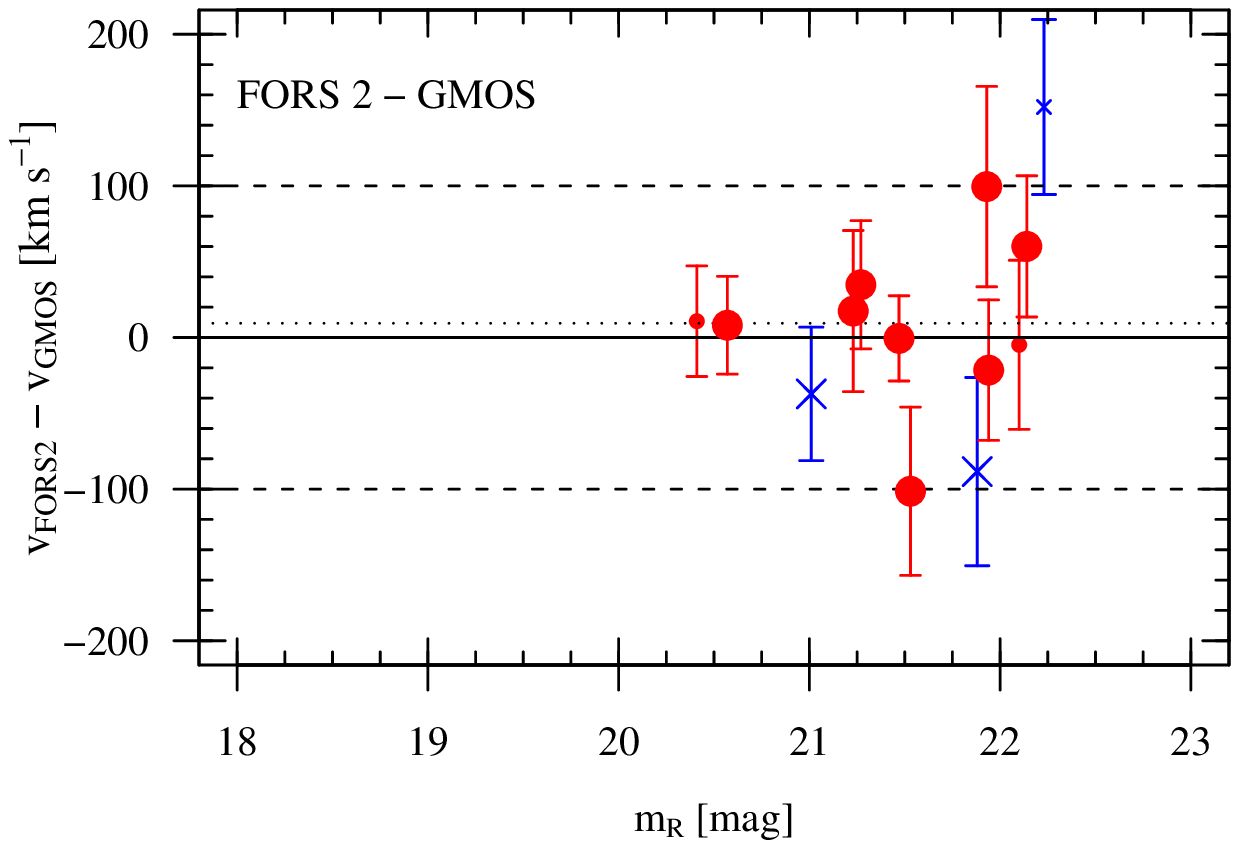}}
{\includegraphics[width=0.48\textwidth]{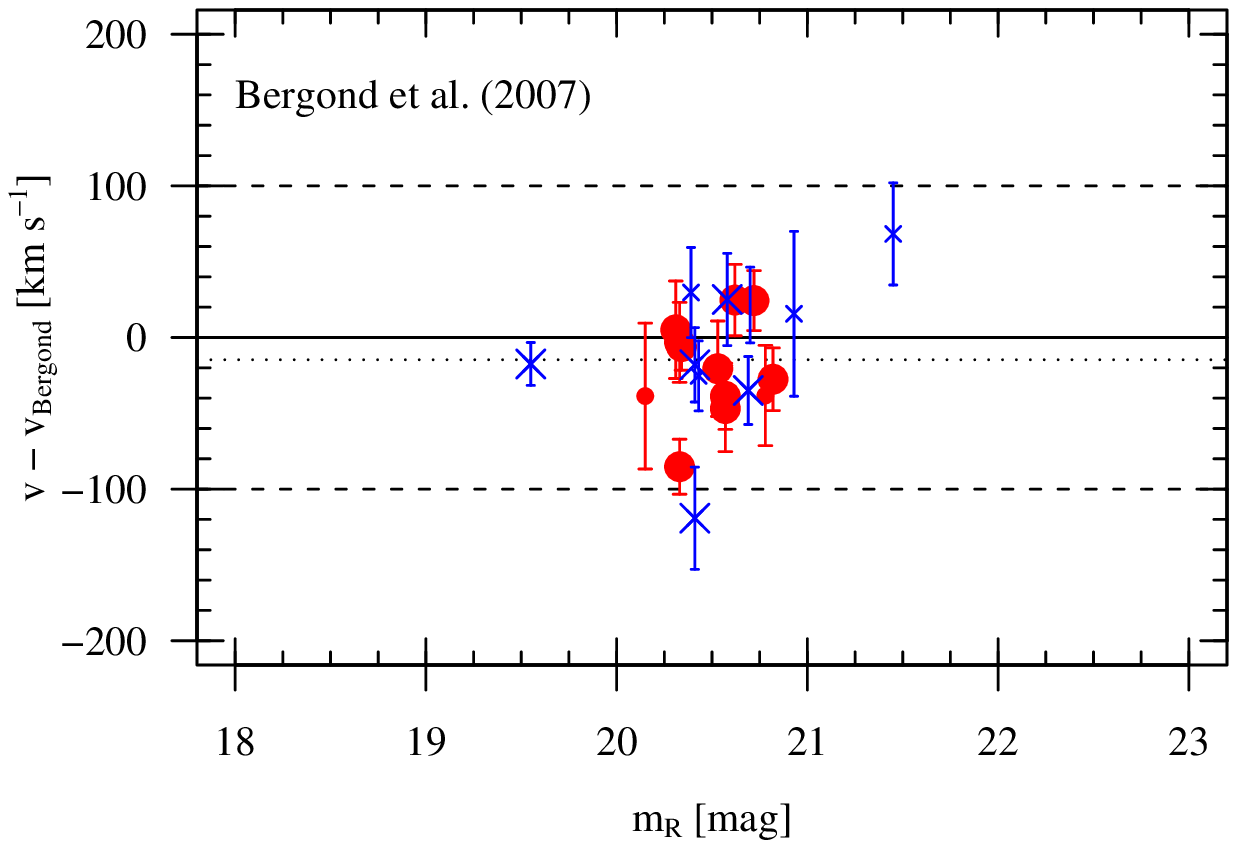}}
\caption{Duplicate measurements: Velocity differences vs.~apparent
magnitude. Crosses and dots are blue and red GCs,
respectively. Foreground stars are shown as diamonds. The error--bars
are the uncertainties of the two velocity measurements added in
quadrature. Small symbols indicate objects where at least one of the
spectra is classified as `Class\,B'.  In all panels, the dashed lines
are drawn at $\pm100\,\rm{km\,s}^{-1}$. \emph{Upper left:} The GMOS
observations labelled in~Table\,\ref{tab:gmos} ($\star,\dagger$) were
carried out using the same masks.  The scatter
($\rm{r.m.s.}\approx55\,\rm{km\,s}^{-1}$) is of the order of the
velocity uncertainties, and the offset, which is
$\sim\!10\,\rm{km\,s}^{-1}$ (dotted line), is very small.\emph{Upper
right:} FORS\,2 data comparison: Difference between the velocities
measured for common objects on different masks.  The inset shows the
histogram of the deviations, which follow a normal distribution with
$\sigma \simeq 70\,\rm{km\,s}^{-1}$.  The two outliers with deviations
of $\approx300\,\rm{km\,s}^{-1}$ are re-classified as Class\,B (see
text for details). \emph{Bottom left:} The same for objects measured
using both GMOS and VLT. The offset is small
($\sim\!10\,\rm{km\,s}^{-1}$), and the r.m.s.~is
$\approx70\,\rm{km\,s}^{-1}$.  \emph{Bottom right:} Comparison to the
FLAMES velocity measurements by \cite{bergond07}. The error--bars show
the uncertainties of our and their velocity measurements added in
quadrature. The scatter is small ($\rm{r.m.s.}=41\,\rm{km\,s}^{-1}$),
and the offset, which is $-15\,\rm{km\,s}^{-1}$ (dotted line), is
negligible.}
\label{fig:velcomp}
\end{figure*}

 For the GMOS dataset, two masks (marked by an asterisk/dagger in
Table\,\ref{tab:gmos}) were exposed during both observing campaigns.
As can be seen from the upper left panel of Fig.~\ref{fig:velcomp} where
we plot the velocity differences versus the $R$--magnitude, the
velocities agree fairly well within the uncertainties.  The offset of
$10\,\textrm{km\,s}^{-1}$ is negligible and the r.m.s.~is about $55\,\textrm{km\,s}^{-1}$.

\subsubsection{Objects observed on different  FORS\,2 masks }
The upper right panel of Fig.~\ref{fig:velcomp} compares the velocities measured for
{82} objects present on different FORS\,2 masks. The agreement
is good, and the differences are compatible with the velocity
uncertainties.  The two outliers, GCs with deviations of the order
$300\,\textrm{km\,s}^{-1}$ (objects 89:92$=$90:94, $m_R=21.43$\,mag, and
81:8$=$82:22, $m_R=21.0$\,mag) are both from the data set analysed in
Paper\,I. Also in Table\,3 of D+04, the correlation velocities of
these objects differ significantly (by $153$ and
$369\,\textrm{km\,s}^{-1}$, respectively). We therefore assign these
objects (which nominally have `Class\,A' spectra) to Class\,B.

\subsubsection{GMOS and FORS\,2 spectra}
There are {15} objects which were measured with both FORS\,2
and GMOS. All of them are GCs, and for all but two photometry is
available. The bottom left panel of Fig.~\ref{fig:velcomp} shows the
velocity differences against the $R$--magnitude. The offset of
$\approx10\,\textrm{km\,s}^{-1}$ is negligible, and the r.m.s.~is
$\approx70\,\textrm{km\,s}^{-1}$.
\subsubsection{The  measurements by Bergond et al.~(2007)}
Bergond et al.~(\citeyear{bergond07}, B+07 hereafter) used the FLAMES
fibre--spectrograph on the VLT to obtain very accurate ($\Delta \varv
\!\sim\!10\,\rm{km\,s}^{-1}$) velocities for 149 bright GCs in the
Fornax cluster. Of these objects, 24 (21 of which have Washington
photometry) were also targeted in this study, and the velocities are
compared in the bottom right panel of Fig.~\ref{fig:velcomp}. The
offset is $\approx15\,\rm{km\,s}^{-1}$, and the r.m.s.~is
$\approx40\,\rm{km\,s}^{-1}$. With the exception of two outliers
(9:71=gc216.7, 9:43=gc.154.7), the agreement is excellent. {{The
reason for the deviation of these two objects remains unknown.}}

\subsubsection{Accuracy and final velocities}
The repeat measurements of two GMOS masks shows that our results are
reproducible.  The absence of systematic differences/offsets between
the different spectrographs indicates that the instrumental effects
are small. \par For the GCs for which duplicate measurements exist, we
list as final velocity the mean of the respective Class\,A
measurements (using the $\cal{R}_{\rm{TD}}$ values as weights).  In
case all spectra were classified as Class\,B, the weighted mean of
these velocities is used.

\subsection{Separating GCs from foreground stars}
\label{sect:gcstars}
\begin{figure}
\centering
\resizebox{\hsize}{!}
{\includegraphics[width=0.48\textwidth]{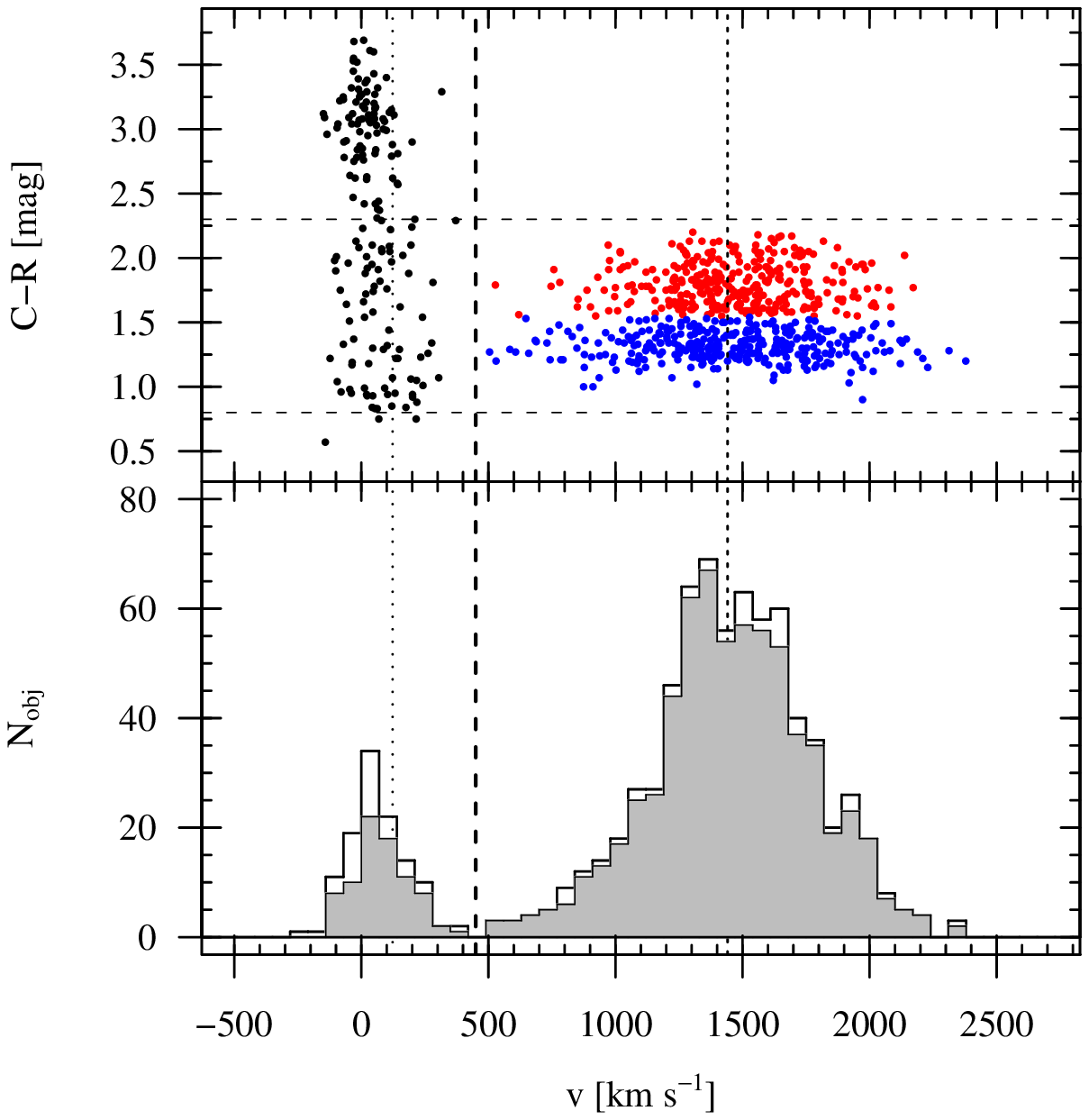}}
\caption{Separating GCs from foreground stars: \emph{Upper panel}:
$C\!-\!R$ colour vs.~heliocentric velocity for objects with velocities
in the range $-500 < \varv < 3000\,\rm{km\,s}^{-1}$. The dashed lines
at $C\!-\!R = 0.8$ and $2.3$ show the colour interval D+03 used to
define GC candidates.  \emph{Lower panel}: The grey histogram shows
the velocities of the objects within the above colour interval. The
unfilled histogram also includes those objects for which no photometry
is available.  In both plots, the short--dashed vertical line at
$1441\,\rm{km\,s}^{-1}$ indicates the systemic velocity of NGC\,1399.
The Galactic foreground stars have a mean velocity of about
$50\,\rm{km\,s}^{-1}$ and exhibit a larger range in colours. The
dotted line shows the GSR velocity vector in the direction of Fornax
($124\,\rm{km\,s}^{-1}$).  The dashed line at $450\,\rm{km\,s}^{-1}$
indicates the limit adopted to separate GCs from foreground stars.}
\label{fig:sampledefine}
\end{figure}

To separate GCs from foreground stars we plot, in the upper panel of
Fig.~\ref{fig:sampledefine}, colours versus heliocentric velocity.
The data points fall into two regions: The highest concentration of
objects is found near the systemic velocity of NGC\,1399 ($1441\,\pm 9
\rm{km\,s}^{-1}$,  Paper\,I). These are the GCs, and we
note that all of them have colours well within the interval used by
D+03 to identify GC candidates (horizontal dashed lines). The second
group of objects, galactic foreground stars, is concentrated towards
zero velocity and occupies a much larger colour range
\footnote{The data set includes foreground stars which do not fulfil
the colour criteria for GC candidates because we extracted all spectra
of a given mask, including those from the `positioning slits' used for
the MXU mask alignment.}. 
The velocity histograms shown in the lower
panel of Fig.~\ref{fig:sampledefine} illustrate that the total sample,
including those objects for which no photometry is available (unfilled
histogram), exhibits the same velocity structure as the one found for the
objects with MOSAIC photometry (grey histogram).
Most importantly, the domains of GCs and foreground stars are
separated by a gap of $\sim\!100\,\rm{km\,s}^{-1}$.  Guided by
Fig.~\ref{fig:sampledefine}, we therefore regard all objects within
the velocity range $450 < \varv_{\rm{helio}} < 2500\,\rm{km\,s}^{-1}$
as bona fide NGC\,1399 GCs.

\subsection{The final velocity catalogue}

We determined \texttt{fxcor} velocities for a total of {{1036}}
spectra.  {{These objects are foreground stars and GCs. Background
galaxies were discarded at an earlier stage of the data analysis, and
there were no ambiguous cases, since there is a substantial velocity
gap behind the Fornax cluster \citep{drinkwaterdwarfs}}}.  \par Our
final velocity catalogue comprises {{908}} unique objects, {{830}} of
which have MOSAIC photometry. This database contains {693\,(656 with
photometry)} GCs, { 210\,(174 with photometry)} foreground stars and
five Fornax galaxies (NGC\,1404, NGC\,1396, FCC\,208, FCC\,222, and
FCC\,1241).  Of the GCs, {471} have velocities classified as Class\,A,
the remaining {222} have Class\,B measurements.  The velocities for
the GCs and the foreground stars are available in electronic form only.  
For each spectrum, these tables give the coordinates,
the \texttt{fxcor}--velocity measurement, the
$\mathcal{R}_{\mathrm{TD}}$--parameter, and the quality flag. The
final velocities and the cross--identifications are also given.

\section{Properties of the globular cluster sample}
\label{sect:sample}
\subsection{Colour and luminosity distributions}
\label{sect:colour}

\begin{figure*}
\centering

{\includegraphics[width=0.48\textwidth]{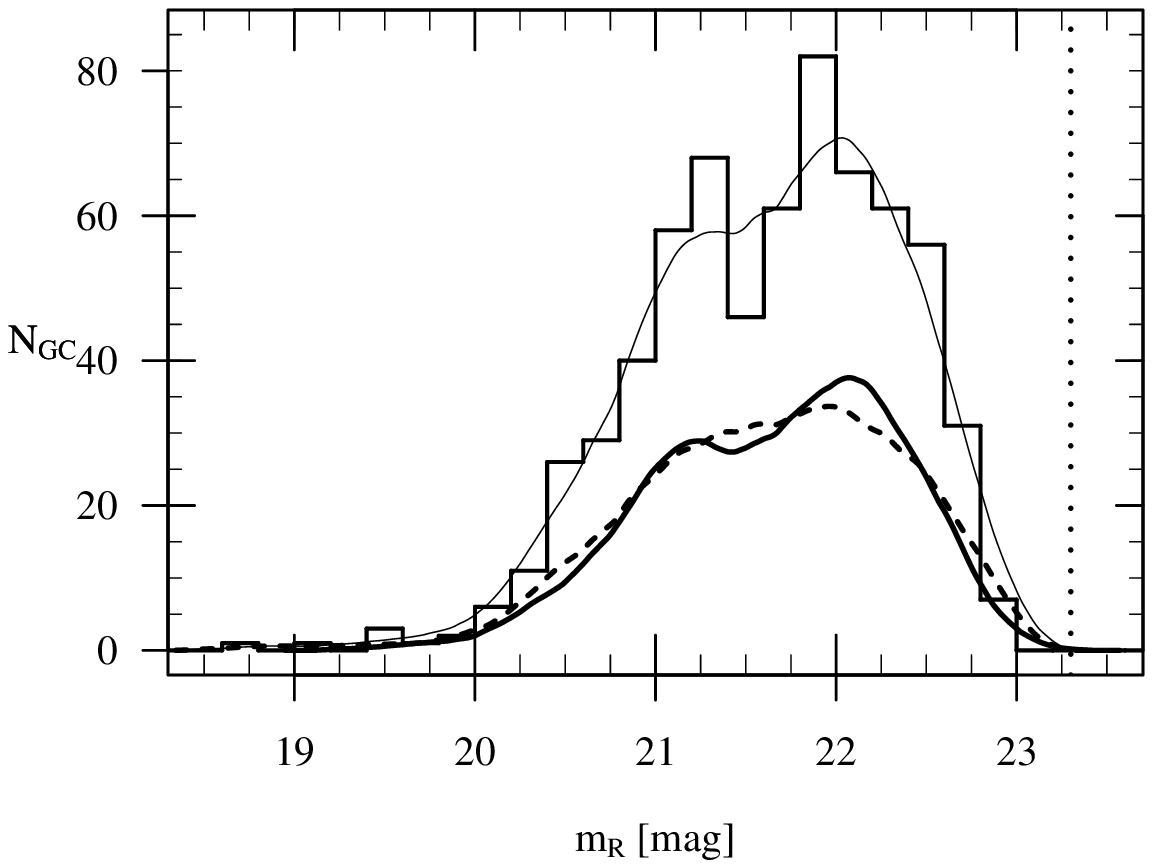}}
{\includegraphics[width=0.48\textwidth]{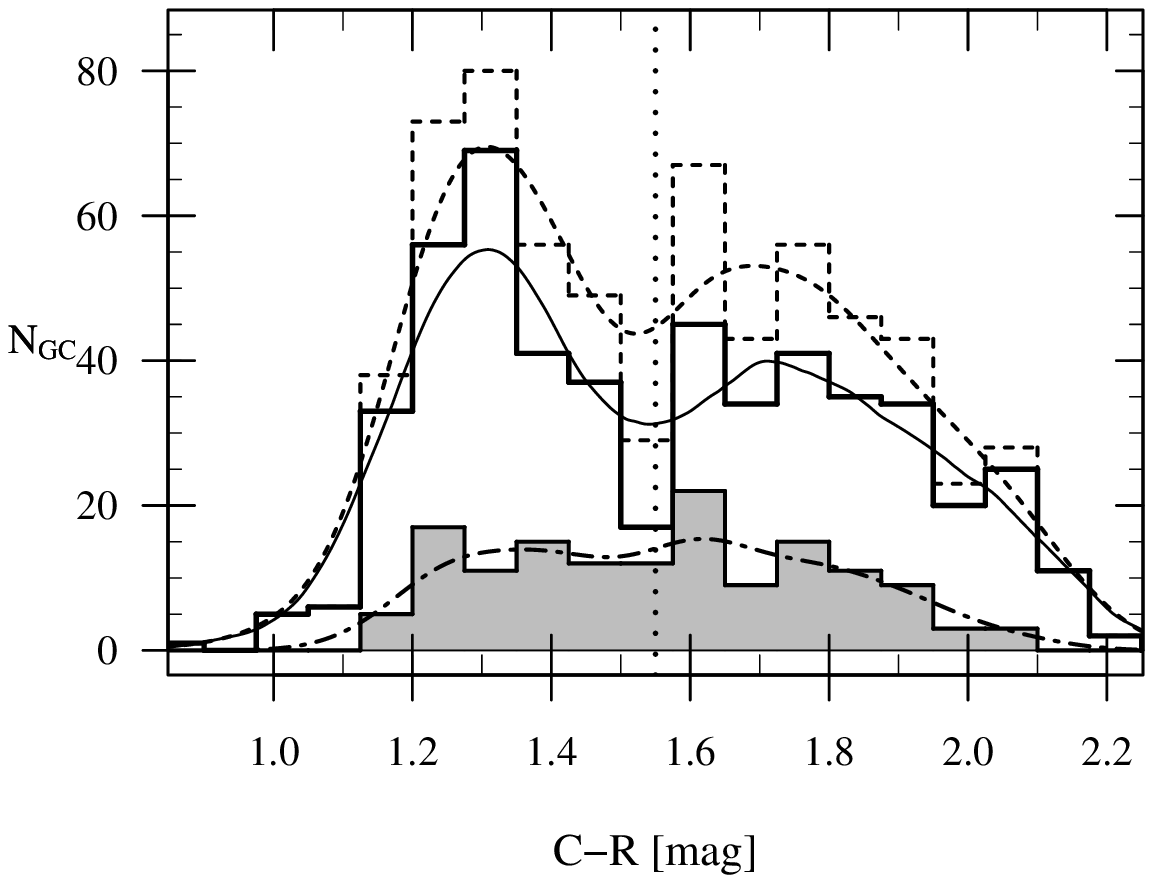}}
{\includegraphics[width=0.48\textwidth]{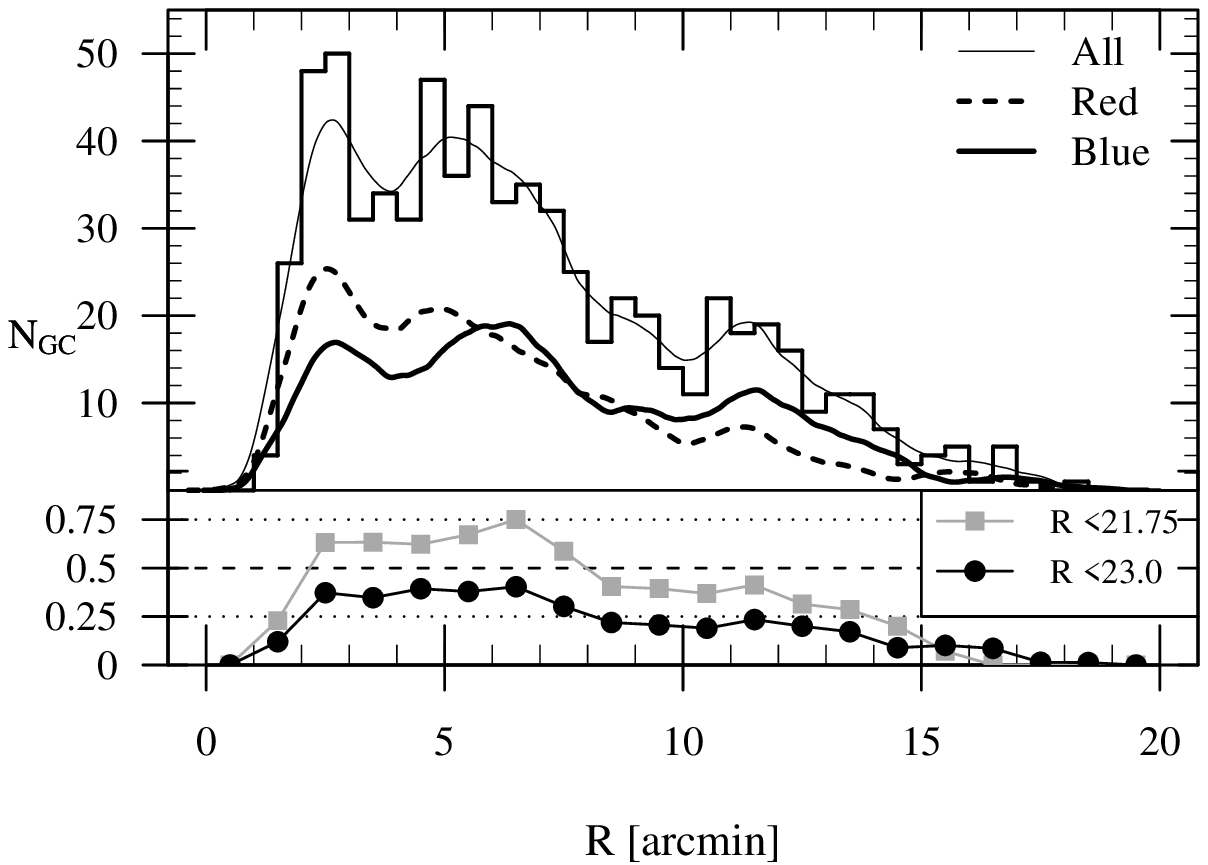}}
{\includegraphics[width=0.48\textwidth]{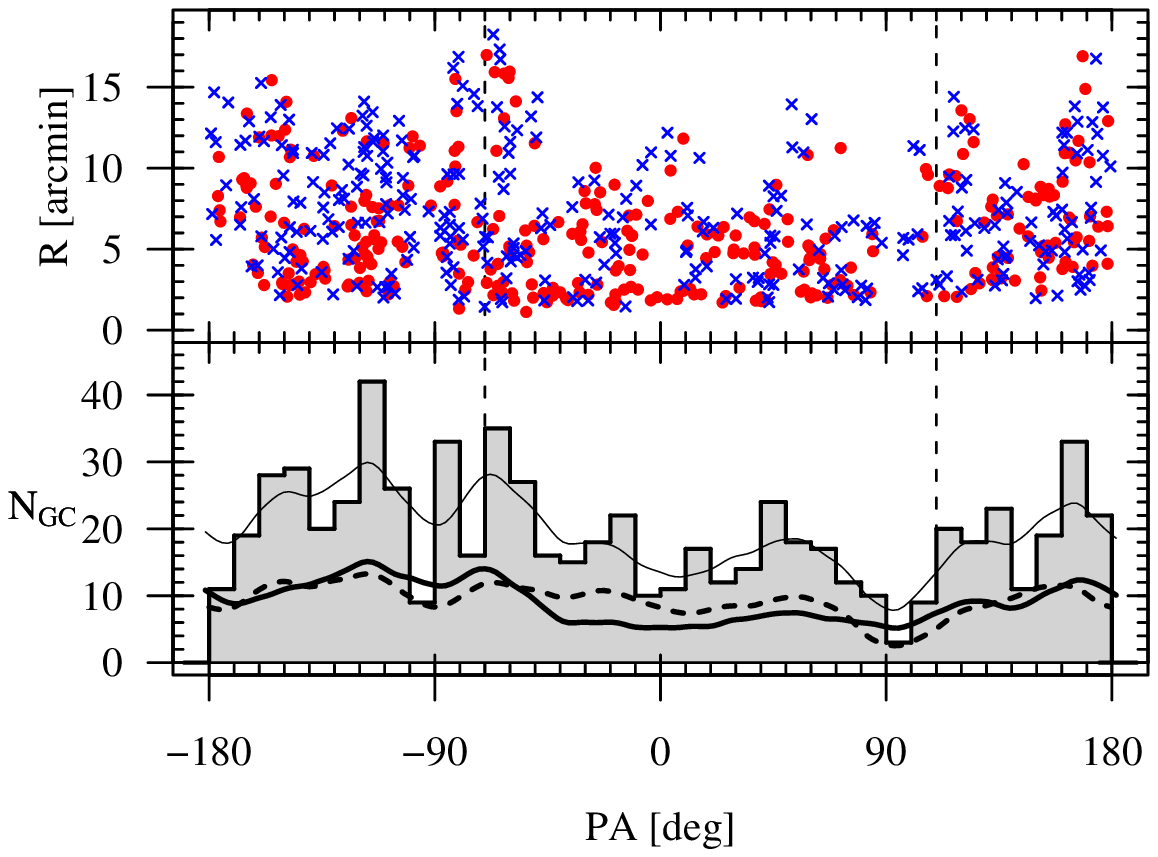}}
\caption{NGC\,1399 spectroscopic GC sample: Photometric properties and
spatial distribution.  \emph{Upper left}: GC luminosity distribution
(bin width = 0.2 mag). The dotted line at $m_R=23.3$ indicates the
turn--over magnitude of the GCS. The thin solid, thick solid and
dashed lines show the kernel density estimates for all, the blue, and
the red GCs, respectively.  \emph{Upper right panel:} Colour
distribution. The dashed histogram shows the distribution of all
{{656}} velocity--confirmed GCs for which MOSAIC photometry is
available, and the solid (unfilled) histogram are the GCs fainter than
$m_R=21.1$, and the grey histogram shows the distribution of the {{144}}
brightest ($m_R < 21.1$) GCs.
The dashed, solid and dot--dashed curves shows the respective kernel
density estimate for the same data; a bandwidth of $0.075$\,mag was used
(same as histogram bins).   
The dotted line at $C\!-\!R=1.55$ indicates the limit
dividing blue from red GCs.
\newline
\emph{Lower left:} Radial distribution. The histogram (\emph{upper
sub--panel}) shows all {{693}} GCs with velocity measurements. The
thin solid, thick solid and dashed lines show the kernel density
estimates for all, the blue and the red GCs, respectively. The radial
completeness, i.e.~the number of GCs with velocity measurements with
respect to the number of GC candidates ($0.9<C\!-\!R<2.2$) from the
D$+03$ photometry is shown in the \emph{lower sub--panel}. The black
dots (grey squares) show the values for a faint--end magnitude limit
of 23.0 (22.75).  \emph{Lower right:} Azimuthal distribution of the
GCs.  The position angle (PA) is measured North over East, and the
dashed vertical lines indicate the photometric major axis of NGC\,1399
($110^\circ$). The \emph{upper sub--panel} shows the radial distance
from NGC\,1399 vs.~PA. Crosses and dots represent blue and red GCs,
respectively.  The histogram (\emph{lower sub--panel}) has a bin width
of $10^{\circ}$. The thin solid, thick solid and dashed lines show
the kernel density estimates for all, the blue and the red GCs,
respectively.}
\label{fig:colhist}
\end{figure*}

The luminosity distribution of our GC sample is plotted in the upper
left panel of Fig.\,\ref{fig:colhist}. The turn-over magnitude (TOM)
of $m_R=23.3$ (D+03) is shown for reference. It illustrates that our
spectroscopic study only probes the bright part of the globular
cluster luminosity function (GCLF).  The distributions for red and
blue GCs, shown as kernel density estimates (e.g.~\citealt{kerndens}),
are very similar. The median $R$--magnitude of the GCs is
{$m_R=21.75$}. The brightest (faintest) cluster has a magnitude of
{{$m_R=18.8$ (22.97)}}. \par

Can our spectroscopic dataset, which contains {{656}} GCs with
known colours, be regarded as a photometrically representative
subsample of the NGC\,1399 GCS?  As mentioned in
Sect.~\ref{sect:MOSAIC}, the colours and magnitudes of our GC sample
are taken from the D+03 photometric study. These authors found a
bimodal colour distribution for GCs in the magnitude range
\mbox{$21 <  m_R < 23$}. The brightest GCs, however, were discovered
to have a unimodal distribution, peaking at an intermediate colour of
$C\!-\!R \simeq 1.55$.\par Figure\,\ref{fig:colhist} (upper right
panel) shows the colour distribution of our sample. The main features
described by D+03 are also found for the spectroscopic data set: The
distribution is clearly bimodal, and the blue GCs show a peak near
$C\!-\!R \simeq 1.3$.  Following D+03 (and Paper\,I), we adopt
$C\!-\!R = 1.55$ as the colour dividing blue from red GCs. 

\par Further, the brightest clusters ($m_R\leq 21.1$, grey histogram) do
not seem to follow a bimodal distribution. 

{Given that the
brightest GCs show no signs of colour bimodality, the division of
these GCs into `blue' and `red' is be somewhat
{arbitrary/artificial}. As will be shown in
Sect.~\ref{sect:sampledefine}, the brightest  GCs 
form indeed   a kinematically distinct subgroup.}

\subsection{Spatial distribution}
The lower left panel of Fig.~\ref{fig:colhist} shows the radial
distribution of the GCs with velocity measurements. Within the central
5\arcmin there are more red than blue GCs, which is a
consequence of the steeper number--density profile of the former (see
Sect.~\ref{sect:powerlaws}).  At large radii, there are slightly more
blue than red GCs.  The median (projected) distance from NGC\,1399 is
$5.\!\arcmin9$, $6.\!\arcmin5$, and $5.\!\arcmin4$ for all, the blue 
and the red GCs, respectively. The galactocentric distances of the GCs
lie in the range $1\farcm1 \leq R \leq 18\farcm3$, i.e. $6 \lesssim R
\lesssim 100\,\rm{kpc}$.  
For comparison,
the data set   analysed in Paper\,I  covered
the range \mbox{$1\farcm1\leq R\leq 9\farcm2$}. 
The lower sub--panel 
plots an estimate for the radial completeness of our spectroscopic
sample: Dots and squares show the number of GCs with velocity
measurements divided by the number of GC candidates from the D$+03$
photometric catalogue for two different faint--end magnitude
limits. Considering the brighter half of our GC sample (grey
squares), the completeness lies above 50 per cent for radii between
2\arcmin{} and 8\arcmin{}, and at about 14\arcmin{}, it drops below 25
per cent.  \par

The bottom right panel in Fig.~\ref{fig:colhist} shows the azimuthal
distribution of the GCs (the position angle (PA) is measured North
through East).  The upper sub--panel plots radial distance versus PA,
and the azimuthal completeness decreases drastically beyond $\sim
8\arcmin$. The lower sub--panel shows the histogram of the azimuthal
distribution. 
The paucity of GCs around
$-90^{\circ}, 0^{\circ}, 90^{\circ}$ and $180^{\circ}$ results from
the choice of mask positions on the plane of the sky
(cf.~Fig.~\ref{fig:map}). 

\par

\section{Sample definition and interloper Removal}
\label{sect:interlopers}
\begin{figure}
\centering
\resizebox{\hsize}{!}
{\includegraphics[width=0.48\textwidth]{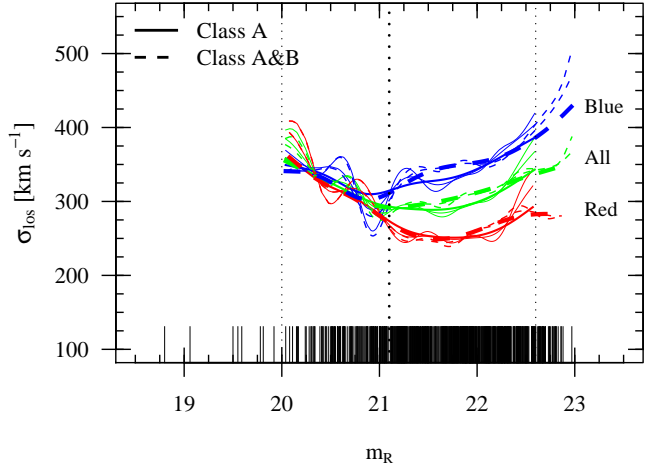}}
\caption{Velocity dispersion as function of $R$--magnitude.  For all
GCs with $m_R> 20$, the dispersion is calculated using Gaussian kernels
with a width of $\sigma_{\mathrm{mag}} \in
\{0.15,0.2,0.3\}\,\textrm{mag}$. Dashed and solid curves show the
values for the full and the `Class\,A' samples, respectively. The
thick lines are the curves for $\sigma_{\mathrm{mag}}=0.3$.  The
dotted line at $22.6\,\textrm{mag}$ indicates the magnitude limit for
Class\,A spectra.  The dashed line at $21.1\,\textrm{mag}$ shows the
division  we adopt between bright and faint GCs. The ticks at the bottom
of the panel show the $R$--magnitudes of the GCs.}
\label{fig:GaussRmag}
\end{figure}
\subsection{GC subpopulations}
\label{sect:subpop}
The principal division is the distinction between red and blue GCs,
which, as shown in Paper\,I and D+03, respectively, behave differently
with regard to kinematics and spatial distribution. However, one also
has to consider that the brightest GCs seem to have a unimodal colour
distribution (cf.~Sect\,\ref{sect:colour}, D+03). To see how the
kinematic properties of the GC populations change with the luminosity
we plot in Fig.~\ref{fig:GaussRmag} the velocity dispersion as
 a function of $R$--magnitude. We divide the spectroscopic GC sample into
blue and red GCs and calculate the line--of--sight velocity dispersion
as a function of $R$--magnitude using a Gaussian window function (note
that GCs brighter than $m_R=20.0$ are omitted because of their sparse
spacing along the $x$-axis). For all three kernels
($\sigma_{\rm{mag}}=0.15, 0.2, 0.3$), the results are similar: For GCs
with $ 20 \lesssim m_R \lesssim 21$, the dispersions of `blue' and `red'
GCs are indistinguishable. For fainter GCs, down to about $m_R\simeq
22$, the blue and red GCs become well separated, and the respective
dispersions do not appear to depend on the magnitude. For GCs fainter
than $m_R\simeq22$, however, the dispersions increase towards fainter
magnitudes -- probably a result of the larger velocity uncertainties
(see Sect.~\ref{sect:uncertainties}). The quality selection (i.e.~the
inclusion of Class\,B measurements, shown as dashed curves in
Fig.~\ref{fig:GaussRmag}), does not have any impact on the detected
features.\par Guided by Fig.~\ref{fig:GaussRmag}, we define the
following samples for the dynamical analysis:
\begin{itemize}
\item{ B(lue)   and $m_{R}\geq 21.1$ and 
$C\!-\!R\leq 1.55$  (256 GCs)}

\item{  R(ed) and $m_{R}\geq 21.1$ and 
$C\!-\!R > 1.55$  (256 GCs) }

\item{  F(aint)  and $m_{R}\geq 21.1$  
(512 GCs) }

\item{ BR(right)  and $m_{R}< 21.1$  (144 GCs)}
\item{The full sample contains A(ll) 693 GCs with radial velocity
measurements}
\end{itemize}
The main quantity we extract from our dataset is the line--of--sight
velocity dispersion $\sigma_{\rm{los}}$. Since dispersion measurements
react very severely to the presence of outliers and sampling
characteristics, it is important to remove possible contaminants. \par
As can already be seen from Fig.~\ref{fig:map}, our sample probably
contains a number of GCs belonging to NGC\,1404
($\varv_{\rm{helio}}=1947\,\rm{km\,s}^{-1}$). Our approach to identify
these objects is detailed in Sect.~\ref{sect:1404inter}.\par
\begin{figure*}
\centering
{\includegraphics[width=0.48\textwidth]{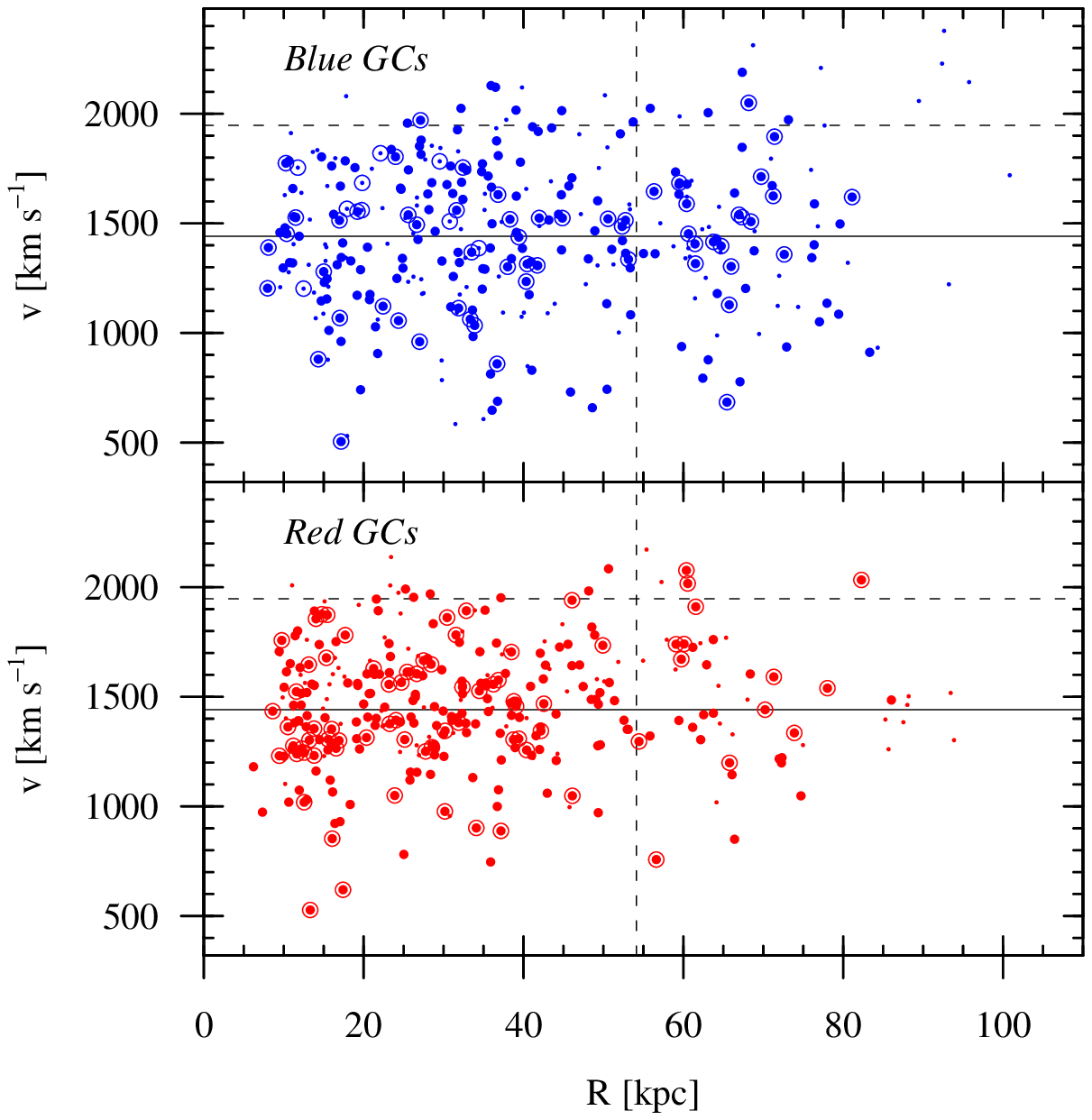}}
{\includegraphics[width=0.48\textwidth]{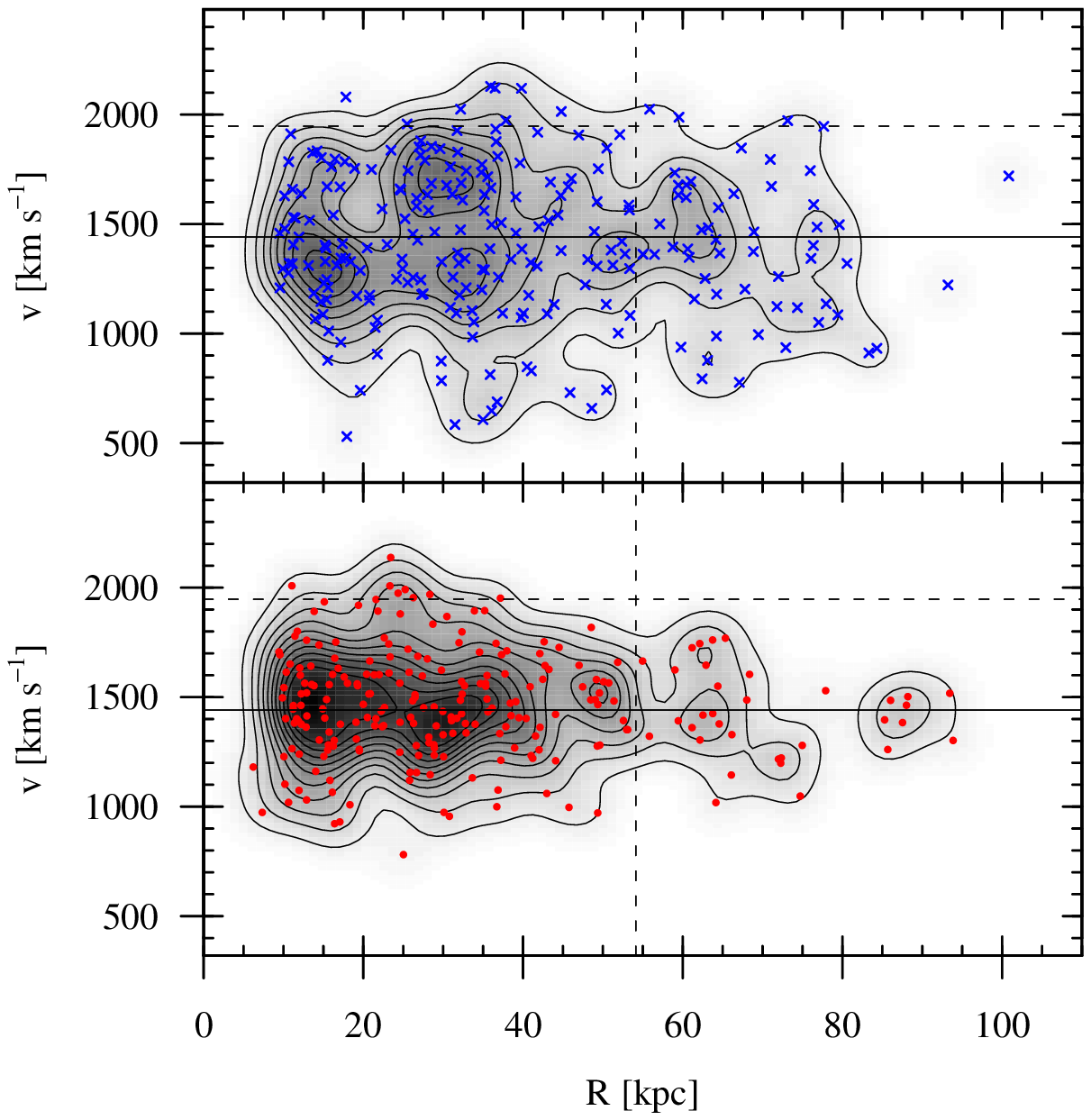}}
\caption{NGC\,1399 GC velocities. \emph{Left panel:} Velocities
  vs.~galactocentric distance for the blue and red GC sample (shown in
  the upper and lower sub--panel, respectively). Large and small dots
  show GCs with `Class\,A' and `B' velocity measurements,
  respectively.  Objects brighter than $R=21.1$ are marked by circles.
  The dashed line at 54\,kpc shows the projected distance of
  NGC\,1404. The solid and dashed horizontal lines indicate the
  systemic velocities of NGC\,1399 and NGC\,1404, respectively.
  \emph{Right:} Densities for the data points for the blue and red
  samples (\textsc{Biii} and \textsc{Riii}
  cf.~\ref{sect:sampledefine}) shown as grey--scale and contour
  plot. The blue subsample appears to exhibit substructure which is
  not present in the red.}

\label{fig:sampleall}
\end{figure*}

 Secondly,
the presence of GCs with high relative velocities at large
galactocentric radii (see Fig.~\ref{fig:sampleall}, left panel)
potentially has a large impact on the derived mass profiles. The
treatment of these GCs with extreme velocities is discussed in
Sect.~\ref{sect:extremeint}. \par In Sect.~\ref{sect:sampledefine}, we
label the subsamples obtained from the samples defined above after the
interloper removal.
 
\subsection{NGC\,1404 GC interloper removal}
\label{sect:1404inter}
In the sky, NGC\,1404 is the closest (giant) neighbour of NGC\,1399.
It lies $9.\!\arcmin8$ southeast of NGC\,1399, which corresponds to a
projected distance of only 54\,kpc. Its systemic velocity is
$1947\,\rm{km\,s}^{-1}$ (NED\footnote{NASA/IPAC Extragalactic Database
\newline \texttt{http://nedwww.ipac.caltech.edu}}), which is inside
the velocity range of GCs belonging to NGC\,1399.  Although NGC\,1404
is reported to have an unusually low specific frequency of only
$\rm{S_N}\simeq2$ \citep{richtler92,forbes98}, it is probable that GCs
found in its vicinity belong to NGC\,1404 rather than to NGC\,1399.
These GCs contaminate the NGC\,1399 sample and would -- if unaccounted
for -- lead us to overestimate the line--of--sight velocity
dispersion and influence the measurement of the higher moments of the
velocity distribution.\par In Fig.~\ref{fig:histgrey} (left panel) we
show the radial velocities of the GCs versus the projected distance from
NGC\,1404. Within 3\arcmin{} from NGC\,1404, the velocity distribution
is skewed towards higher velocities, with two thirds of the GCs having
velocities within $220\,\rm{km\,s}^{-1}$ of the NGC\,1404 systemic
velocity. The mean and median velocity of all GCs in this area are
$1785$ and $1936\,\rm{km\,s}^{-1}$, respectively (the corresponding
values are indicated by the dashed and solid arrows).  Beyond
3\arcmin{} from NGC\,1404, the velocity field is clearly dominated by
NGC\,1399 GCs, and for the GCs within 3\arcmin--5\arcmin{} of
NGC\,1404, we find a mean (median) velocity of $1510$
($1475$)\,$\rm{km\,s}^{-1}$.  Guided by these findings, we exclude
{all} {23 (13 red and 10 blue)} GCs within {{$3\arcmin$
($\simeq16.6\,\rm{kpc}$) of NGC\,1404}} from our analysis of the
NGC\,1399 GCS. 

To distinguish the resulting sub--samples, we use Roman numerals as
labels, assigning `\textsc{i}' to the unaltered samples listed in
Sect.~\ref{sect:subpop}, and `\textsc{ii}' to the data excluding the
GCs in the vicinity of NGC\,1404.

\par Beyond the scope of this paper remains the  question whether the velocity
structure in the vicinity of NGC\,1404 is a superposition in the plane
of the sky, or if it traces a genuine interaction between the galaxies
and their GCSs (as modelled by \citealt{bekki03}). 
A more complete spatial coverage of the NGC\,1404
region as well as a larger number of GC velocities in this area are
required to uncover the presence of possible tidal structures.

\subsection{Velocity diagrams and extreme velocities}

\label{sect:extremeint}
Having dealt with the identification of GCs likely to belong to the
neighbouring galaxy, we now study the velocity diagrams for the blue
and red GC subpopulations. The main difference between red and blue
GCs shown in Fig.~\ref{fig:sampleall} is that the red GCs are more
concentrated towards the systemic velocity of NGC\,1399, while the
blue GCs have a larger range of velocities, implying a higher velocity
dispersion for the latter.  The red GCs occupy a wedge--shaped region
in the diagram, suggesting a declining velocity dispersion. \par
We further find in both sub-panels GCs with large relative velocities
which appear to deviate from the overall velocity distribution.  The
derived mass profile hinges on the treatment of these objects. As can
be seen from Table\,5 in \cite{schuberth06}, the inclusion of only two
interlopers at large radii is enough to significantly alter the
parameters of the inferred dark matter halo.

The right panel of Fig.~\ref{fig:sampleall} shows the densities of the
data points for the samples which will be used in the dynamical
analysis. One notes that the (low--level) contours for the blue GCs
are more irregular than those of the red GCs. Also, for radii between
20 and 40\,kpc, the blue GCs seem to avoid the systemic velocity of
NGC\,1399. The velocity structure found for the blue GCs appears to be
much more complex than that of red GCs. 

The very concept of interlopers is inherently problematic: If we
regard interlopers as an \emph{unbound population}, we are able to
identify only the extreme velocities and leave many undetected.  Here
we are primarily interested in a practical solution.  To locate
possible interlopers, we therefore chose an approach similar to the
one described by \cite{pom90}, who studied the effect of unbound
particles on the results returned by different mass estimators (namely
the virial mass estimator and the projected mass estimator as defined
in \citealt{htb85}). Perea et al.~used a statistical `Jacknife'
technique to identify interlopers in their N--body simulations of
galaxy clusters. Since the spatial distributions of the GC
subpopulations around NGC\,1399 do not follow the mass distribution
(which is at variance with the underlying assumption of the virial
mass estimator), we choose the {\emph{tracer mass estimator}} (TME,
\citealt{evans03}) which can be generalised to the case where the
number density of the tracer population is different from the overall
mass density. \par
\begin{figure*}
\centering
{\includegraphics[width=0.48\textwidth]{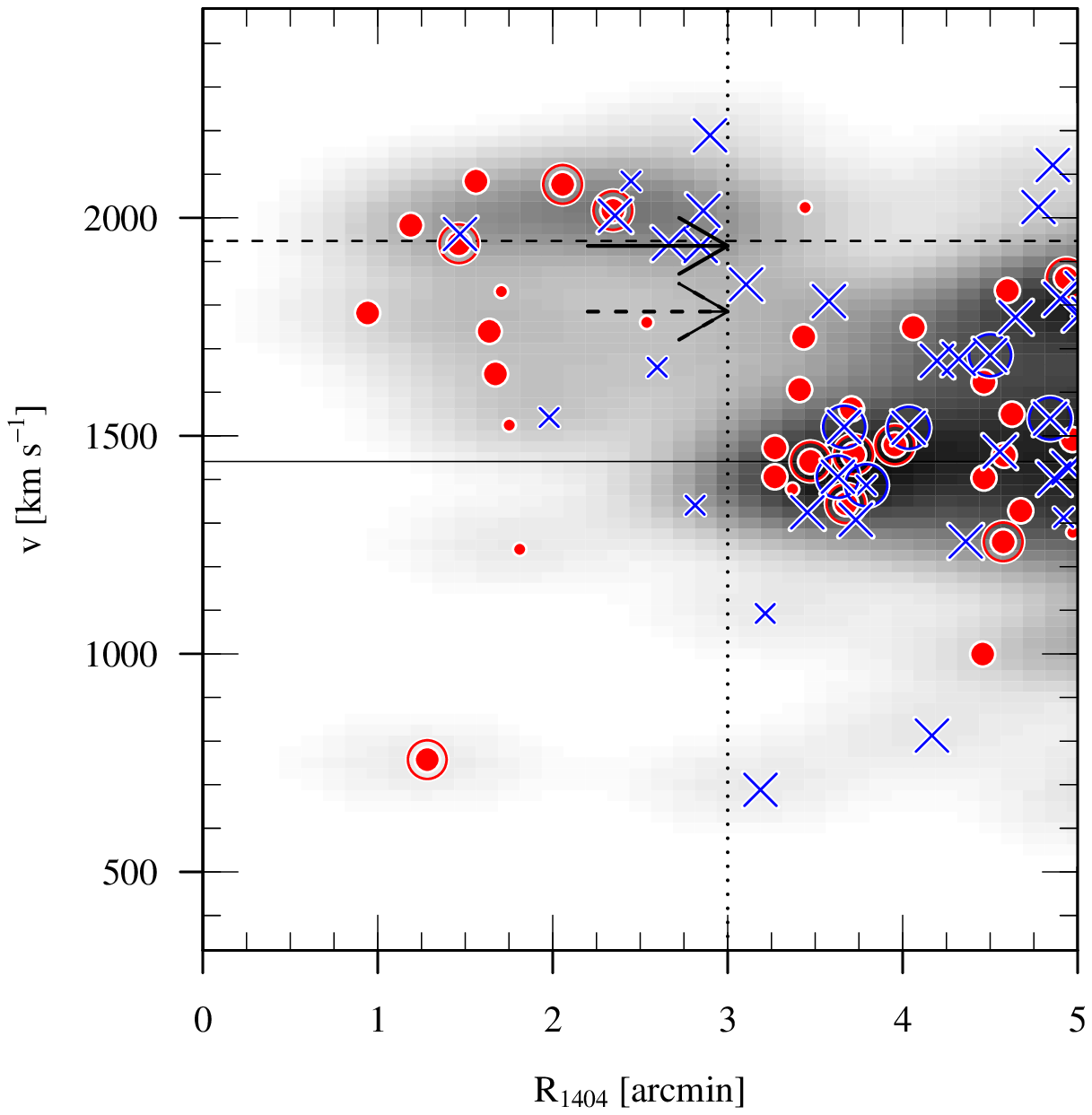}}
{\includegraphics[width=0.48\textwidth]{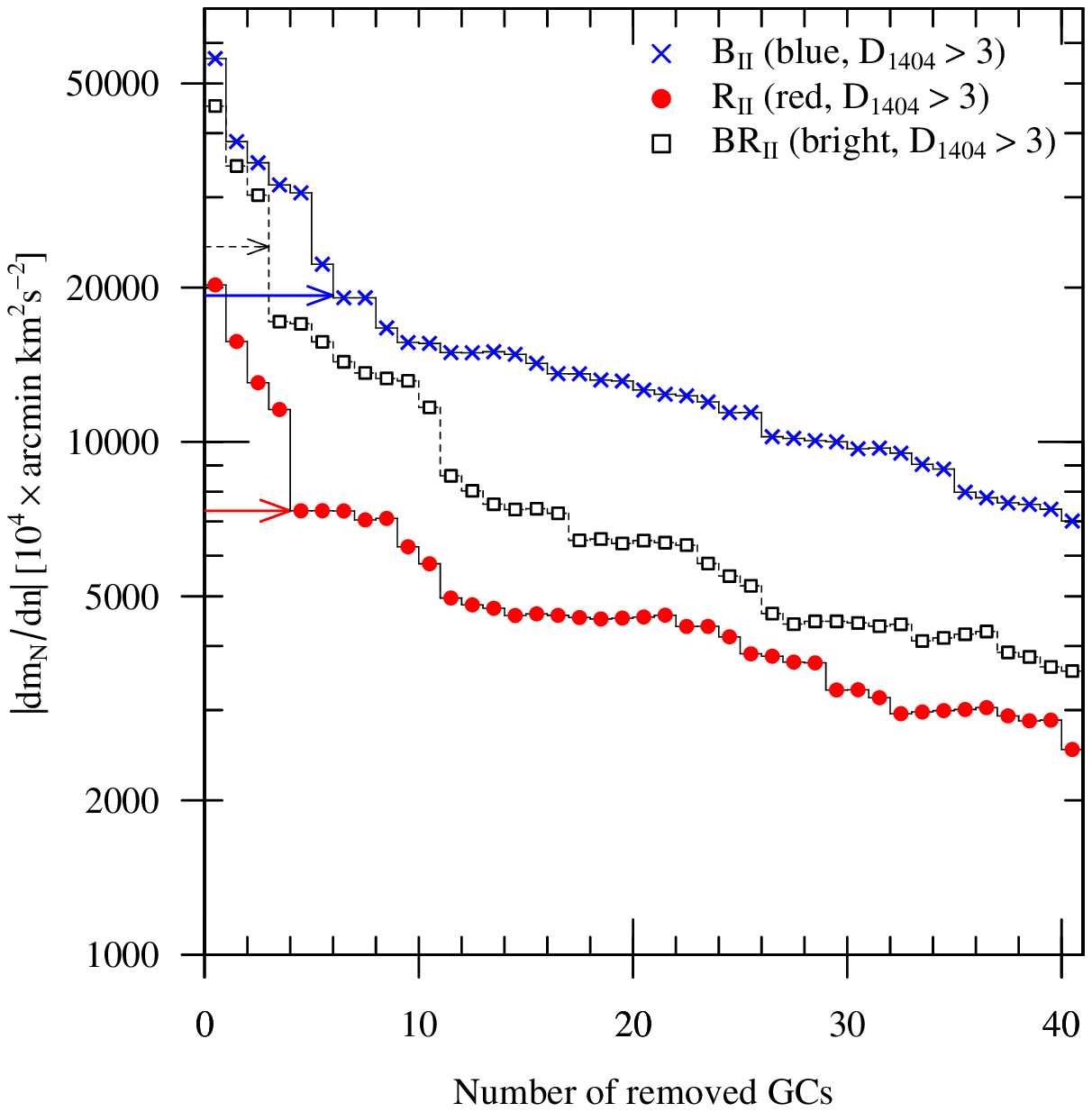}}
\caption{Interloper removal.  \emph{Left:} Velocities vs.~distance
from NGC\,1404. Crosses and dots represent blue and red GCs,
respectively. Large and small symbols refer to Class\,A and B velocity
measurements, respectively. Bright GCs ($m_R\leq 21.1$\,mag) are marked
by a circle. The systemic velocities of NGC\,1399 and NGC\,1404 are
shown as solid and dashed lines, respectively. The dotted line at
3\arcmin{} is the limit we adopt to separate NGC\,1404 GCs from the
ones belonging to NGC\,1399.  The dashed and solid arrow indicate the
mean and median velocity of the GCs within 3\arcmin{}, respectively.
The grey-scale shows the density of the data points smoothed with a
Gaussian (${\sigma_{x}=0\farcm5}$, $\sigma_{y}=75\,\rm{km\,s}^{-1}$).
\emph{Right:} Interloper removal using the tracer mass estimator
(TME). The magnitude of the derivative of $m_N$
(cf.~Eq.~\ref{eq:outrej1}) as function of $n$, the number of
eliminated interloper candidates.  Crosses and dots show red and blue
GCs ($R> 21.1$, $D_{1404}>3\arcmin$), respectively. The bright GCs are
shown as rectangles.  The arrows indicate the number of rejected GCs
for the different samples (B\textsc{ii}: 6, R\textsc{ii}: 4, and
BR\textsc{ii}: 3 GCs). }
\label{fig:outrej}
\label{fig:histgrey}
\end{figure*}

 The method we use works as follows: For a set of
$N$ GCs, we first calculate the quantity $m_N$:
\begin{equation}
m_N = \langle\varv^2 \cdot R \rangle = C_{\rm{proj}}\cdot  M_{\mathrm{TME}} \; ,
\label{eq:outrej1}
\end{equation}
which is proportional to $M_{\rm{TME}}$, the tracer mass estimate, and
where ${\varv}$ is the line--of--sight velocity relative to the
NGC\,1399 systemic velocity and $R$ is the projected distance from
NGC\,1399.  We then select, as potential outlier, the GC which has the
largest contribution i.e.~$\max (\varv^2\cdot R)$, remove it from the
list and calculate 

\begin{equation}
m_i=\frac{1}{(N-1)}\sum_{j\neq i}^N {\varv_{j}}^{2}\cdot R_j \; ,
\end{equation}
the expression for the remaining $(N-1)$ GCs.  This procedure is
repeated and, in the right panel of Fig.~\ref{fig:outrej}, we plot the
magnitude of the derivative of $m_N$ with respect to $n$, the number
of eliminated GCs.\par Obviously, selectively culling the high
relative velocities at large distances drastically lowers the estimate
for the total mass, resulting in large values of
$|\rm{d}m_N/\rm{d}n|$. Once the algorithm  with increasing $n$ starts
removing objects from the overall velocity field, the mass differences
between steps becomes smaller. \par As can be seen from the right panel
of Fig.~\ref{fig:outrej}, a convergence is reached after removing four
GCs from the red subsample \textsc{Rii}. For the blue GCs
(\textsc{Bii}), the situation is not as clear, and we decide to remove
six GCs. For the bright GCs, we remove three GCs.\par As opposed to
the constant velocity cuts (at $\varv_{\mathrm{helio}}=800$ and
$2080\,\textrm{km\,s}^{-1}$) used in Paper\,I, this algorithm does not
introduce a de facto upper limit on line--of--sight velocity
dispersion. \par Compared to schemes which use a jackknife to search
for $3\sigma$ deviations in a local ($\sim15$ neighbours) velocity
field (which may be ill--defined in the sparsely sampled outer
regions), our approach also works when the rejected GC is not the only
deviant data point in its neighbourhood.

\subsection{Defining the subsamples}
\label{sect:sampledefine}
To assess the impact of the interloper removal discussed in the
preceding sections, we assign the following labels: The primary label
indicates the parent sample (All, BRight, Faint, Blue, and Red,
cf.~Sect.~\ref{sect:subpop}), and the Roman numerals \textsc{i--iv}
refer to the \emph{sequence} of interloper removal
applied:\begin{description}
\item[\textsc{i}:]{Full data set, no interloper removal}
\item[\textsc{ii}:]{GCs within 3\arcmin{ of NGC\,1404 removed (Sect.~\ref{sect:1404inter})}}
\item[\textsc{iii}:]{Extreme velocities removed via TME algorithm (Sect.~\ref{sect:extremeint})}
\item[\textsc{iv}:]{Restriction to Class\,A velocity measurements }
\end{description}

To expand the radial coverage, we define the `extended' samples
\textsc{Av}, \textsc{Bv}, and \textsc{Rv} where we include the GC
velocities presented by B+07 (with the Washington
photometry given in \citealt{schuberth08}).  From the list of B+07, we
use all Class\,A GCs classified as intra--cluster GCs (ICGS), plus the
GCs in the vicinity of NGC\,1399 while excluding GCs within 3\arcmin
of NGC\,1404. These GCs are shown as large triangles
in the upper panels of Fig.~\ref{fig:dispersion} where we plot the GC
velocities versus the distance from NGC\,1399.  Since most GCs in the
list of B+07 are brighter than $m_R=21.1$, no bright--end magnitude
cutoff is applied to the `extended' samples. \par The basic
statistical properties of the samples defined above are listed in
Table\,\ref{tab:losvd} and discussed in the following section.

\section{The line--of--sight velocity distribution}
\label{sect:losvd}
\begin{table*}
\caption{Descriptive statistics of the line--of--sight velocity
distribution. Below we summarise the basic statistical properties for
the velocity samples defined in Sect.~\ref{sect:sampledefine}. The
samples are identified in the first column while the second column
gives the corresponding number of GC velocities. Col.~3 lists the mean
velocity, and the standard deviation is given Col.~4. The dispersion
$\sigma_{\rm{los}}$, as returned by the \cite{pm93} estimator, is
given in Col.~5. The third and fourth moments of the velocity
distributions, the skewness and the (reduced) kurtosis $\kappa$, are
listed in Cols.~6 and 7, respectively, and the uncertainties were
estimated using bootstrap resampling. Columns\,8 through 10 give the
$p$--values returned by the Anderson--Darling, Lilliefors
(Kolmogorov--Smirnov), and the Shapiro--Wilks (W--test) test for
normality, respectively. 
{{The Roman numerals refer to the sequence of subsamples defined in Sect.~\ref{sect:sampledefine}, where 
\textsc{(i)} is {the full sample, prior to  interloper removal}. 
\textsc{(ii)}: {GCs within 3\arcmin{} of NGC\,1404 removed,}
\textsc{(iii)}: {extreme velocities removed via TME algorithm (Sect.~\ref{sect:extremeint});}
\textsc{(iv)}: {restriction to Class\,A velocity measurements}, and \textsc{(v)}
includes the data from B+07}}.}
\centering
\begin{tabular}{llr@{$\pm$}lcr@{$\pm$}lr@{$\pm$}lr@{$\pm$}lll l l} \hline \hline
ID & \multicolumn{1}{c}{$N_{GC}$}& \multicolumn{2}{c}{$\langle \varv
\rangle$}& \multicolumn{1}{c}{sd}&
\multicolumn{2}{c}{$\sigma_{\rm{los}}$}& \multicolumn{2}{c}{skew}&
\multicolumn{2}{c}{$\kappa$}& \multicolumn{1}{c}{$p_{\rm{AD}}$}&
\multicolumn{1}{c}{$p_{\rm{KS}}$}& \multicolumn{1}{c}{$p_{\rm{SW}}$}\\

\multicolumn{1}{l}{(1)}&
\multicolumn{1}{c}{(2)}&
\multicolumn{2}{c}{(3)}&
\multicolumn{1}{c}{(4)}&
\multicolumn{2}{c}{(5)}&
\multicolumn{2}{c}{(6)}&
\multicolumn{2}{c}{(7)}&
\multicolumn{1}{c}{(8)}&
\multicolumn{1}{c}{(9)}& 
\multicolumn{1}{c}{(10)} 
\\ \hline

\multicolumn{14}{l}{No colour selection, {{no magnitude limit}}}\\ 
\hline

A{\sc{i}}   &  693  &   1457  &   12  &   312  &  309  &  9  &  -0.16  &  0.09  &  0.12  &  0.15  &  0.10  &  0.08  &  0.10  \\ 

A{\sc{ii}}  &  670  &   1446  &   12  &   306  &  302  &  9  &  -0.18  &  0.09 &  0.21  &  0.16  &  0.07  &  0.07  &  0.08  \\ 

A{\sc{v}}   &  729  &   1438  &   11  &   294  &  290  &  8  &  -0.33  &  0.08  &  0.16  &  0.16  &  0.01  &  0.06  &  0.01  \\ 
\hline
\multicolumn{14}{l}{{Faint ($m_R<21.1$) GCs, no colour selection}}\\ 
\hline

F{\sc{ii}}  &  493  &   1449  &   14  &   309  &  304  &  10  &  -0.07  &  0.10  &  0.12  &  0.17  &  0.45  &  0.46  &  0.69  \\ 

F{\sc{iv}}   &  297  &   1432  &   17  &   288  &  285  &   12  &  -0.23  &  0.11  &  -0.07  &  0.18  &  0.28  &  0.37  &  0.14  \\

\hline
\multicolumn{14}{l}{Bright ($m_R<21.1$) GCs }\\ \hline
BR{\sc{i}}   &  144  &   1435  &   25  &   307  &  305  &   18  &  -0.50  &  0.18  &  0.44  &  0.39  &  0.05  &  0.03  &  0.02  \\ 
BR{\sc{ii}}  &  140  &   1428  &   25  &   294  &  292  &   18  &  -0.59  &  0.20  &  0.61  &  0.45  &  0.03  &  0.03  &  0.01  \\ 

BR{\sc{iii}}   &  137  &   1424  &   24  &   280  &  279  &   17  &  -0.67  &  0.21  &  0.72  &  0.53  &  0.02  &  0.04  &  0.01  \\ 
BR{\sc{iv}}   &  127  &   1412  &   25  &   283  &  282  &   18  &  -0.63  &  0.22  &  0.67  &  0.52  &  0.03  &  0.02  &  0.01  \\ 

\hline
\multicolumn{14}{l}{Blue GCs $(m_R\geq21.1)$}\\ \hline

B{\sc{i}}   &  256  &   1452  &   23  &   362  &  358  &   16  &  -0.06  &  0.12  &  -0.35  &  0.17  &  0.65  &  0.50  &  0.67  \\ 

B{\sc{ii}}   &  246  &   1435  &   23  &   356  &  351  &   16  &  -0.04  &  0.12  &  -0.27  &  0.18  &  0.72  &  0.55  &  0.81  \\ 

B{\sc{iii}}   &  240  &   1415  &   22  &   337  &  333  &   16  &  -0.21  &  0.11  &  -0.42  &  0.16  &  0.22  &  0.14  &  0.09  \\ 

B{\sc{iv}}   &  139  &   1426  &   29  &   338  &  334  &   20  &  -0.27  &  0.14  &  -0.49  &  0.20  &  0.10  &  0.06  &  0.07  \\

B{\sc{v}}   &  336  &   1420  &   17  &   316  &  312  &   12  &  -0.3  &  0.11  &  -0.14  &  0.17  &  0.18  &  0.39  &  0.05  \\ 

\hline
\multicolumn{14}{l}{Red GCs $(m_R\geq21.1)$}\\ \hline

R{\sc{i}}   &  256  &   1472  &   16  &   259  &  256  &   12  &  0.02  &  0.14  &  0.02  &  0.21  &  0.48  &  0.70  &  0.70  \\ 
R{\sc{ii}}   &  247  &   1463  &   16  &   254  &  251  &   12  &  0.02  &  0.14  &  0.09  &  0.22  &  0.37  &  0.60  &  0.63  \\ 
R{\sc{iii}}   &  243  &   1463  &   16  &   242  &  239  &   11  &  0.04  &  0.13  &  -0.08  &  0.20  &  0.55  &  0.80  &  0.66  \\ 
R{\sc{iv}}   &  158  &   1437  &   19  &   238  &  234  &   13  &  -0.02  &  0.14  &  -0.14  &  0.25  &  0.80  &  0.63  &  0.76  \\ 
R{\sc{v}}   &  329  &   1453  &   14  &   254  &  251  &   10  &  -0.2  &  0.15  &  0.33  &  0.36  &  0.24  &  0.09  &  0.12  \\ 

\hline 
\hline 
\label{tab:losvd}
\end{tabular}
\end{table*}
We expect the line--of--sight velocity distribution (LOSVD) to be
nearly Gaussian.  Deviations from Gaussianity may be caused by orbital
anisotropies, the presence of interlopers, strong variations of the
velocity dispersion profiles with radius, or rotation.  Our sample,
being the largest of its kind so far, allows us to test the
statistical properties of the LOSVD in detail. \par
Table\,\ref{tab:losvd} summarises the statistical properties of the
velocity samples defined in Sect.~\ref{sect:sampledefine}. Besides the
first four moments of the distributions (i.e.~mean, dispersion,
skewness, and kurtosis), we list the $p$--values returned by the
Anderson--Darling (AD) and Kolmogorov--Smirnov (KS), and the
Shapiro--Wilks (SW) tests for normality.  
With the exception of the dispersion which we
calculate using the expressions given in \cite{pm93}, the data listed
in Table\,\ref{tab:losvd} were obtained using the functions of the
\texttt{e1071} and \texttt{nortest} packages in {\sf{R}}--statistics
software\footnote{{\sf{R}} Development Core Team \citep{R}
\newline\texttt{http://www.r-project.org}} for the higher moments and
the normality tests, respectively. \par
{{The SW test
\citep{shapirowilk,royston:1982a} is one of the most popular tests for
normality which also works for small samples.  The Anderson--Darling
\citep{stephens74} test is a variant of the KS--test tailored to be
more sensitive in the wings of the distribution.  The SW and
AD tests are among the tests recommended by \cite{dagostinoAD}; note
that these authors caution against the use  of the KS test which  has a
much smaller statistical power.  A detailed
discussion of the application of the AD test in the context of galaxy
group dynamics is given in \cite{hou09}. These authors compare the
performance of the AD test to the more commonly used KS and $\chi^2$
tests and find that the AD test is the most powerful of the
three. They conclude that it is a suitable statistical tool to detect
departures from normality, which allows the identification of dynamically
complex systems.}}    \par The velocity histograms for
the entire, the `bright', the red and the blue sample are shown in
Fig.~\ref{fig:velhist}. In all four sub-panels, we plot the
corresponding samples prior to any interloper removal (unfilled
histograms).  The solid histogram bars show the data after removing
GCs in the vicinity of NGC\,1404 and the GCs identified by the
$M_{\mathrm{TME}}$--algorithm (in panel\,\emph{a}, the solid histogram
also excludes the bright GCs). The dashed bars show these
samples when restricted to `Class\,A' velocity measurements.  The
striking difference between the velocity distributions of red and blue
GCs is that the red GCs seem to be well represented by a Gaussian
while the blue GCs appear to avoid the systemic velocity.  Also the
distribution of the bright GCs seems to be double--peaked.

Below we examine whether these distributions are
consistent with being Gaussian.
\subsection{Tests for normality}
Adopting $p \leq 0.05$ as criterion for rejecting the Null hypothesis of
normality, we find that \emph{none of the bright} subsamples
(\textsc{BRi--BRiv}) is consistent with being drawn from a normal
distribution. 

The fact that the full sample (\textsc{Ai, Aii, Av}) deviates from
Gaussianity (at the $\sim90$ per level) appears to be due to the
presence of the bright GCs: the velocity distribution of objects
fainter than 21.1 (sample \textsc{Fii} is the union of \textsc{Bii}
and \textsc{Rii}) cannot be distinguished from a Gaussian.

\par For the {red} subsamples (\textsc{Ri} through \textsc{Riv}), a 
Gaussian seems to be a valid description. Only the `extended' sample
(\textsc{Rv}) performs worse (probably because it encompasses bright
GCs), where the hypothesis of normality is rejected at the 91\% level
by the KS--test.\par While the full blue data set and the sample after
removing the GCs near NGC\,1404 (\textsc{Bi} and \textsc{Bii}) are
consistent with being Gaussian, the outlier rejection (\textsc{Biii})
and restriction to the `Class\,A' velocities (\textsc{Biv}) lead to
significantly lower $p$--values. In the case of the latter, all tests
rule out a normal distribution at the 90\% level, which appears
reasonable given that the distribution has a pronounced dip
(cf.~Fig.~\ref{fig:velhist}, panel \emph{(c)}, dashed histogram).

Below
we address in more detail the deviations from Gaussianity, as quantified 
by the higher moments of the LOSVD.
\subsection{Moments of the LOSVD}

\subsubsection{Mean}
Assuming that the GCs are bound to NGC\,1399, the first moment of the
LOSVD is expected to coincide with the systemic velocity of the
galaxy, which is $1441\,\pm 9\,\rm{km\,s}^{-1}$ (Paper\,I). As can be
seen from the third column in Table\,\ref{tab:losvd}, this is, within
the uncertainties, the case for all subsamples, with the exception of
R{\sc i}, where the GCs in the vicinity of NGC\,1404 are responsible
for increasing the mean velocity.
\subsubsection{Dispersion}
The second moments (Table\,\ref{tab:losvd}, Col.~5), the dispersions,
are calculated using the maximum--likelihood method presented by
\cite{pm93}, where the individual velocities are weighted by their
respective uncertainties. 
Note that these estimates do not differ from the standard deviations
given in Col.~4 by more than $5\,\textrm{km\,s}^{-1}$.  The interloper
removal, by construction, lowers the velocity dispersion. The largest
decrease can be found for the blue GCs, where the difference between
B{\sc i} and B{\sc iii} is $25\,\rm{km\,s}^{-1}$. For the red sample,
the corresponding value is $17\,\rm{km\,s}^{-1}$. We only consider at
this stage the full radial range of the sample. As will
be shown later, the effect of the interloper removal on the dispersion
calculated for individual radial bins can be much larger. \par The
main feature is that the dispersions for the blue and red GCs differ
significantly, with values of $333\pm 16$ and
$239\pm11\,\rm{km\,s}^{-1}$ for the blue and red samples
(\textsc{Biii} and \textsc{Riii}), respectively. The corresponding
values given in Paper\,I (Table\,2) are
$\sigma_\mathrm{blue}=291\pm14$ and $\sigma_\mathrm{red}=255\pm13$.
The agreement for the red GCs is good, and the discrepancy found for
the blue GCs is due to the different ways of treating extreme
velocities: In Paper\,I such objects were culled from the sample by
imposing radially constant velocity cuts (at $800$ and
$2080\,\textrm{km\,s}^{-1}$ ), while the method employed here (see
Sect.~\ref{sect:extremeint}) does not operate with fixed upper/lower
limits, which leads to a larger dispersion.
\subsubsection{Skewness}
Coming back to the issue of Gaussianity, we calculate the third moment
of the LOSVD: The skewness is a measure of the symmetry of a
distribution (the Gaussian, being symmetric with respect to the mean,
has a skewness of zero):
\begin{equation}
{\rm{skew}}= \frac{1}{N} \sum_{j=1}^{N} \left[ \frac{x_j -
\bar{x}}{\sigma} \right]^3  \,.
\end{equation}
The uncertainties were estimated using a bootstrap (with 999
 resamplings). We consistently find a \emph{negative} skewness for the
 bright subsamples (\textsc{BRi--BRiv},
 Fig.~\ref{fig:velhist}\,\emph{b}), i.e.~there are more data points in
 the low--velocity tail of the distribution. \par  The blue GCs also
 show a \emph{negative} skewness, which is significant for the samples
 \textsc{Biii} and \textsc{Biv}
 (cf.~Fig.~\ref{fig:velhist}\,\emph{c}).  A similar finding was
 already described in Paper\,I. \par The distribution of the red GCs 
 is symmetric with respect to the systemic velocity, as is to be expected for a
 Gaussian distribution (cf.~Fig.~\ref{fig:velhist}\,\emph{d}).
\subsubsection{Kurtosis}
 In the following, $\kappa$ denotes the \emph{reduced kurtosis}, or
\emph{kurtosis excess} (to assign the value zero to a normal
distribution), i.e.:
\begin{equation}
\kappa = \frac{1}{N} \sum_{j=1}^{N} \left[ \frac{x_j -
\bar{x}}{\sigma} \right]^4 - 3 \,.
\end{equation}
The reduced kurtosis $\kappa$ for the various subsamples of our data
is given in Col.~7 of Table\,\ref{tab:losvd}.  The uncertainties
$\Delta \kappa$ were again estimated using a bootstrap.  The brightest
GCs have (marginally significant) positive kurtosis values, meaning
that the distribution is more `peaked' than a Gaussian. \par The blue
samples (\textsc{Bi--Biv}), on the other hand, have a negative
kurtosis, indicative of a more flat--topped distribution which might
be the result of a tangential orbital bias.\par The red GCs are
consistent with $\kappa=0$, which would be expected in the case of
isotropic orbits.
\subsection{Summary of the statistical tests}
We  briefly summarise the main findings from the above
statistical considerations.  The velocity dispersions of the red and
blue subsamples are, as in Paper\,I, significantly different, with the
blue GCs showing a larger dispersion.  We note that the dispersions of
the bright $(R > 21.1)$ subsamples have values in between those found
for the red and blue GCs.  \par The entirety of the GCs
(\textsc{Ai,Aii}) is not Gaussian, which is most likely due to the
pronounced skewness of the brightest GCs.\par The blue subsamples,
although formally consistent with being Gaussian, show (after the
outlier rejection) significantly negative values for both skewness and
kurtosis.\par The red GCs are well represented by a Gaussian, which
would be expected in the case of isotropic orbits.

\begin{figure}
\centering
\resizebox{\hsize}{!}
{\includegraphics[width=0.48\textwidth]{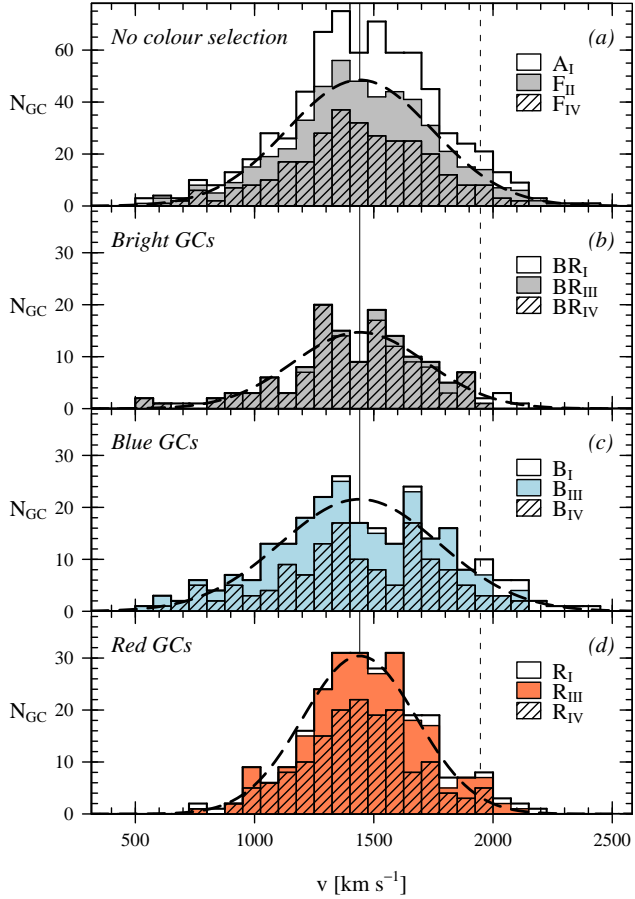}}
\caption{Velocity histograms.  \emph{(a):} The unfilled histogram
shows all 693 GCs (A{\textsc{i}}). The solid histogram bars are the
faint ($R>21.1$) GCs outside the 3\arcmin{} radius around NGC\,1404
(i.e.~sample F{\textsc{ii}}), and the dashed histogram is the same for
the `Class\,A' velocity measurements after outlier rejection (sample
F{\textsc{iv}}).  \emph{(b):} The bright ($R<21.1$) GCs,
BR{\textsc{i}}, BR{\textsc{iii}}, and {\textsc{BRiv}}. Panels
\emph{(c)} and \emph{(d)} show the same for blue and red GCs,
respectively.  In all panels, the solid and dashed vertical lines show
the systemic velocity of NGC\,1399 and NGC\,1404, respectively. The
dashed curve is a Gaussian with the width as given in
Table\,\ref{tab:losvd} for the data represented as solid histogram
(i.e.~F\textsc{ii}, BR\textsc{iii}, \textsc{Biii}, and \textsc{Riii},
respectively). Note that the range of the $y$--axis has been adjusted
for easier comparison, with panel (a) having twice the range of the
subsequent plots.
}
\label{fig:velhist}
\end{figure}

\section{Rotation}
\label{sect:rot}
The amount of rotation found in a GCS and among its subpopulations
might be indicative of the host galaxy merger history.  The presence
of rotation (rather than the absence thereof, since one only
observes projected rotation), might help constrain GCS formation
scenarios. \par
\begin{figure}
\centering
\resizebox{\hsize}{!}
{\includegraphics[width=0.48\textwidth]{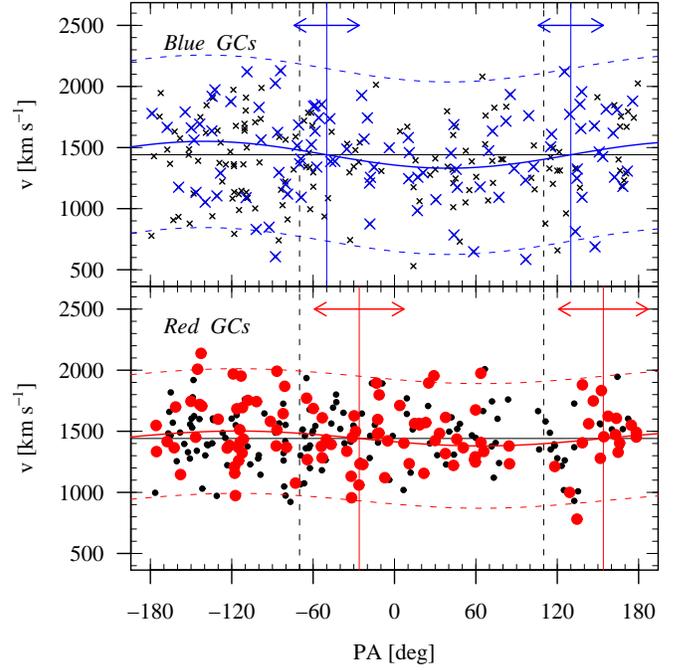}}
\caption{Rotation of the NGC\,1399 GCS.  Velocities versus position
angle. The \emph{upper panel} shows the data for the blue sample
\textsc{Biii} (i.e. the data after the removal of the GCs in the
vicinity of NGC\,1404 and the outlier rejection) for the azimuthally
complete radial interval $R\leq 8\farcm0$. Large symbols are GCs in
the `outer' $(4<R \leq 8\farcm0)$ region for which a significant
rotation signal is found. The solid curve shows the best--fit
sine--curve for this sample, the dashed curves indicate the $\pm
2\cdot \sigma_{\mathrm{los}}$ boundaries, where
$\sigma_{\mathrm{los}}=353\pm26\,\textrm{km\,s}^{-1}$. The vertical
solid lines and arrows indicate $\Theta_0$ and $\Delta\Theta_0$,
respectively. The \emph{lower panel} shows the same for the red GCs
(sample \textsc{Riii}). The velocity dispersion of the red GCs in the
`outer' interval is
$\sigma_{\mathrm{los}}=255\pm19\,\textrm{km\,s}^{-1}$. In both panels,
the vertical dashed lines indicate the position angle (photometric
major axis) of NGC\,1399 ($110^\circ$). The respective fit parameters
are listed in Table\,\ref{tab:rotation}. }
\label{fig:rotall}
\end{figure}
In Paper\,I, we did not find any signature of rotation, with the
exception of the outer ($\textrm{R} > 6 \arcmin$) metal--poor
GCs, for which a marginal rotation amplitude of $68 \pm
60\,\textrm{km\,s}^{-1}$ was detected (for a position angle of
$\Theta_0 = 140^{\circ} \pm 39^{\circ}$). Our revised and extended
data set provides the opportunity to re--examine this finding.
\par We apply the method used in Paper\,I, which consists of fitting a
sine--curve to the velocities plotted versus the position angle (as
shown in Fig.~\ref{fig:rotall}):
\begin{equation}
\varv_{\rm{rot}}(\Theta) = \varv_{\rm{sys}} + A_{\rm{rot}} \sin(\Theta
-\Theta_0) \; .
\label{eq:sinus}
\end{equation}
Further, to ensure that the rotation signal is not merely an artifact
of the inhomogeneous spatial and azimuthal coverage, we calculate the
rotation amplitude for 1000 Monte Carlo (MC) realizations, in which we
keep the positions of our original dataset but permutate the
velocities. From this we determine the fraction $f$ of MC runs for
which a rotation amplitude \emph{larger} than the observed one was
found: i.e.~the smaller the value of $f$, the higher the significance
of the rotation signal.\par Table~\ref{tab:rotation} lists the values
for various GC subsamples. Since, as can be seen from
Fig.~\ref{fig:colhist} (bottom right panel), the azimuthal coverage
becomes very patchy for galactocentric distances beyond about
$8\farcm0{}\, (\simeq44\,\textrm{kpc})$, we also give the results for
the radial range $0\arcmin\leq R\leq8\farcm0$, where the spatial
coverage is more uniform and the completeness
(cf.~Fig.~\ref{fig:colhist}, bottom left panel) is relatively
high. This selection is then further divided into an `inner'
($0\arcmin\leq R\leq 4\farcm0$) and an `outer'
($4\farcm0<R\leq8\farcm0$) region.\par
\subsection{Rotation of the brightest GCs}
The brightest GCs ($m_R\leq 21.1$), show no sign of rotation.  For all
selections (\textsc{BRi}--\textsc{BRiv}) and radial intervals
considered in Table\,\ref{tab:rotation}, the rotation amplitude is
consistent with being zero.
\subsection{Rotation of the metal--rich GCs}
For the metal--rich GCs, 
only the sample {\textsc{Riii}} (i.e.~the data set obtained after
masking the GCs in the vicinity of NGC\,1404 and applying the outlier
rejection algorithm) shows a weak rotation signal
($A=61\pm35\,\textrm{km\,s}^{-1}, \Theta=154\pm33^\circ$) for the
`outer sample'. The data and the fit are shown in the lower panel of
Fig.~\ref{fig:rotall}. The corresponding $f$--value of $0.27$ shows
that a rotation signal of this magnitude arises from the randomised
data with a probability of 27 per cent, meaning no significant
rotation is detected for the red GCs.  Moreover, this weak rotation
signal vanishes completely when only the `Class\,A' velocity
measurements are considered (\textsc{Riv}).
\subsection{Rotation of the metal--poor GCs}
Within the central $4\farcm0\, (\simeq22\,\rm{kpc})$, none of the blue
GC subsamples shows any rotation signal. For the `outer samples',
however, the rotation amplitudes lie in the range
110--126\,$\textrm{km\,s}^{-1}$, and the values obtained for
$\Theta_0$ agree within the uncertainties.  For the blue GCs
\textsc{Bii} (after removal of GCs near NGC\,1404), the amplitude is
$A=110\pm53\,\textrm{km\,s}^{-1}$, and $\Theta_0=130\pm24^\circ$ (see
Fig.~\ref{fig:rotall}, upper panel), which agrees with the rotation
values quoted in Paper\,I. After further restricting this sample to `Class\,A'
velocities, the results agree very well, although the smaller number
of data points leads to larger uncertainties. We
caution, however, that the probability of such an amplitude resulting
from the randomised data still is about 10 per cent.

\subsection{NGC\,1399 and the rotation of its GCS}
As stated above, we find no rotation for the red (metal--rich) GC
population. In this respect, the red GCs appear to reflect the
properties of the stellar body of NGC\,1399 for which the
spectroscopic study by \cite{saglia00} found only a small ($\lesssim
30\,\rm{km\,s}^{-1}$) rotation along the major axis (an equally small
signal was found for the slightly off--centre slit position parallel
to the minor axis).\par For the blue GCs, a significant rotation
signature is found for the radial range between 4 and 8
arcminutes. The axis of rotation is consistent with the photometric
major axis of NGC\,1399 ($\rm{PA}=110^\circ$). Due to the very patchy
angular coverage for radii beyond about 8\arcmin{}, no statement about
the rotation of the GCs beyond $\sim44\,\rm{kpc}$ can be made.\par In
the literature the amount of rotation is quantified in terms of the
parameter $\varv_{\mathrm{rot}}/\sigma_{\mathrm{los}}$, (i.e. the
ratio of rotational velocity to velocity dispersion). For NGC\,1399
we find $\varv/\sigma\simeq0$ for the metal--rich GCs and
$\varv/\sigma\lesssim 0.3$ for the metal--poor GCs.

\begin{table*}
\caption{Rotation amplitudes and angles for the different
subsamples. The first column lists the sample identifiers.  Columns 2
through 4 give the number of GCs, the parameters obtained by
fitting Eq.~\ref{eq:sinus} to the data over the full radial
range. Column 5 gives the fraction $f$ of MC runs for which
$A_{\rm{MC}} \geq A_{\rm{rot}}$ (see text for details).  Columns 6--9,
10--13, and 14--17 give the GC numbers and parameters obtained when
restricting the radial range to $0\leq R\leq8\farcm0$, $0\leq R
\leq4\farcm0$, and $4< R\leq8\farcm0$, respectively.}
\centering

\resizebox{0.99\textwidth}{!}{
\begin{tabular}{ll r@{$\pm$}l  r@{$\pm$}l l |l r@{$\pm$}l r@{$\pm$}ll | l r@{$\pm$}l r@{$\pm$}ll | l r@{$\pm$}l r@{$\pm$}ll }\hline \hline
& \multicolumn{6}{|c|}{\underline{$\quad 0\arcmin \leq R \leq R_{max}\quad$}}
& \multicolumn{6}{|c|}{\underline{$\quad 0\arcmin \leq R \leq 8\farcm0\quad$}}
& \multicolumn{6}{|c|}{\underline{$\quad 0\arcmin \leq R \leq 4\farcm0\quad$}}
& \multicolumn{6}{|c}{\underline{$\quad 4\arcmin < R \leq 8\farcm0\quad$}}
\\

ID & \multicolumn{1}{|c}{$N_{GC}$}& 
\multicolumn{2}{c}{$\Theta_0$} & 
\multicolumn{2}{c}{$A_{\rm{rot}}$} & 
\multicolumn{1}{c|}{$f$}

&
\multicolumn{1}{|c}{$N_{GC}$}& 
\multicolumn{2}{c}{$\Theta_0$} & 
\multicolumn{2}{c}{$A_{\rm{rot}}$} & 
\multicolumn{1}{c|}{$f$}
&
\multicolumn{1}{|c}{$N_{GC}$}& 
\multicolumn{2}{c}{$\Theta_0$} & 
\multicolumn{2}{c}{$A_{\rm{rot}}$} & 
\multicolumn{1}{c|}{$f$}
&
\multicolumn{1}{|c}{$N_{GC}$}& 
\multicolumn{2}{c}{$\Theta_0$} & 
\multicolumn{2}{c}{$A_{\rm{rot}}$} & 
\multicolumn{1}{c}{$f$}\\


(1) & \multicolumn{1}{|c}{(2)}& \multicolumn{2}{c}{(3)} 
& \multicolumn{2}{c}{(4)} & \multicolumn{1}{c|}{(5)} 
& \multicolumn{1}{|c}{(6)}& \multicolumn{2}{c}{(7)} 
& \multicolumn{2}{c}{(8)} & \multicolumn{1}{c|}{(9)} 

& \multicolumn{1}{|c}{(10)}& \multicolumn{2}{c}{(11)} 
& \multicolumn{2}{c}{(12)} & \multicolumn{1}{c|}{(13)} 

& \multicolumn{1}{|c}{(14)}& \multicolumn{2}{c}{(15)} 
& \multicolumn{2}{c}{(16)} & \multicolumn{1}{c}{(17)}

\\ \hline

\textsc{Ai} &  693  &  109  &  19  &  48  &  16  &  0.02 
&  476  &  132  &  27  &  40  &  19  &  0.12  
  &  193  &  152  &  123  &  14  &  29  &  0.88 
  &  283  &  130  &  24  &  59  &  25  &  0.06
  \\

\textsc{Aii}  &  670  &  134  &  26  &  36  &  16  &  0.10 
&  472  &  143  &  29  &  39  &  19  &  0.17  
  &  193  &  152  &  123  &  14  &  29  &  0.84  
  &  279  &  142  &  26  &  56  &  25  &  0.12  \\

\textsc{Fii} &  493  &  141  &  27  &  42  &  19  &  0.09 
  &  347  &  143  &  27  &  47  &  22  &  0.10 
  &  142  &  170  &  110  &  17  &  32  &  0.87
&  205  &  139  &  24  &  71  &  30  &  0.05  \\ 

\textsc{Fiv} &  297  &  127  &  48  &  27  &  24  &  0.56  
 &  218  &  146  &  31  &  46  &  28  &  0.19 
  &  88  &  207  &  128  &  18  &  39  &  0.87  
  &  130  &  136  &  26  &  73  &  38  &  0.12  \\

\hline

 \textsc{BRi}&  144  &  91  &  57  &  36  &  35  &  0.6 
&  102  &  160  &  170  &  14  &  41  &  0.91  
  &  45  &  139  &  473  &  8  &  74  &  0.99 
  &  57  &  169  &  150  &  20  &  48  &  0.94  \\


\textsc{BRii}&  140  &  100  &  85  &  24  &  34  &  0.79 
 &  102  &  160  &  170  &  14  &  41  &  0.98 
  &  45  &  139  &  473  &  8  &  74  &  1 
  &  57  &  169  &  150  &  20  &  48  &  0.95  \\


\textsc{BRiii}  &  137  &  109  &  77  &  26  &  32  &  0.71 
  &  102  &  160  &  170  &  14  &  41  &  0.94  
  &  45  &  139  &  473  &  8  &  74  &  1 
  &  57  &  169  &  150  &  20  &  48  &  0.95  \\

\textsc{BRiv} &  127  &  107  &  114  &  18  &  35  &  0.89 
 &  93  &  230  &  173  &  14  &  46  &  0.95 
  &  39  &  259  &  124  &  36  &  85  &  0.93  
  &  54  &  144  &  327  &  9  &  49  &  1  \\ 

\hline

\textsc{Bi}  &  256  &  113  &  18  &  93  &  32  &  0 
 &  161  &  118  &  24  &  87  &  39  &  0.12  
  &  59  &  136  &  108  &  32  &  55  &  0.91 
  &  102  &  118  &  21  &  126  &  53  &   0.039  \\ 
 
\textsc{Bii} &  246  &  131  &  24  &  75  &  31  &  0.03 
 &  157  &  131  &  28  &  76  &  38  &  0.11  
  &  59  &  136  &  108  &  32  &  55  &  0.8  
  &  98  &  130  &  24  &  110  &  53  & 0.097  \\


 \textsc{Biii} &  240  &  116  &  29  &  59  &  31  &  0.1 
 &  157  &  131  &  28  &  76  &  38  &  0.13  
  &  59  &  136  &  108  &  32  &  55  &  0.9
  &  98  &  130  &  24  &  110  &  53  &   0.105    \\

\textsc{Biv}  &  139  &  136  &  54  &  41  &  42  &  0.57  
 &  92  &  148  &  32  &  79  &  50  &  0.31 
&  34  &  192  &  117  &  33  &  69  &  0.85
  &  58  &  140  &  26  &  123  &  72  &  0.172   \\ 

\hline

\textsc{Ri}  &  256  &  108  &  70  &  18  &  22  &  0.66  
 &  190  &  164  &  50  &  30  &  26  &  0.46  
  &  83  &  222  &  142  &  15  &  39  &  0.9
  &  107  &  154  &  44  &  47  &  36  &  0.46  \\ 

\textsc{Rii}  &  247  &  182  &  89  &  14  &  23  &  0.85 
  &  190  &  164  &  50  &  30  &  26  &  0.45 
 &  83  &  222  &  142  &  15  &  39  &  0.95 
  &  107  &  154  &  44  &  47  &  36  &  0.440  \\ 

\textsc{Riii}  &  243  &  159  &  61  &  20  &  21  &  0.69  
  &  189  &  162  &  40  &  37  &  26  &  0.37  
  &  83  &  222  &  142  &  15  &  39  &  0.88  
  &  106  &  154  &  33  &  61  &  35  &   0.272  \\

\textsc{Riv} &  158  &  102  &  94  &  16  &  27  &  0.83 
  &  126  &  132  &  76  &  22  &  32  &  0.76  
  &  54  &  246  &  275  &  11  &  46  &  0.97  
  &  72  &  119  &  56  &  41  &  39  &  0.61  \\ 
\hline
\hline
\end{tabular}
}
\label{tab:rotation}
\end{table*}
\section{Radial velocity dispersion profiles}
\label{sect:dispersion} 
The line--of--sight velocity dispersion as a function of the projected
radius is the quantity we aim to reproduce with the Jeans models
described in Sect.~\ref{sect:jeans}.  The histograms displayed in
Fig.~\ref{fig:velhist} and the data given in Table\,\ref{tab:losvd}
show that the blue and red subpopulations have significantly different
velocity distributions, as is already known from Paper\,I. We calculate
the dispersion profiles separately for both subpopulations.
\label{sect:pm}
First, we determine the dispersion values using the same annular bins
for both subpopulations and divide our data into radial bins covering
the full radial range (starting at $1.0,3.5,5.5,7.5,9.5,12.5,15.5$,
{and} $30.0$ arcmin).  The limits of the bins are shown as dotted
lines in the upper and middle panels of Fig.~\ref{fig:dispersion}.

The middle panels of Fig.~\ref{fig:dispersion} show the radial
dispersion profiles for the red and blue GCs.  Circles show the values
obtained for the GCs fainter than $m_R=21.1$, prior to any interloper
removal (samples \textsc{Ri} and \textsc{Bi} for red and blue GCs,
respectively). The corresponding profiles obtained after removing the
GCs in the vicinity of NGC\,1404 and the most extreme velocities
(samples \textsc{Riii} and \textsc{Biii}) are shown as filled squares.
For the extended samples \textsc{Rv} and \textsc{Bv}, shown as
diamonds, the number of GCs in a bin is given in parenthesis.  The
dispersion profiles for the fixed radial bins for the samples
\textsc{Ri} through \textsc{Rv} and \textsc{Bi}--\textsc{Bv} are given
in Table\,\ref{tab:fixbin}.

\par Next, to obtain data points with similar statistical
uncertainties, we fix the number of GCs per bin (and thereby allow for
a larger range in radial extent of the bins). In the bottom panels of Fig.~\ref{fig:dispersion}, we
use a moving window containing 35 GCs to plot the samples
\textsc{Riii} and \textsc{Biii} together with the extended samples
containing the B+07 measurements \textsc{Rv} and \textsc{Bv}. The
corresponding values are listed in Table\,\ref{tab:dispersion}
\subsection{Dispersion of the  metal--rich subpopulation}

For \textsc{Riii} and \textsc{Rv}, shown in the middle left panel of
Fig.~\ref{fig:dispersion}, we find that the velocity dispersion
declines for radii beyond $\sim\!30\,\rm{kpc}$, while the values for
the sample \textsc{Ri} (prior to any interloper rejection) even show
an increasing dispersion.  The removal of the GCs in the vicinity of
NGC\,1404 mostly affects the $4^{\rm{th}}$ bin where the dispersion
drops from $260\pm35$ to $216\pm30\,\rm{km\,s}^{-1}$.  Removing the
three potential outliers in the $5^{\rm{th}}$ bin by means of the
TME--algorithm lowers the dispersion by about
$90\,\rm{km\,s}^{-1}$.\par The bottom left panel of
Fig.~\ref{fig:dispersion} compares the moving--bin results for the
samples \textsc{Riii} and \textsc{Rv}. Both samples show again a
smooth decline of the dispersion for larger radii. The marginally
higher values found for the \textsc{Rv} data which include the B+07
velocities is likely due to the inclusion of brighter ($R\leq
21.1\,\rm{mag}$) clusters which have a higher velocity dispersion than
the fainter red GCs.  The $3^{\rm{rd}}$ bin for \textsc{Riii} has a
surprisingly high dispersion of $305\pm44\,\rm{km\,s}^{-1}$ which is
50 (67) $\rm{km\,s}^{-1}$ higher than the value found in the previous
(next) bin.  This rise is caused by a number of GCs at high velocities
(cf.~Fig.~\ref{fig:sampleall}), leading to a mean velocity which
exceeds the systemic velocity of NGC\,1399 by
$\sim120\,\rm{km\,s}^{-1}$. In the upper panel of
Fig.~\ref{fig:kappa}, where we show the dispersion in a
semi--logarithmic plot, the deviation of this data point becomes even
more evident. In the modelling presented in
Sect.~\ref{sect:massmodels} this data point (marked by an asterisk in
Table\,\ref{tab:dispersion}) is omitted.

\subsection{Dispersion of the  metal--poor subpopulation}
\par
\begin{figure*}
\centering
\resizebox{\hsize}{!}{
\includegraphics[width=0.98\textwidth]{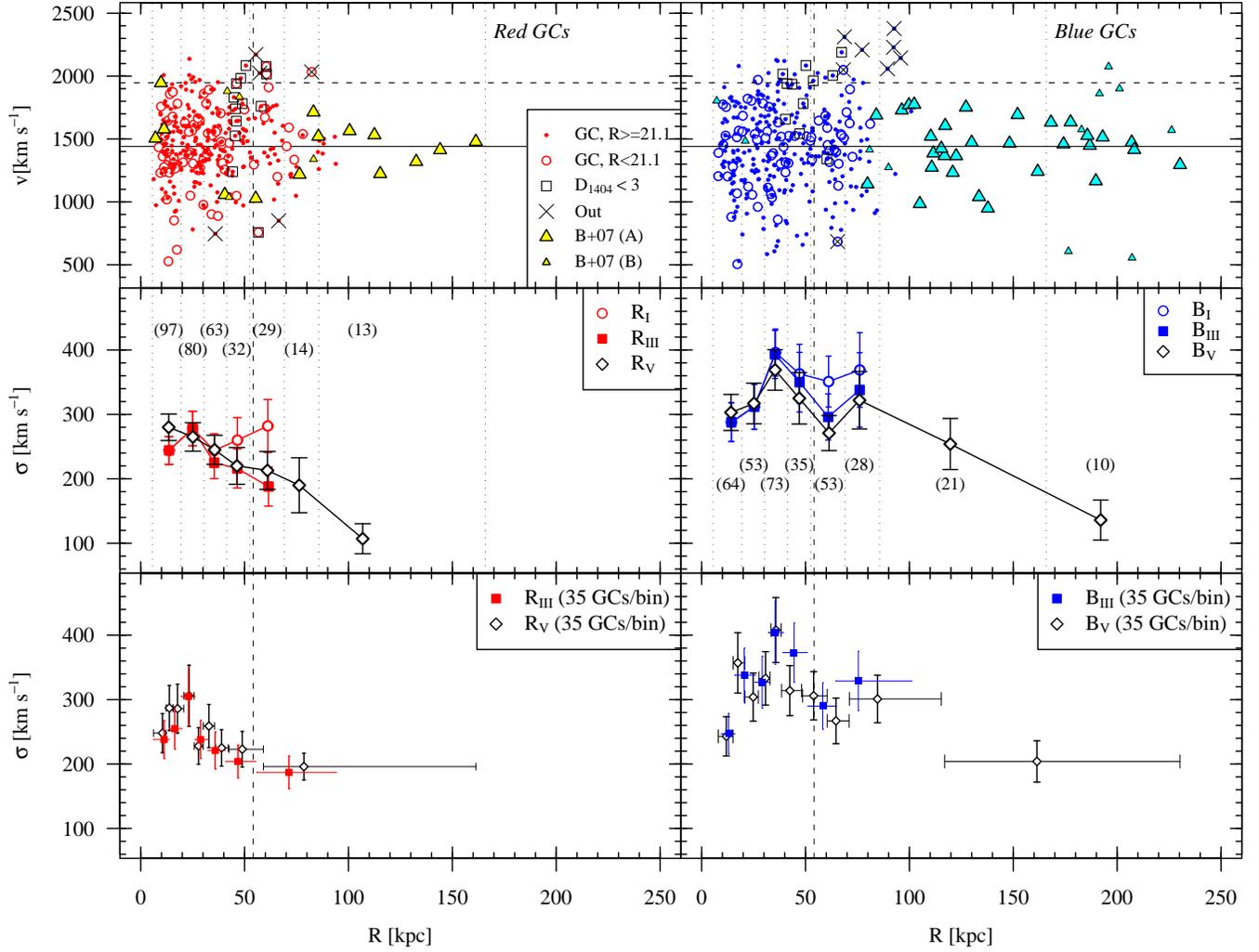}
}
\caption{Velocity diagrams and line--of--sight velocity dispersion
profiles.  \emph{Upper left}: radial velocity vs.~galactocentric
distance for the red GCs. Dots and circles show faint and bright ($m_R
< 21.1$\,mag) GCs, respectively. Large and small triangles are the
data points from \cite{bergond07} with Class\,A and B velocity
measurements, respectively (see text for details). Squares mark GCs
rejected on account of their proximity to NGC\,1404; crosses are the
GCs removed by the TME rejection algorithm. The solid and dashed
horizontal lines are the systemic velocities of NGC\,1399 and
NGC\,1404, respectively. The vertical dashed line shows the
(projected) distance of NGC\,1404. The dotted vertical lines show the
radial bins for which the dispersion shown in the middle panels is
calculated. \emph{Upper right:} The same for the blue
GCs. \emph{Middle panels:} The dispersion profiles for fixed radial
bins. Circles are the values obtained for the full samples
(\textsc{Ri} and \textsc{Bi}) prior to any interloper removal. Filled
squares show the values after removing GCs in the vicinity of
NGC\,1404 and the outliers identified by the TME algorithm (samples
\textsc{Riii} and \textsc{Biii}). The dispersion profiles for the
extended samples (\textsc{Rv} and \textsc{Bv}) including the
velocities from \cite{bergond07} are shown as diamonds. The labels in
parentheses give the number of GCs in a given bin. \emph{Bottom
panels:} Filled squares are the profiles for \textsc{Riii} and
\textsc{Biii} obtained for a moving bin comprising 35 GCs. Diamonds
show the same for the samples \textsc{Rv} and \textsc{Bv}.  These
moving--bin profiles which are used for the modelling in
Sect.~\ref{sect:massmodels} are listed in
Table\,\ref{tab:dispersion}.}
\label{fig:dispersion}
\end{figure*}

 For the blue GCs (shown in the right panels of
Fig.~\ref{fig:dispersion}) the situation is more complicated in so far as
 the dispersion profile is not as smooth.  For the full sample
\textsc{Bi} as shown in the middle right panel of
Fig.~\ref{fig:dispersion}, the dispersion rises quite sharply by over
$100\,\rm{km\,s}^{-1}$ for the range between $\sim10$ and
$35\,\textrm{kpc}$ where it reaches almost $400\,\rm{km\,s}^{-1}$ before
levelling out at $\sim360,\rm{km\,s}^{-1}$.  The effect of the
interloper removal is strongest for the fifth bin where removing five
GCs (four of which are in the vicinity of NGC\,1404 and one GC is
rejected by the TME algorithm, sample\textsc{Biii}) decreases the
dispersion by more than $50\,\rm{km\,s}^{-1}$. For the sixth bin,
however, the dispersion is again somewhat higher (although the values
agree within the uncertainties). For the sample \textsc{Bi} and
\textsc{Biii} no clear trend for the behaviour of the GCs towards
larger radii is discernible.\par Taking into account the velocities
for the outer GCs by B+07 shows however,  that also the blue GCs have
a declining velocity dispersion profile (sample \textsc{Bv}, shown as
diamonds).

The lower right panel of Fig.~\ref{fig:dispersion} shows the velocity
dispersion profiles for the samples \textsc{Biii} and \textsc{Bv} for
a moving window containing 35 GCs. Compared to the red GCs these
profiles appear less continuous. This behaviour is likely due to
inhomogeneities and substructures in the velocity field as illustrated
in the right panel of Fig.~\ref{fig:sampleall}.

\subsection{Comparison to the stellar data}
\label{sect:himo}
\begin{figure}
\centering
\resizebox{\hsize}{!}
{\includegraphics[width=0.48\textwidth]{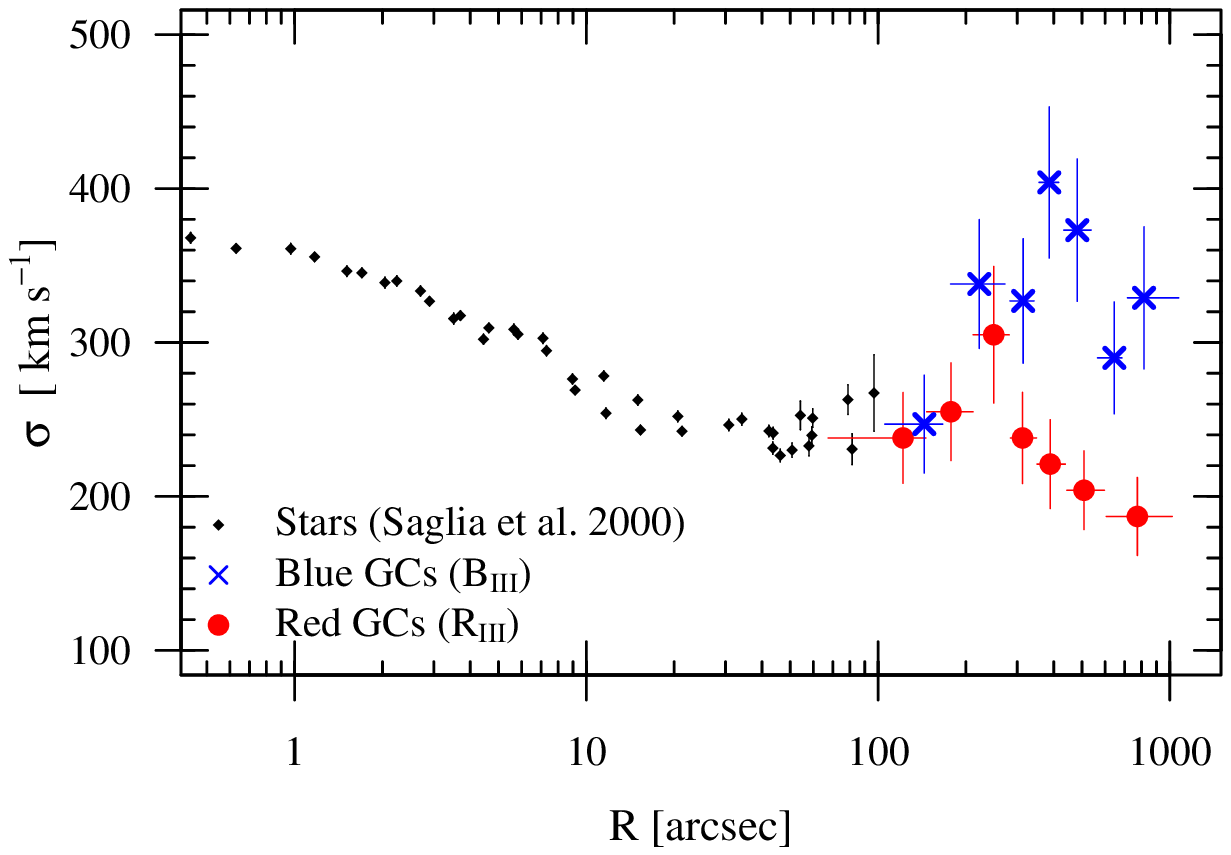}}
\resizebox{\hsize}{!}
{\includegraphics[width=0.489\textwidth]{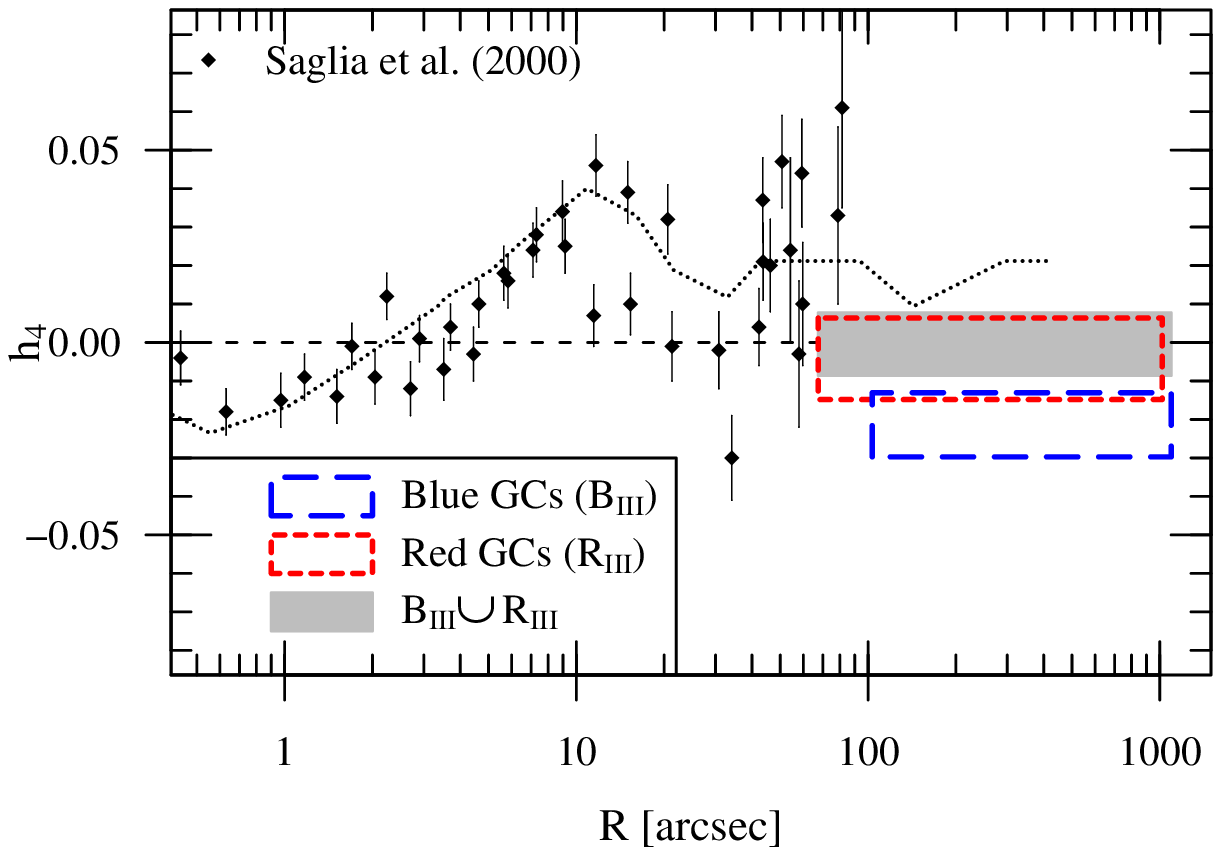}}
\caption{ Comparison to the stellar kinematics presented by
\cite{saglia00}.  \emph{Upper panel}: Line--of--sight velocity
dispersion profiles. Diamonds show the stellar kinematics, dots and
crosses are our measurements of the red ({\sc Riii}) and blue ({\sc
Biii}) GCs, respectively.  \emph{Lower panel}: Gauss--Hermite
parameter $h_4$ (Eq.~\ref{eq:kappa}) vs.~projected radius. The values
for the blue and red GCs are shown as long--dashed and dashed
rectangle, respectively. The solid grey rectangle shows the value for
the combined red and blue sample. The values derived from
absorption--line spectroscopy by \cite{saglia00} are shown as
diamonds, and the dashed line represents their best fit model
(cf.~their Fig.~5).}
\label{fig:kappa}
\end{figure}

In Fig.~\ref{fig:kappa} we compare our values to the results
presented by \cite{saglia00}. {{These authors used
Gauss--Hermite polynomials to analyse absorption--line spectra of the
stellar body of NGC\,1399 out to $97\arcsec{}(\simeq9\,\rm{kpc})$}}.
The radial ranges covered by both data sets only have a marginal overlap.
\paragraph{Stellar velocity dispersion profile:}
In the upper panel of Fig.~\ref{fig:kappa} we plot the velocity
dispersion profiles for the blue and red GCs (\textsc{Biii, Riii})
obtained for moving bins of 35 GCs (same as the bottom panels in
Fig.~\ref{fig:dispersion}). With the exception of the third data
point, the red GCs appear to follow the trend shown by the stars in
NGC\,1399. Only the innermost blue data point is comparable to the
stellar data, all subsequent bins show a much higher dispersion.
\paragraph{The fourth moment:}
In the lower panel of Fig.~\ref{fig:kappa} we compare Gauss--Hermite
$h_4$ values given by Saglia et al.~to the corresponding values for
the GCs which were converted from the $\kappa$ values given in
Table\,\ref{tab:losvd} using the following relation
\citep{marelfranx93}:
\begin{equation}
\kappa \simeq 8 \sqrt{6}\cdot h_4
\label{eq:kappa}
\end{equation}
We find that the value for the blue GCs (sample \textsc{Biii}) lies
significantly below the stellar values, while the value for the red
GCs (\textsc{Riii}) seems to agree with the $h_4$ values of the
stars. This figure also demonstrates that $\kappa$ is not additive,
which is illustrated by the grey area indicating the value obtained
when combining the blue (\textsc{Biii}) and red (\textsc{Riii})
sample.

\section{Jeans Models}
\label{sect:jeans}
The mass profile of NGC\,1399 is estimated by comparing the values
derived from spherical, non--rotating Jeans models to the observed
line--of--sight velocity dispersions.  The assumption of spherical
symmetry is justified by the near--spherical appearance of the
galaxy. To account for the degeneracy between mass and  orbital
anisotropy, which can only be broken using much larger datasets than
ours, we calculate models for different (constant) values of the
anisotropy parameter $\beta$.
\subsection{The Jeans equation }
The spherical, non--rotating Jeans equation reads:
\begin{equation}
\frac{\mathrm{d}\left(\ell(r)\, \sigma_{r}^{2}(r)\right)}{\mathrm{d}r} 
+2\, \frac{\beta(r)}{r}\,\ell(r)\,\sigma_{r}^{2}(r)= - \ell(r) \,\frac{G\cdot M(r)}{r^{2}} ,
\label{eq:jeans}
\end{equation}
\begin{displaymath}
\textrm{with}\qquad \beta \equiv 1 - \frac{{\sigma_{\theta}}^2}{{\sigma_{r}^2}}\;.
\end{displaymath} 

Here, $r$ is the {{true 3D}} radial distance from the centre and $\ell$ is the
spatial (i.e.,~three--dimensional) density of the GCs; ${\sigma_r}$
and ${\sigma_\theta}$ are the radial and azimuthal velocity
dispersions, respectively. $\beta$ is the anisotropy parameter, $M(r)$
the enclosed mass and $G$ is the constant of gravitation.\par For our
analysis, we use the expressions given by e.g.~\cite{mamonlokas05},
see also \cite{marelfranx93}.  Given a mass distribution $M(r)$, a
three--dimensional number density of a tracer population $\ell(r)$,
and a \emph{constant} anisotropy parameter $\beta$, the solution to
the Jeans equation (Eq.~\ref{eq:jeans}) reads:
\begin{equation}
\ell(r)\, {\sigma_{r}^{2}(r)} =  \mathrm{G} \int_r^\infty \ell(s) \, M(s) \frac{1}{s^{2}} \left( \frac{s}{r} \right)^{2 \beta} \mathrm{d} s .
\label{eq:sigl}
\end{equation}
This expression is then projected using the following integral:
\begin{equation}
\sigma_{\mathrm{los}}^{2}(R)= \frac{2}{N(R)} \left[ \int_{R}^{\infty}  \frac{\ell \sigma_{r}^{2}\, r\, \mathrm{d}r}{\sqrt{r^{2}-R^{2}}}
- R^{2}\int_{R}^{\infty}  \frac{\beta \ell \sigma_{r}^{2}\,  \mathrm{d}r}{r \sqrt{r^{2}-R^{2}}} \right]\,,
\label{eq:siglos}
\end{equation}
where $N(R)$ is the projected number density of the tracer population,
and $\sigma_{\rm{los}}$ is the line--of--sight velocity dispersion, to
be compared to our observed values. In the following, we discuss the
quantities required to determine $\sigma_{\rm{los}}(R)$.
\subsection{Globular cluster number density profiles}
\begin{table}
\caption{Globular cluster number density profiles: Fitting a cored
power--law (Eq.~\ref{eq:corepl}) to the data given in Table\,2 of
\cite{bassino06} yields the parameters listed below (the profile for
`all' GCs was obtained by adding the number counts of the red and
blue GCs).  } \centering
\begin{tabular}{lr@{$\pm$}lr@{$\pm$}lr@{$\pm$}l r@{.}l } 
\hline  \hline
& \multicolumn{2}{c}{$N_0$} & \multicolumn{2}{c}{$R_{0}$} 
& \multicolumn{2}{c}{$\alpha$} 
& \multicolumn{2}{c}{$\mathcal{B}\left(\frac{1}{2},\alpha \right)$} 
\\ 

& \multicolumn{2}{c}{[$\rm{GCs\, arcmin}^{-2}$]} & \multicolumn{2}{c}{[arcmin]} 
& \multicolumn{2}{c}{} 
& \multicolumn{2}{c}{}
\\ \hline
Red & 28.43&7.82 &1.63 &0.34  & 1.02&0.04 & 1&98 \\
Blue &8.39 &1.05 &2.91 & 0.42 & 0.79&0.03  & 2&32\\
All & 35.54&6.13 &1.74 & 0.27 & 0.84 & 0.02 & 2&22 \\
\hline \hline
\end{tabular}
\label{tab:coef}
\end{table}
\label{sect:powerlaws}
The GCS of NGC\,1399 has been the target of two wide--field
photometric studies. The work presented by \cite[hereafter B+06]{bassino06} extends the earlier D+03 study
upon which the analysis presented in Paper\,I was based. 
For our analysis, we
use a cored power--law profile (Reynolds--Hubble law) to fit the data
from B+06:
\begin{equation}
N(R) = N_{0} 
\left( 1 + \left(\frac{R}{R_{\mathrm{0}}}\right)^2 \right)^{-\alpha}\,.
\label{eq:corepl}
\end{equation}
Here, $R_0$ is the core radius, and $2\cdot\alpha$ is the slope of the
power--law in the outer region.  For the above expression, the Abel
inversion has an analytical solution (i.e.~Eq.~\ref{eq:corepl} can be
deprojected exactly, e.g.~\citealt{saha96}), and the
three--dimensional number density profile reads:
\begin{equation}
\ell(r) = \frac{N_{0}}{R_0} 
\frac{1}{\mathcal{B} \left(\frac{1}{2},\alpha \right)} 
\cdot
\left( 1 + 
\left(\frac{r}{R_{\mathrm{0}}}\right)^2 \right)^{-\left(\alpha+\frac{1}{2}\right)}\,,
\label{eq:depropl}
\end{equation}
where $\mathcal{B}$ is the Beta function. Table\,\ref{tab:coef} lists
the values obtained by fitting Eq.~\ref{eq:corepl} to the number
density profiles given in Table\,2 of B+06. For the red GCs, the
fit was performed for $R<35\arcmin$, and $R< 45\arcmin$ for the
blue/all GCs.  \par
\begin{figure}
\resizebox{\hsize}{!}
{\includegraphics[width=0.48\textwidth]{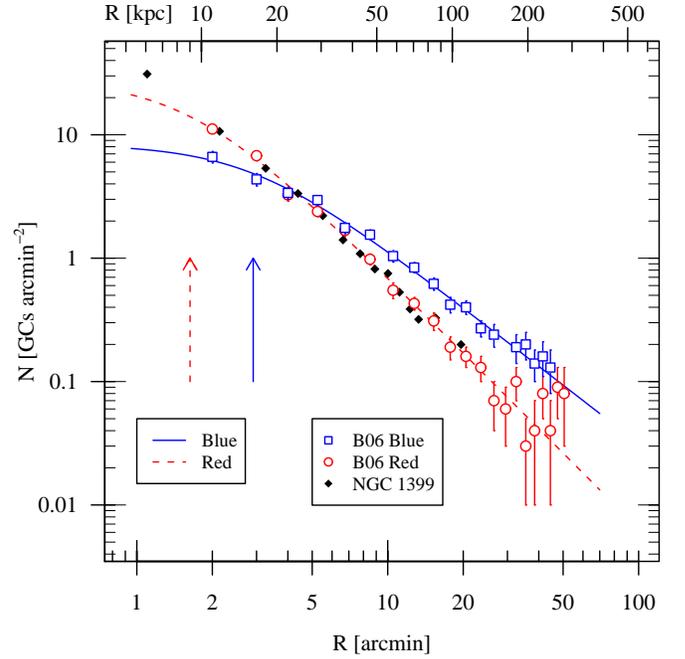}}
\caption{GC number density profiles. The data from B+06 are shown
together with the fits to Eq.~\ref{eq:corepl}. Circles and squares
represent the red and blue GCs, respectively. The fits to the red and
the blue number density distributions are shown as dashed and solid
line, respectively. The coefficients are listed in
Table\,\ref{tab:coef}. The dashed and solid arrows indicate the core
radii $R_0$ as returned from the fits for the red and blue GCs,
respectively. Diamonds show the surface brightness profile of
NGC\,1399, as given in Table\,4 of D+03 (the data points have been
shifted for comparison with the red GC number density profile).  }
\label{fig:numdens}
\end{figure}

 Figure\,\ref{fig:numdens} shows the number density
profiles for the blue (open squares) and red GCs (circles) as given in
B+06, together with the corresponding fits. One clearly sees that the
red GCs have a steeper number density profile than the blue
GCs. For reference, the surface brightness profile of NGC\,1399
(as given by D+03, but scaled to match the red GC profile) is shown
with diamonds. Note that in the region of overlap ($2\arcmin\lesssim
R \lesssim 20\arcmin$) the slope of the densities of the stars and
the red GCs are indistinguishable.

\par
\subsection{Luminous matter}
\label{sect:luminous}
In order to assess the dark matter content of NGC\,1399, we first need
 a model for the luminous matter. Moreover, the spatial distribution
 of the stars is needed to model the stellar velocity dispersion
 profile (see Sect.~\ref{sect:saglia}).

\subsubsection{Deprojection of the surface brightness profile}
D+03 give the following fit to the $R$--band surface brightness profile
of NGC\,1399:
\begin{equation}
\mu(R)= 2.125 \log \left[ 1+ \left(\frac{R}{0\farcm055}
\right)^2\right] + 15.75 \; ,
\label{eq:lumprof}
\end{equation}
where $R$ is the projected radius.  Assuming an absolute solar
luminosity of $M_{\odot,R} = 4.28$, the surface brightness profile in
units of $L_{\odot}\,\rm{pc}^{-2}$ reads:
\begin{equation}
I(R) = 1.10\times 10^4 \left[1+ \left( \frac{R}{0\farcm055}\right)^2
\right]^{-0.85} \; ,
\label{eq:surfb}
\end{equation}
At a distance of 19\,Mpc, $0\farcm055$ correspond to
$304\,\rm{pc}$. Using Eq.~\ref{eq:depropl} we obtain the
deprojected profile (in units of $L_{\odot}\,\rm{pc}^{-3}$):
\begin{equation}
j(r)\left[ \frac{L_{\odot}}{\rm{pc}^{3}} \right] = 16.33 \left[1+ \left( \frac{r}{304\,\rm{pc}}\right)^2 
\right]^{-1.35} \; . 
\label{eq:mylum}
\end{equation}
We now compare this to the deprojected profile given in D+03. These
authors obtained the three--dimensional luminosity distribution by
numerically projecting a cored power-law and compare the result to
the fitted surface brightness profile. The luminosity density profile
they obtained using this procedure reads:
\begin{equation}
L_{\rm{T_1}}\left[ \frac{L_{\odot}}{\rm{pc}^{3}} \right]= 101 \left[ 1+ \frac{r}{221\,\rm{pc}} \right]^{-2.85} \; .
\label{eq:dirschlum}
\end{equation}
One notes that the central luminosity density of this profile is more
than six times higher than what we find in
Eq.~\ref{eq:mylum}. Further, the profile presented by D+03 is slightly
steeper.\par
\begin{figure}
\resizebox{\hsize}{!}
{\includegraphics[width=0.48\textwidth]{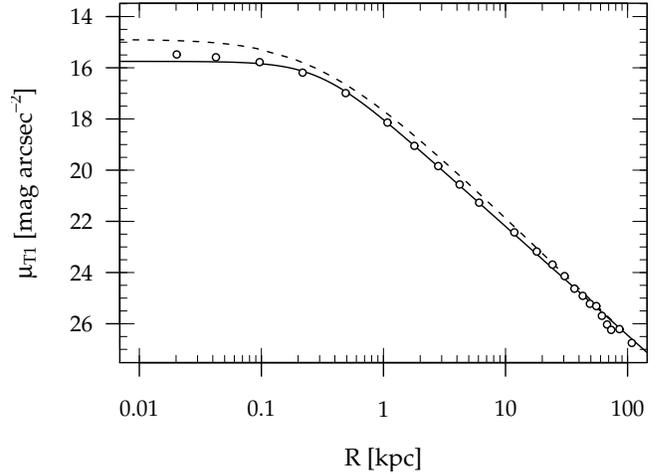}}
\caption[]{Comparison between different deprojections of the surface
brightness profile of NGC\,1399: The dashed curve is the
re--projection of the luminosity density profile as given in
D+03 (Eq.~\ref{eq:dirschlum}). The solid curve shows the fit to the data
(Eq.~\ref{eq:lumprof}), which corresponds to the re--projection of
Eq.~\ref{eq:mylum}. }
\label{fig:deprorepro}
\end{figure}

To show the difference between the two deprojections, we plot in
Fig.~\ref{fig:deprorepro} the surface brightness profile of NGC\,1399
as given in Table\,4 of D+03, together with the fit
(Eq.~\ref{eq:lumprof}, solid line) and the re--projection of their
luminosity density profile (dashed line).  The latter overestimates
the measured surface brightness of NGC\,1399 and therefore results in
a larger stellar mass. 

\subsubsection{Stellar mass--to--light ratio:}
In Paper\,I, Richtler et al.~quote an $R$--band stellar
mass--to--light ratio of $M/L_R=5.5$, which they obtained by
converting the $B$--band value derived by \cite{saglia00}. Recently,
\cite{gebhardt07} used the Hubble Space Telescope to
obtain kinematics for the very central part of NGC\,1399 in order to
measure the mass of the nuclear black hole. From their dynamical
models, these authors find a best--fit $M/L_R=5.2\pm0.4$ for a
distance of $21.1\,\rm{Mpc}$. At a distance of $19.0\,\rm{Mpc}$, as
assumed in this study, this corresponds to $M/L_R=5.77\pm0.4$, which
is in excellent agreement with the value given in Paper\,I. We thus
adopt a stellar mass--to--light ratio of $\Upsilon_{\star}=M/L_R=5.5$.
\subsubsection{Enclosed stellar mass}
We obtain the enclosed stellar mass by integrating Eq.~\ref{eq:mylum}:
\begin{equation}
M_{\rm{stars}}(R) = 4 \pi \Upsilon_{\star} \int_{0}^{R} j(r) r^2
\rm{d}r\:, \end{equation} which
can be expressed in terms of
the hypergeometric function:
\begin{equation}
M(R) = C_1 \cdot \frac{3}{4} \pi \Upsilon_{\star}\, R^3 
\,_{2}F_{1}\left(\left[\frac{3}{2},\gamma \right],\left[ \frac{5}{2}\right],
-\frac{R^2}{B^2} \right),
\label{eq:hypgeo}
\end{equation}
where $C_1=16.33$ is the constant, $B=304\,\rm{pc}$ the core radius,
and $\gamma=1.35$ is the exponent from
Eq.~\ref{eq:mylum}. $_2F_1(a,b;c;z)$ is the hypergeometric function as
defined in e.g. \cite{AS}.\par In practice, the
hypergeometic function can be calculated to arbitrary precision
(e.g.~\citealt{NUMREC}), but the computation of the integral
over $M(R)$ becomes very time--consuming. In the programme calculating
the Jeans--models, we therefore apply a piece--wise definition of the
mass profile, using only standard functions: For radii smaller than
3430\,pc, we use a $6^{\rm{th}}$-order polynomial, and a
$3^{\rm{rd}}$-order polynomial in the range $3430 \leq R \leq
9200$. For $R> 9200\,\rm{pc}$ the following expression is a good
approximation to Eq.~\ref{eq:hypgeo}:
\begin{eqnarray}
\lefteqn{M(R) =
C_2\cdot 4 \pi \Upsilon_{\star} R_{0,\star}^3 \cdot } \nonumber\\ 
&& \left( 
\ln\left(1+ \frac{R}{R_{0,\star} } +  \frac{R^2}{R_{0,\star}^2 }  \right)
- \frac{R}{R_{0,\star}} \left(1+\frac{R^2}{R_{0,\star}^2 }\right)^{-1} 
+ \sqrt{\frac{R}{ R_{0,\star}}}
\, \right)\; , 
\nonumber\\
\end{eqnarray}
where  $R_{0,\star}=1598\,\rm{pc}$, and $C_2=0.106$.
\subsection{The anisotropy parameter $\beta$}
The solutions to the Jeans equation given in Eq.~\ref{eq:sigl} are
defined for \emph{constant} values of the anisotropy parameter. Since $\beta$
cannot be determined from the data itself, we calculate our models for
a set of anisotropy parameters.  The higher moments derived in
Sect.~\ref{sect:himo} suggest that the red cluster population is most
likely isotropic, possibly slightly radial, while the blue GCs might
show a mild tangential bias. We therefore use $\beta = -0.5,0,+0.5$
for modelling the GCs.
\subsection{Dark matter profiles}
Numerical cosmological simulations predict cuspy dark matter density
profiles (e.g.~\citealt{bullock01}). However, the very
inner shape apparently depends on numerical details (e.g.~\citealt{diemand05}), while agreement is reached regarding the outer
profile which declines as $R^{-3}$.  Here we use an NFW \citep{nfw97} halo to represent a cuspy dark component. The
density profile is given by:
\begin{equation}
\varrho_{\rm{NFW}}(r)=  \frac{\varrho_s}{\left( \frac{r}{r_s}\right) 
\left( 1 + \frac{r}{r_s}\right)^2} \, ,
\label{eq:rhonfw}
\end{equation}
and the cumulative mass reads:
\begin{equation}
M_{\rm{NFW}}(r)= 4 \pi \varrho_s r_s^3 \cdot \left( 
\ln \left( 1 + \frac{r}{r_s} \right) -
\frac{\frac{r}{r_s}}{1+ \frac{r}{r_s}}
\right) \ .
\label{eq:nfw}
\end{equation}
However, in low surface brightness galaxies, where dark matter
contributes significantly already in the central regions, most
observations are not compatible with a cuspy halo \citep{gentile07}
but rather with a cored density profile. We therefore also model a
dark matter density profile with a core. One possibility is to use the
density profile, which \cite{burkert95} introduced to represent the
dark matter halo of dwarf galaxies.  For this halo, the density
profile is:
\begin{equation}
\varrho(r) = \frac{\varrho_0}{\left( 
  1+ \frac{r}{r_{0}}\right) \left(1+ \frac{r^2}{{r_0^2}} \right)} \; ,
\label{eq:rhoburkert}
\end{equation}
and the cumulative mass is given by the following expression:
\begin{eqnarray}
\lefteqn{M(r) =} \nonumber\\
  & &  4 \pi \varrho_{\rm{0}} {r_{0}^3} \left( 
\frac{1}{2} \ln \left(  1 + \frac{r}{r_0}\right) +
\frac{1}{4} \ln \left( 1 + \frac{r^2}{r_0^2}\right) -
\frac{1}{2} \arctan \left( \frac{r}{r_0}\right) 
\right) \nonumber \; .
\\
\label{eq:massburkert}
  \end{eqnarray}

\section{Mass models for NGC\,1399}
\label{sect:massmodels}
\begin{table*}
\label{tab:bergond}
\caption{Jeans modelling best fit parameters: {\it{(a)}} using our GC
  dataset (i.e.~samples \textsc{Riii} and \textsc{Biii}), {\it{(b)}}
  using the extended data set including the values from
  \cite{bergond07} (\textsc{Rv} and \textsc{Bv}), and {\it{(c)}} for
  the stellar velocity dispersion profile by \cite{saglia00}.  The
  tracer population is given in the first column, the anisotropy
  parameter $\beta$ in the second. Columns 3 and 4 list the best--fit
  NFW halo parameters. The following columns give the $\chi^2$ and
  $\chi^2/\nu$, where $\nu$ is the number of degrees of freedom
  (i.e. $k-1-p$, where $k$ is the number of data points and $p=2$ the
  number of free parameters). Columns 7--9 give the virial parameters
  of the dark halos, as defined by Bullock et al.~(2001),
  i.e.~assuming a density contrast of $\Delta=337$. The parameters for
  the \cite{burkert95} halo are shown in Cols.~10--12.
} \centering
\begin{tabular}{rlrrllllrrrlr}\hline \hline
&&&\multicolumn{7}{|c|}{NFW dark halo} & \multicolumn{3}{c}{Burkert halo}\\
  & & \multicolumn{1}{r|}{$\beta$} & $r_s$ & $\varrho_S$ & $\chi^2$ &$\chi^2/\nu$&$M_{\rm{vir}}$&
$R_{\rm{vir}}$&\multicolumn{1}{c|}{$c_{\rm{vir}}$} & $r_{0}$ & $\varrho_0$ & $\chi^2$\\ 

&  &  \multicolumn{1}{r|}{} & $[\rm{kpc}]$ & $[M_{\odot}\,\rm{pc}^{-3}]$ & & &$[10^{12}M_{\odot}]$ &$[\rm{kpc}]$& \multicolumn{1}{c|}{}& $[\rm{kpc}]$  & $[M_{\odot}\,\rm{pc}^{-3}]$\\

&   \multicolumn{1}{l}{(1)} & \multicolumn{1}{r|}{(2)} & (3) &(4) &(5) &(6) &(7)& (8)&\multicolumn{1}{c|}{(9)} &(10) & (11) & (12)\\

  \hline

  \multicolumn{10}{l}{{{({\it{a}}) NGC\,1399 GC data (\textsc{Riii} and \textsc{Biii}):}}}\\\hline

\it{a1}  &Red (\textsc{Riii})&$-0.5$ & 14 & 0.055 & 0.72 & 0.24 & 4.7& 430 & 31  & 11 & 0.0896 & 0.49\\
\it{a2}  &Red (\textsc{Riii})& 0     & 16 & 0.042 & 0.85 & 0.28 & 5.2& 450 & 28  & 12 & 0.0776 & 0.57\\
\it{a3}  &Red (\textsc{Riii})&$+0.5$ & 24 & 0.019 & 1.02 & 0.34 & 7.9& 490 & 20  & 16 & 0.0447 & 0.72 \rule[-1.5ex]{0ex}{2ex}\\ 

\it{a4}  &Blue (\textsc{Biii})&$-0.5$ & 101      & 0.00271 & 8.5 & 2.1 & 51 & 950   & 9  & 35 & 0.0213 & 6.9\\
\it{a5}  &Blue (\textsc{Biii})&0      & 174      & 0.00099 & 9.5 & 2.4 & 73 & 1075  & 6  & 46 & 0.0119 & 8.0\\
\it{a6}  &Blue (\textsc{Biii})&$+0.5$ &$\geq 200$& 0.00061 & 12  & 3.0 & 58 & 1000  & 5  & 94 & 0.0031 & 9.9\\ \hline

\it{a7}  &\multicolumn{2}{l}{$\rm{Red}_{0.0}$ and $\rm{Blue}_{0.0}$}   & 63 & 0.0038 & 28& 2.8 & 18& 680 & 11 &25 & 0.0237 & 28\\ 
\it{a8}  &\multicolumn{2}{l}{$\rm{Red}_{+0.5}$ and $\rm{Blue}_{-0.5}$} & 37 & 0.0110 & 29& 2.9 & 13& 610&  16 & 20 &0.0388 & 27 \\ \hline

\it{a9}  &\multicolumn{2}{l}{$\rm{Red}_{0.0}$ and  $\rm{Stars}_{0.0}$}  & 23 & 0.0190 &  8.3& 0.42 &6.1 & 470& 20 & 9 & 0.1343& 14\\ 
\it{a10} &\multicolumn{2}{l}{$\rm{Red}_{0.0}$ and  $\rm{Stars}_{+0.5}$} & 34 & 0.0088 &  7.3& 0.37 &8.0 & 510& 15 & 12 &0.0728& 8.7 \\ \hline

  \multicolumn{10}{l}{{{({\it{b}}) NGC\,1399 GCs, extended data set  including  the Bergond et al.~(\citeyear{bergond07}) GC data (\textsc{Rv} and \textsc{Bv}):}}}\\\hline 
\it{b1}  &Red (\textsc{Rv})&$-0.5$& 12 & 0.093 & 3.8 & 0.63 &5.4& 450& 38 &  9 & 0.166 & 3.4\\ 
\it{b2}  &Red (\textsc{Bv})& 0    & 14 & 0.069 & 4.0 & 0.67 &6.1& 470& 34 & 11 & 0.112  & 3.5\\
\it{b3}  &Red (\textsc{Bv})&$+0.5$& 19 & 0.038 & 4.5 & 0.75 &7.6& 510& 27 & 14 & 0.071&3.9 \rule[-1.5ex]{0ex}{2ex}\\
\it{b4}  &Blue (\textsc{Bv})&$-0.5$ & 36 & 0.0140 & 14 & 2.0 & 17 & 660 &18 & 22 & 0.0416 & 11\\
\it{b5}  &Blue (\textsc{Bv})&0      & 50 & 0.0071  & 16 & 2.3 & 20 & 690 &14 & 28 & 0.0243 & 13\\
\it{b6}  &Blue (\textsc{Bv})&$+0.5$ & 94 & 0.0018  & 19 & 2.7 & 25 & 750 & 8 & 43 & 0.0094 & 16\\ \hline
\it{b7}  &\multicolumn{2}{l}{$\rm{Red}_{0.0}$ and  $\rm{Blue}_{0.0}$ }   & 35 & 0.012 &  29&1.8 &13 &600&17 & 20 & 0.03876 & 27 \\ 
\it{b8}  &\multicolumn{2}{l}{$\rm{Red}_{+0.5}$ and  $\rm{Blue}_{-0.5}$ } & 20 & 0.019 &  27& 1.7&13 &590&20 & 18 & 0.05124 & 24\\  \hline 

\it{b9}  &\multicolumn{2}{l}{$\rm{Red}_{0.0}$ and  $\rm{Stars}_{0.0}$ }  & 30 & 0.013 &  13&0.57&8.9&535&18 & 11 &0.1064  & 19 \\ 
\it{b10}
  &\multicolumn{2}{l}{$\rm{Red}_{0.0}$ and  $\rm{Stars}_{+0.5}$} & 38 & 0.0077 &  14& 0.61&9.5 &545&14 &15 & 0.05653 & 14\\  \hline

  \multicolumn{10}{l}{{{({\it{c}}) Stellar velocity dispersion profile (data from Saglia et al.~\citeyear{saglia00}):}}}\\\hline

\it{c1} & Stars&0        & 23 & 0.019  & 7.3 & 0.52 &   6.1 & 470 &20  & 5  & 0.2996 & 10\\
\it{c2} & {Stars}&$+0.5$ & 47 & 0.0055 & 5.5 & 0.39&  12  & 590 &13  & 8 & 0.12545 & 7.5 \\ \hline

\hline \hline

\label{tab:models}
\end{tabular}
\end{table*}
In this Section, we present the results of the Jeans modelling. First,
we treat the tracer populations separately, i.e.~we fit the velocity
dispersion profiles of the red and blue GC samples using NFW halos. In
a second step, we search for a set of parameters to describe both
populations. The stellar velocity dispersion profile by
\cite{saglia00} is then used to put further constraints on the mass
model of NGC\,1399. Finally, the results for the NFW halos are
compared to cored Burkert halo models.
\subsection{Jeans analysis}
\label{sect:modcomp}
To obtain an estimate for the NGC\,1399 mass profile, we compare the
observed velocity dispersion profiles to Jeans models
(Eq.~\ref{eq:siglos}). 
The stellar mass--to-light ratio is assumed to be constant (see
Sect.~\ref{sect:luminous}), and the (constant) anisotropy parameter
takes the values $\beta = -0.5, 0$, or $+0.5$ (corresponding to a mild
tangential, isotropic, and a slightly radial orbital bias). To find
the best Jeans model, we adjust the parameters of the dark halo. \\
The dark matter halos considered here (NFW and Burkert halos) are
characterised by two parameters, a scale radius $r_{\rm{dark}}$ and a
density $\varrho_{\rm{dark}}$. For a given tracer population and
anisotropy $\beta$, we calculate a grid of models where the density
acts as free parameter while the radii have discrete values,
i.e. $r_{\rm{dark}} \in\{1,2,3,\ldots,200\}\,\rm{kpc}$.  The results
of this modelling are summarised in Table\,\ref{tab:models}, which
quotes the best--fit halo parameters and the corresponding $\chi^2$
values for the different tracer populations and anisotropies. \par The
confidence level (CL) contours are calculated using the definition by
\cite{avni76}, i.e.~using the difference $\Delta\chi^2$ above the
minimum $\chi^2$ value. With two free parameters,
e.g.~$(r_{\rm{dark}},\varrho_{\rm{dark}})$ the 68, 90, and 99\,per
cent contours correspond to $\Delta\chi^2=2.30, 4.61$, and $9.21$,
respectively.\par
\begin{figure*}
\begin{tabular}{|ccc|}\hline
\includegraphics[width=0.3\textwidth]{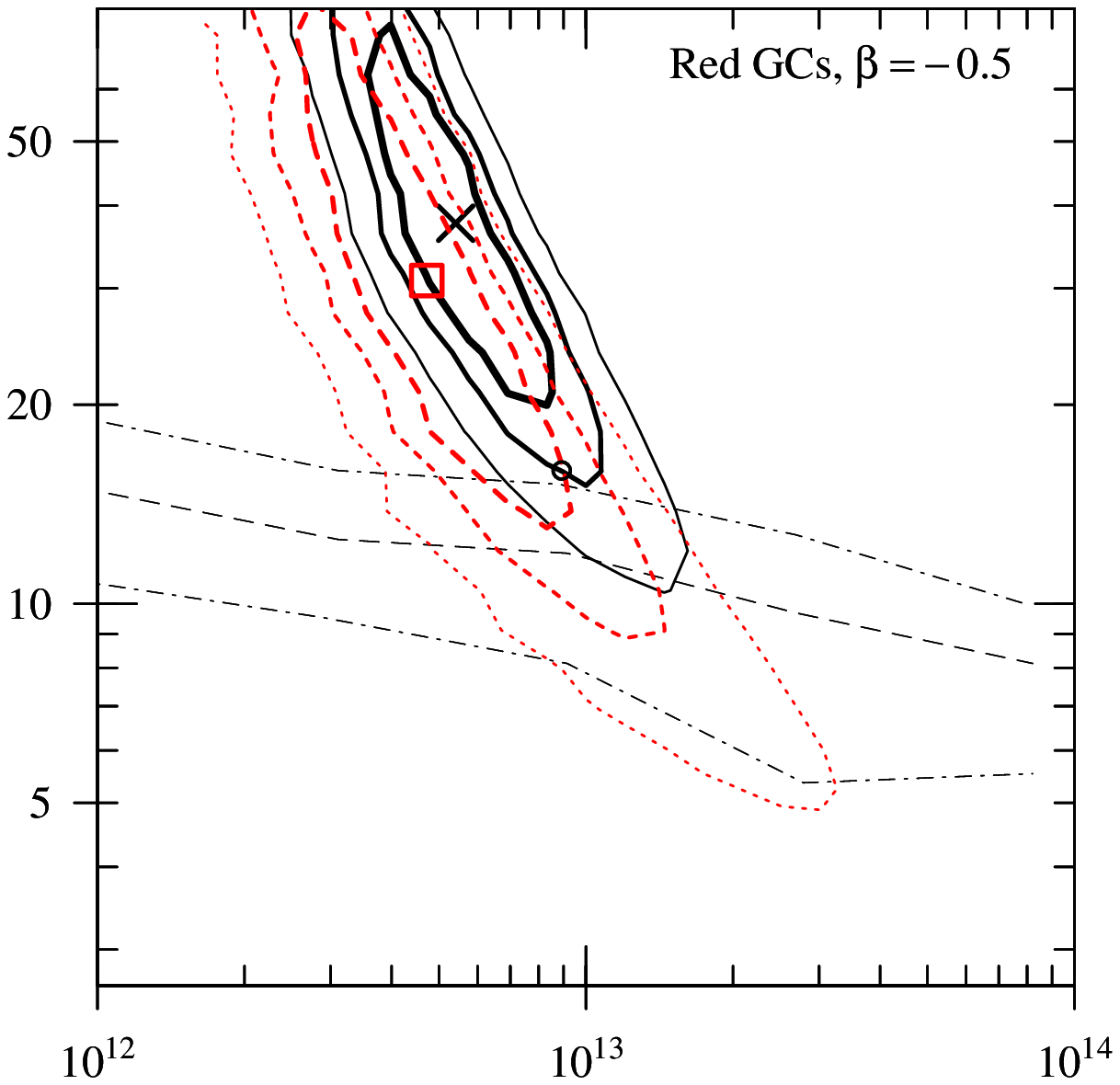}&
\includegraphics[width=0.3\textwidth]{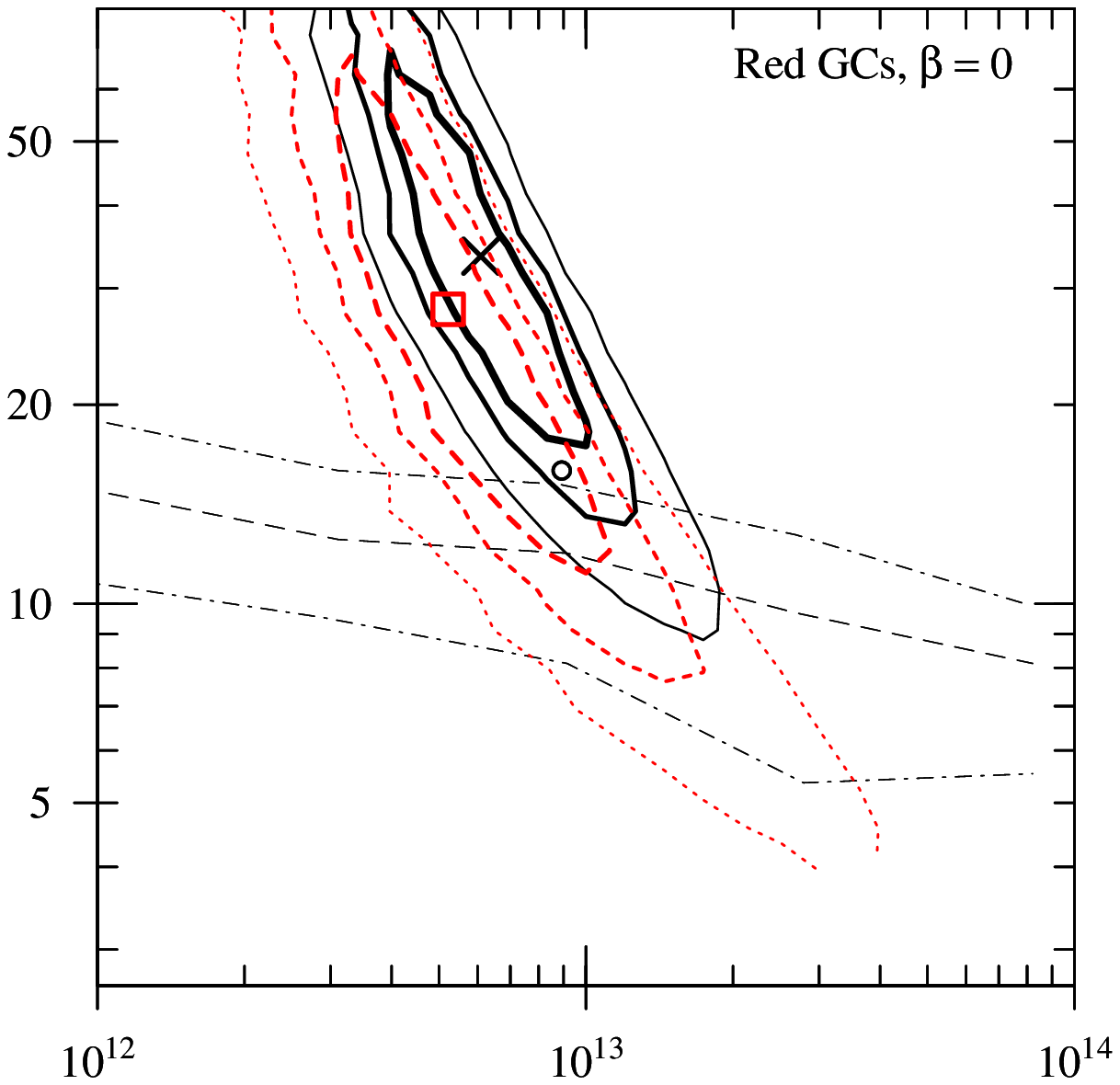}&
\includegraphics[width=0.3\textwidth]{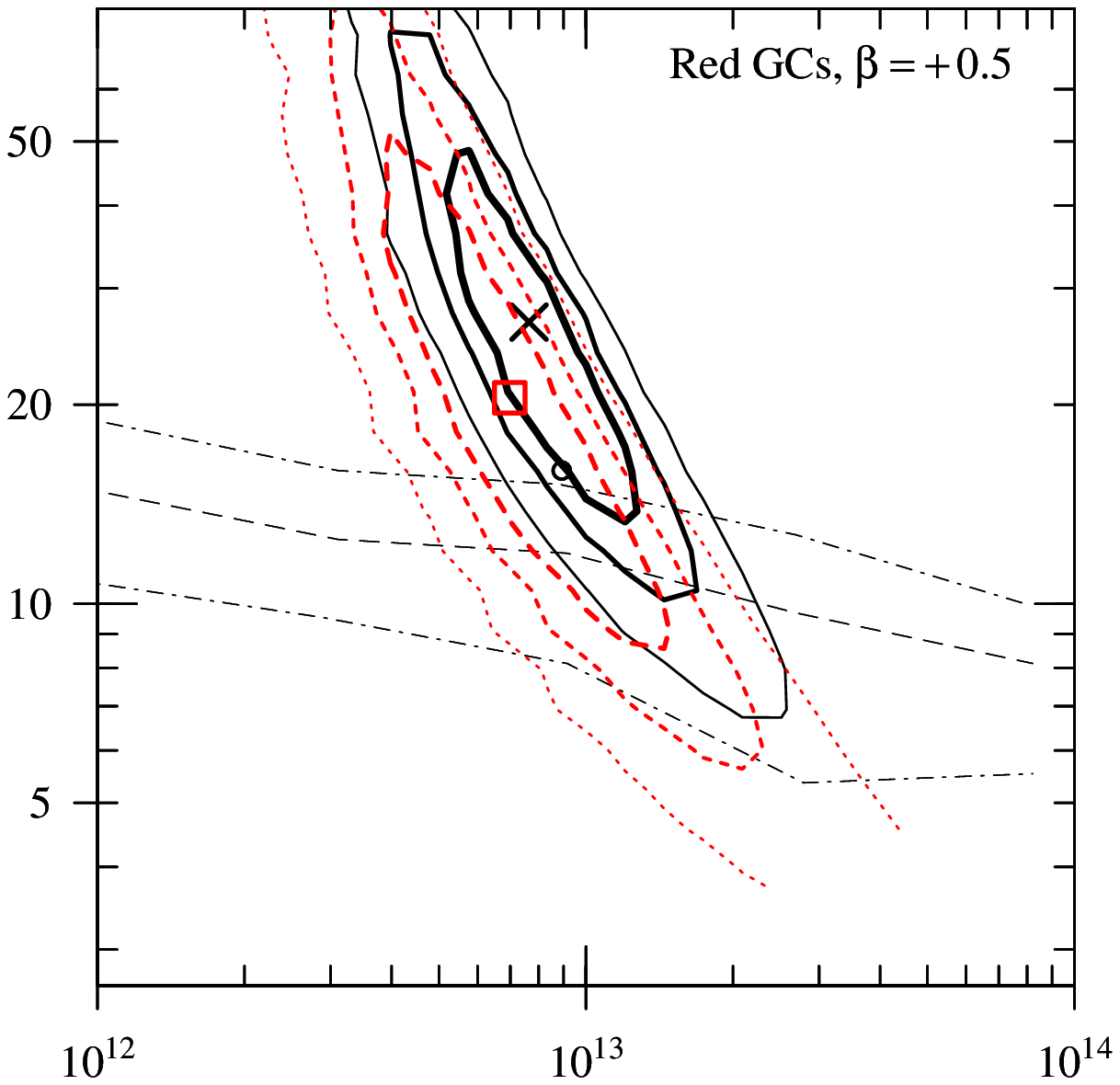}\\
\includegraphics[width=0.3\textwidth]{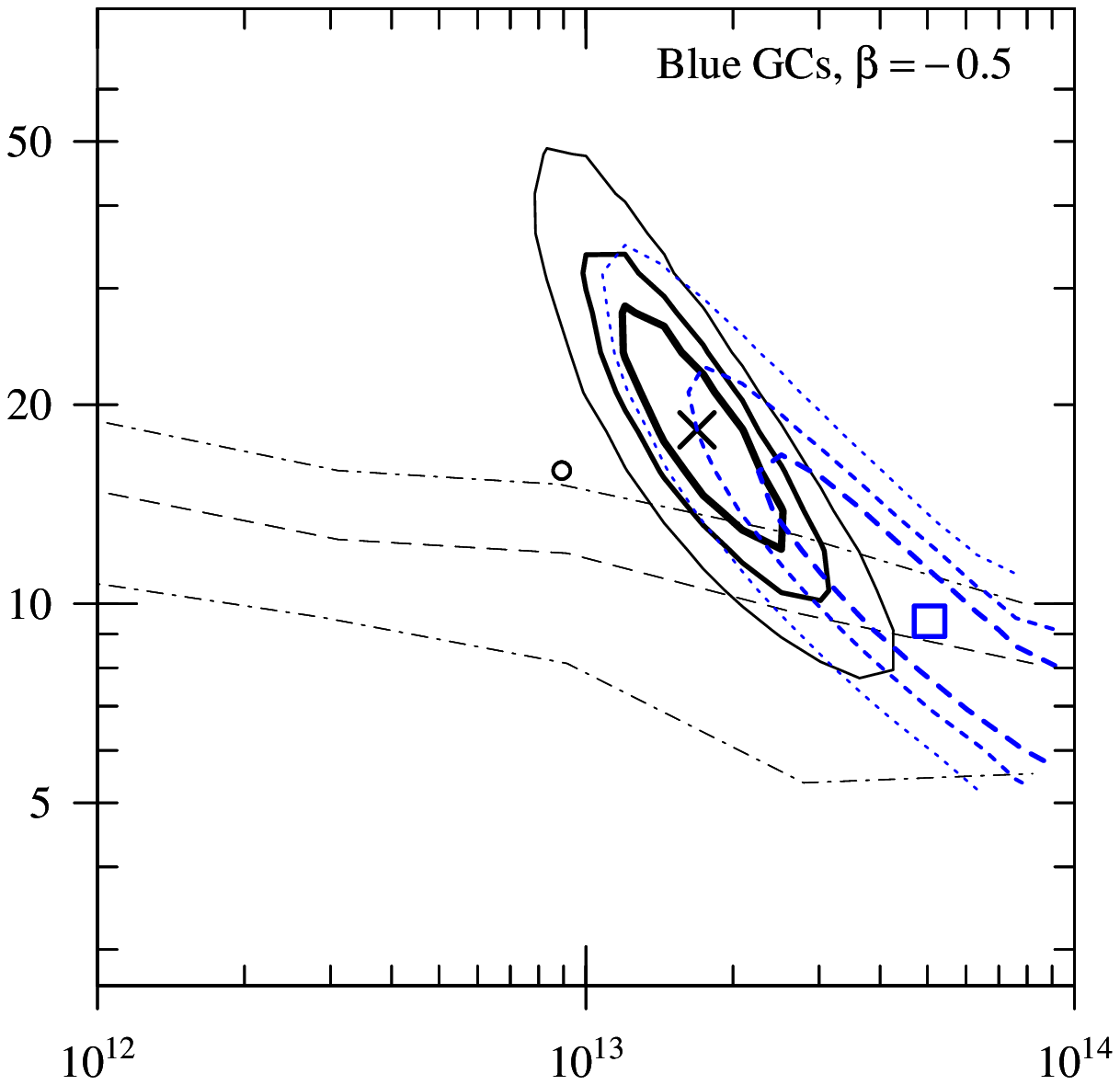}&

\includegraphics[width=0.30\textwidth]{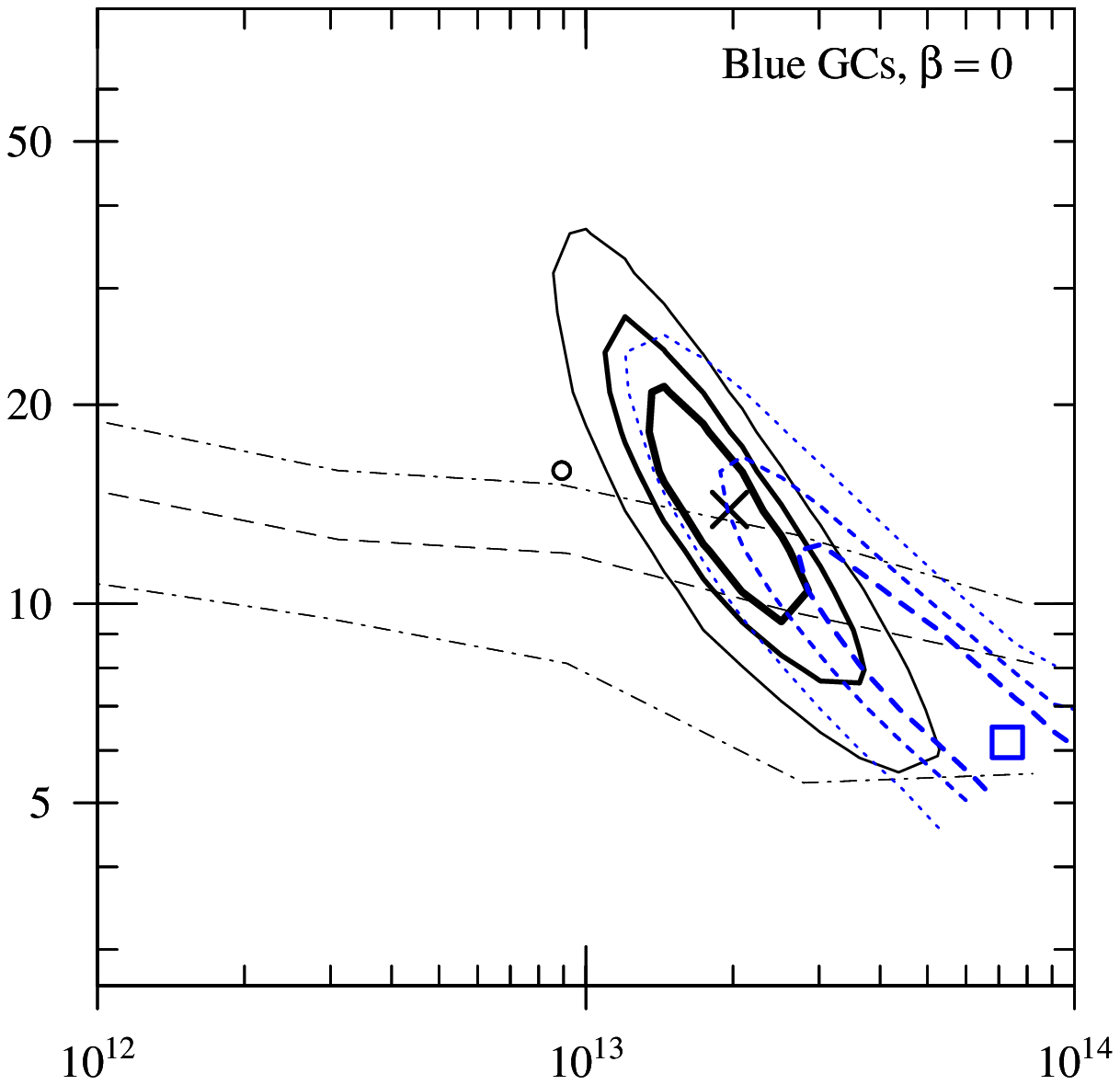}&
\includegraphics [width=0.30\textwidth,]{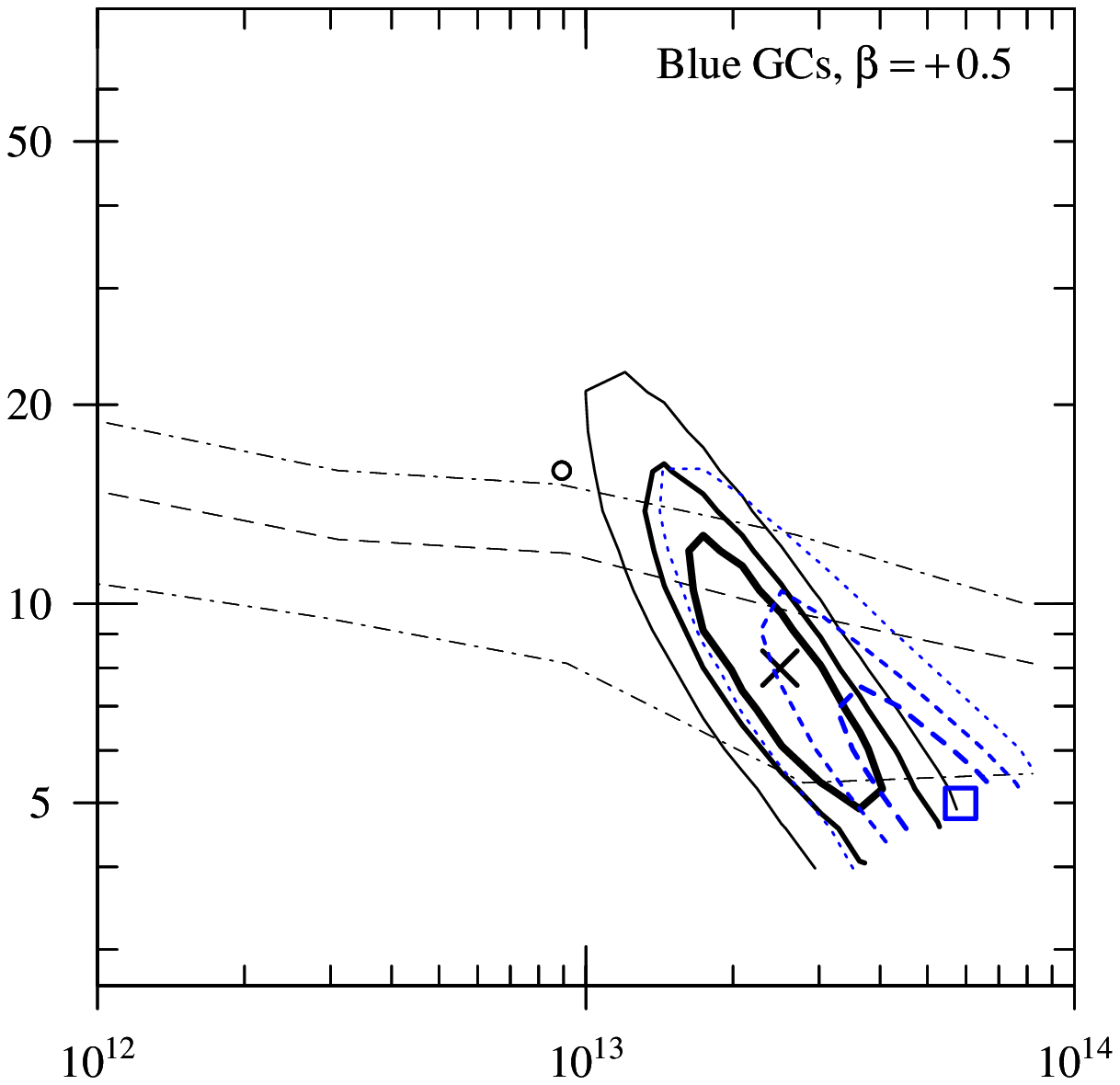}\\ \hline \hline
\multicolumn{1}{|c}{\includegraphics[width=0.30\textwidth]{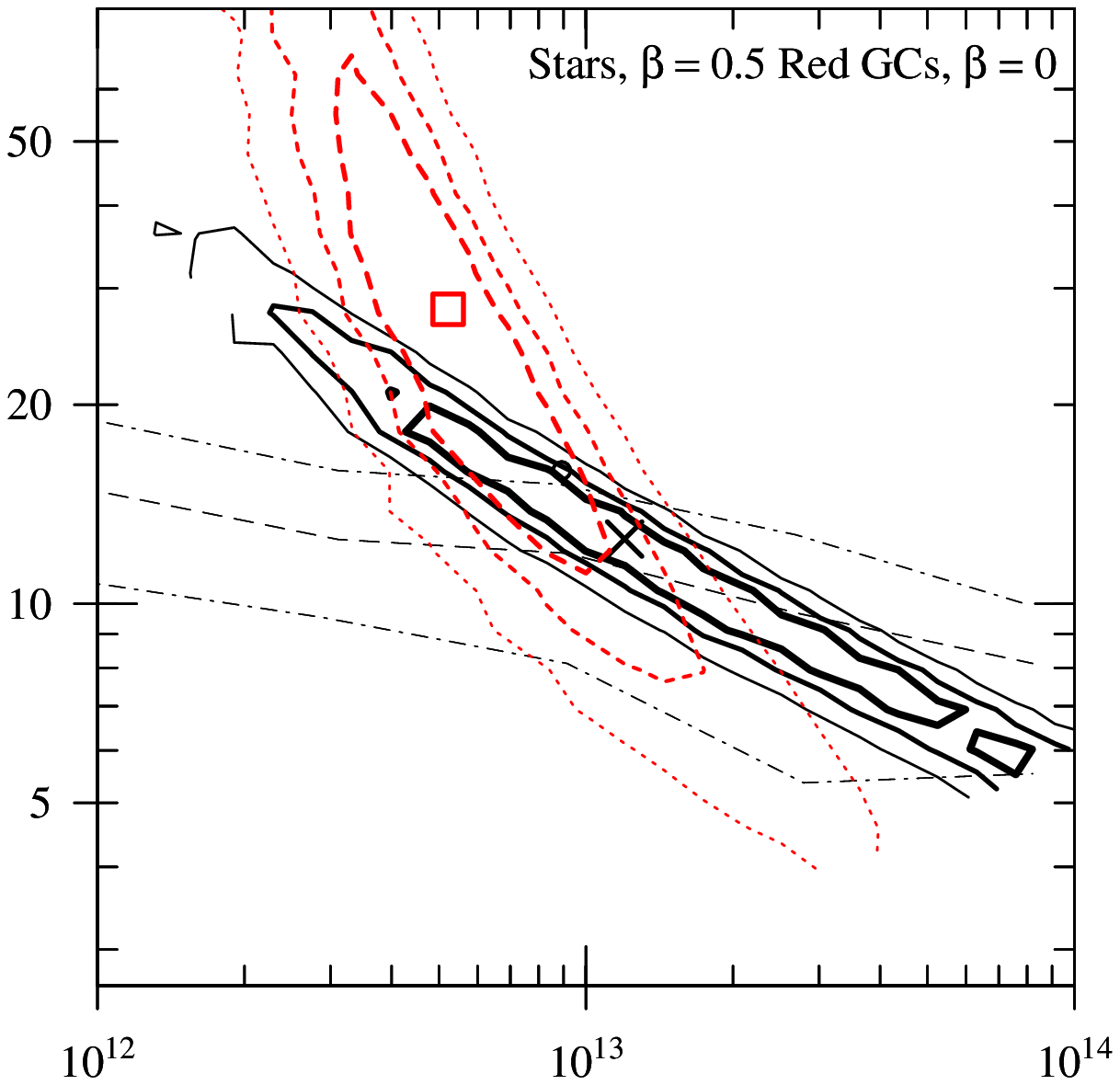}} &
\multicolumn{1}{c|}{\includegraphics[width=0.3\textwidth]{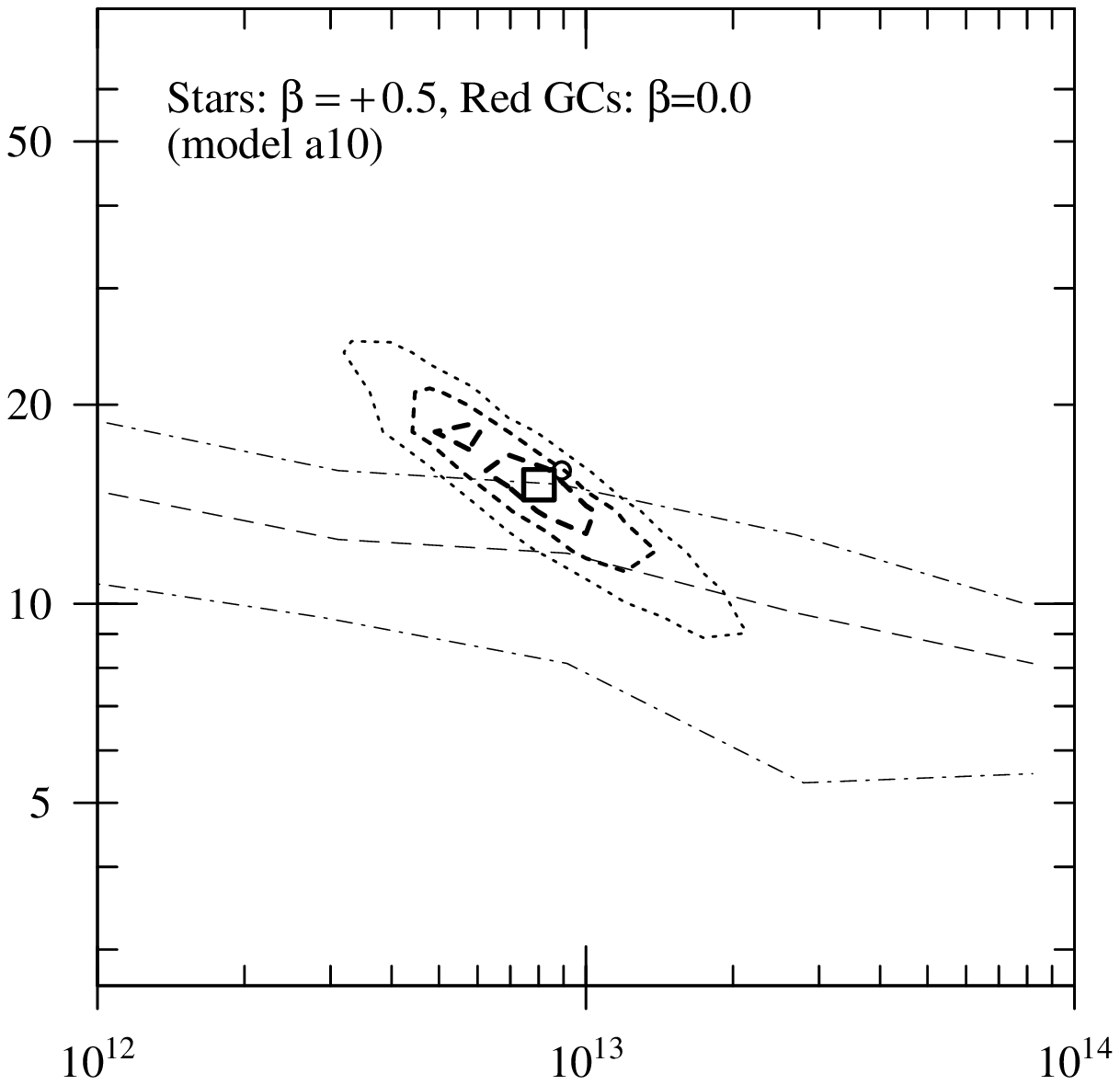}} &
\includegraphics[width=0.30\textwidth]{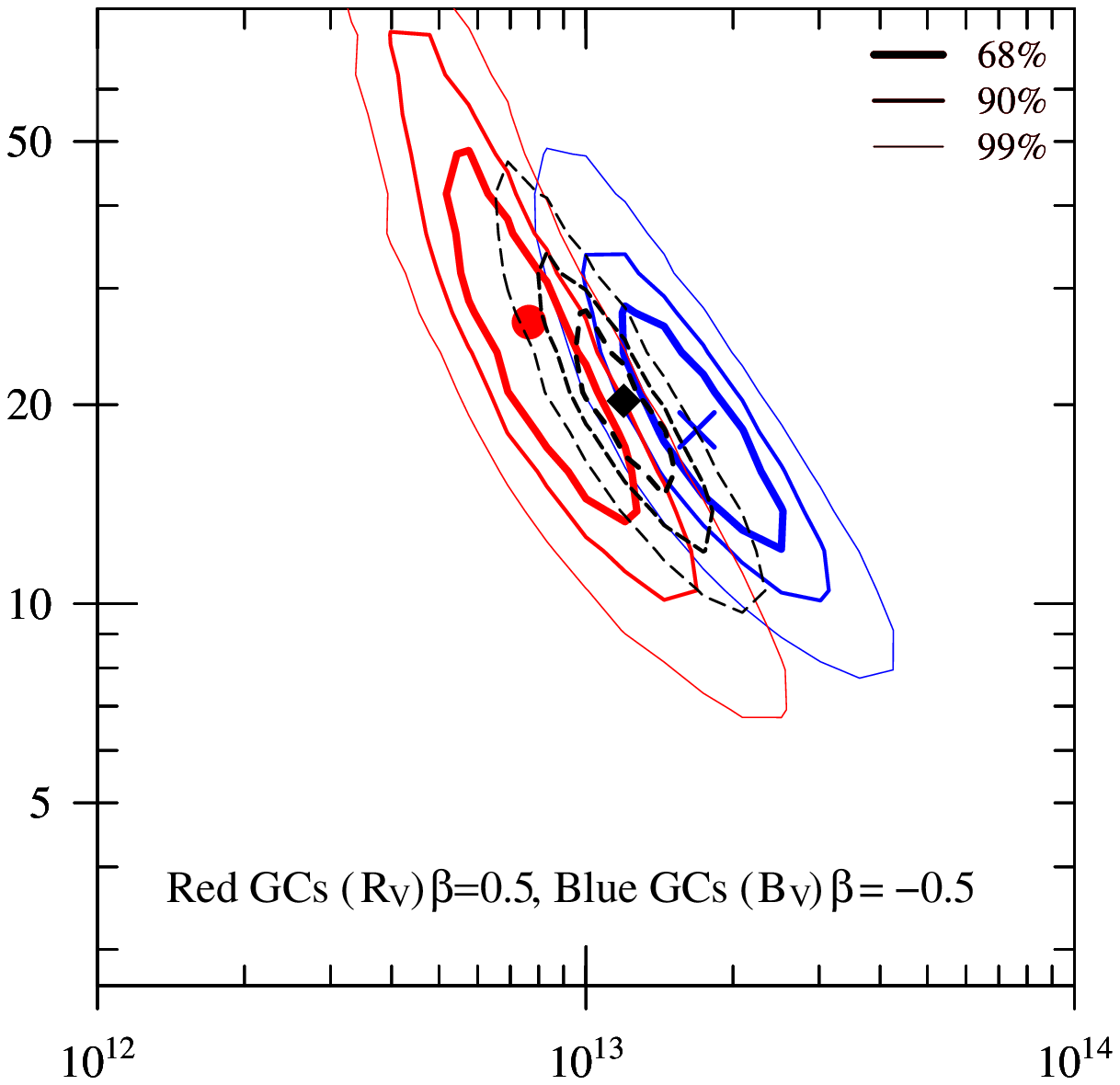}
\\
\hline
\end{tabular}
\caption{Mass models (NFW halo): Confidence level contours (68\%,90\%,
    and 99\%).  The $x$--axis plots the virial mass ($M_{vir}$, in
    units of $M_\odot$) and the $y$--axis is the dimensionless
    concentration parameter $c_{vir}$. \emph{Top row:} Red GCs: From
    left to right, the panels display the results for $\beta=-0.5,\,
    0$, and $+0.5$. Dashed contours show the values obtained for our
    GC data (\textsc{Riii}, models {\it{a1}} through {\it{a3}}), while
    the solid contours show the extended sample (\textsc{Rv}, models
    {\it{b1}} through {\it{b3}}).  The best--fit parameters for
    \textsc{Riii} and \textsc{Rv} are marked by a square and a cross,
    respectively.  \emph{Middle row:} The same for the blue GCs:
    samples \textsc{Biii} (models {\it{a4--a6}}) and \textsc{Bv}
    (models {\it{b4--b6}}).  \emph{Bottom left:} Parameters of the NFW
    halo calculated from the stellar velocity dispersion of NGC\,1399
    as published by Saglia et al.~(\citeyear{saglia00}), assuming
    $\beta=+0.5$ (solid contours, model {\it{c2}}) compared to the red
    GCs \textsc{Riii} for $\beta=0$ (dashed, model {\it{a2}}).  The
    \emph{bottom middle} panel shows model\,{\it{a10}}, the joint
    solution for the isotropic $\beta=0$ red (\textsc{Riii}) and
    $\beta=+0.5$ stellar model shown in the bottom left panel.  The
    \emph{bottom right panel} illustrates the difficulty to find a
    common solution for red and blue GCs. Solid lines are the contours
    for $\beta=0.5$ for red extended sample (\textsc{Rv} centred on
    the dot) and for the blue GCs assuming $\beta=-0.5$ (sample
    \textsc{Bv}, centred on the cross). The diamond indicates the
    joint solution, i.e.~the minimum of $\chi_{red}^2+\chi_{blue}^2$.
    The corresponding contours are shown as dashed lines. Note that
    the best--fit joint solution lies \emph{outside} the respective
    68\% CL contours of the individual GC tracer populations.
\newline In all panels, the long--dashed
    (dot--dashed) lines give the median ($68$ per cent values) for
    simulated halos as found by \cite{bullock01} (cf.~their Fig.~4).
    For comparison, the circle
    ($M_\mathrm{vir}=8.9\times10^{12}\,M_\odot, c_\textrm{vir}=15.9)$
    indicates the parameters of the best--fit NFW halo in Paper\,I.
    Note that the wiggles and gaps in the contours are an artifact of
    the finite grid--size and the re-gridding onto the ($M_{vir},
    c_{vir}$)--plane.}
\label{fig:contours}
\label{fig:solcombine}
\end{figure*}

 To facilitate the comparison of the results for the
NFW halos to values in the literature, we use the equations given in
\cite{bullock01} to express the ($r_{s},\varrho_{s}$) pairs in terms
of the virial parameters ($M_{\rm{vir}},c_{\rm{vir}}$).  These authors
define the virial radius $R_{\rm{vir}}$ such that the mean density
within this radius is $\Delta_{\rm{vir}}=337$ times the mean density
of the universe, and the concentration parameter is defined as
$c_{\rm{vir}}=R_{\rm{vir}}/r_s$. The confidence contours in
Fig.~\ref{fig:contours} are shown in the
($M_{\rm{vir}},c_{\rm{vir}}$) parameter space.

\begin{figure*}
\centering
\begin{tabular}{ll}
{\includegraphics[width=0.32\textwidth]{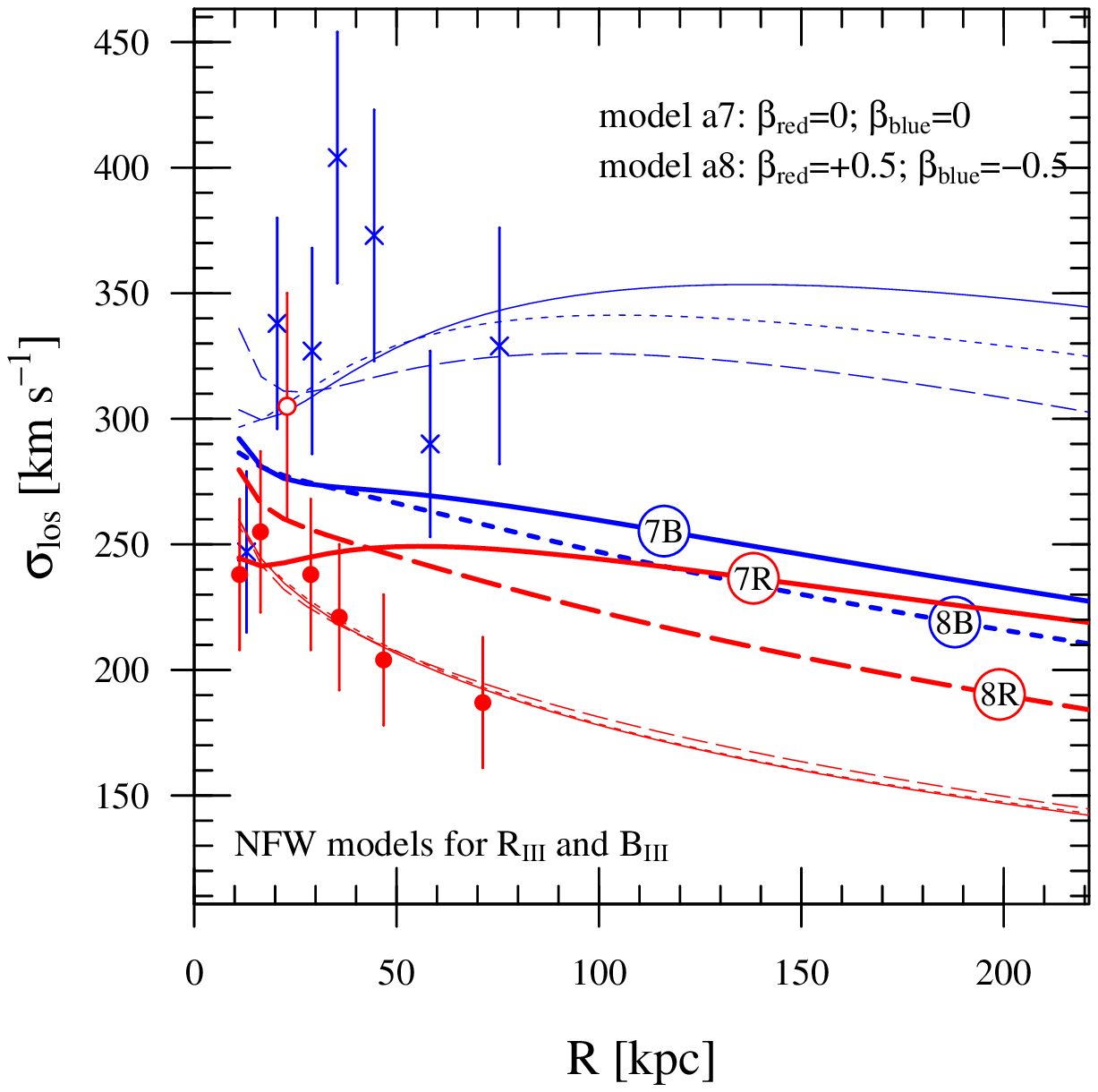}}&
{\includegraphics[width=0.32\textwidth]{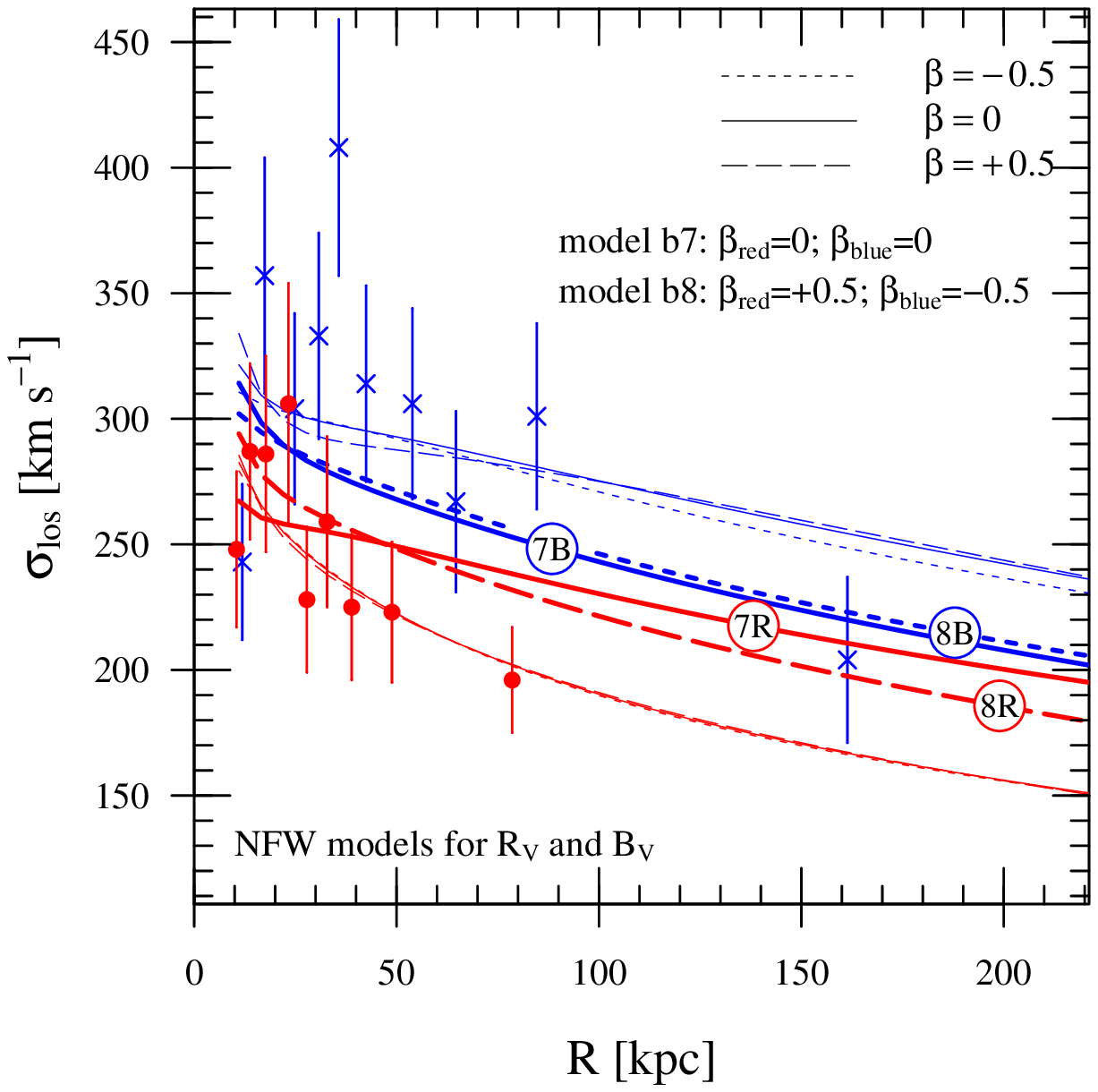}}
{\includegraphics[width=0.32\textwidth]{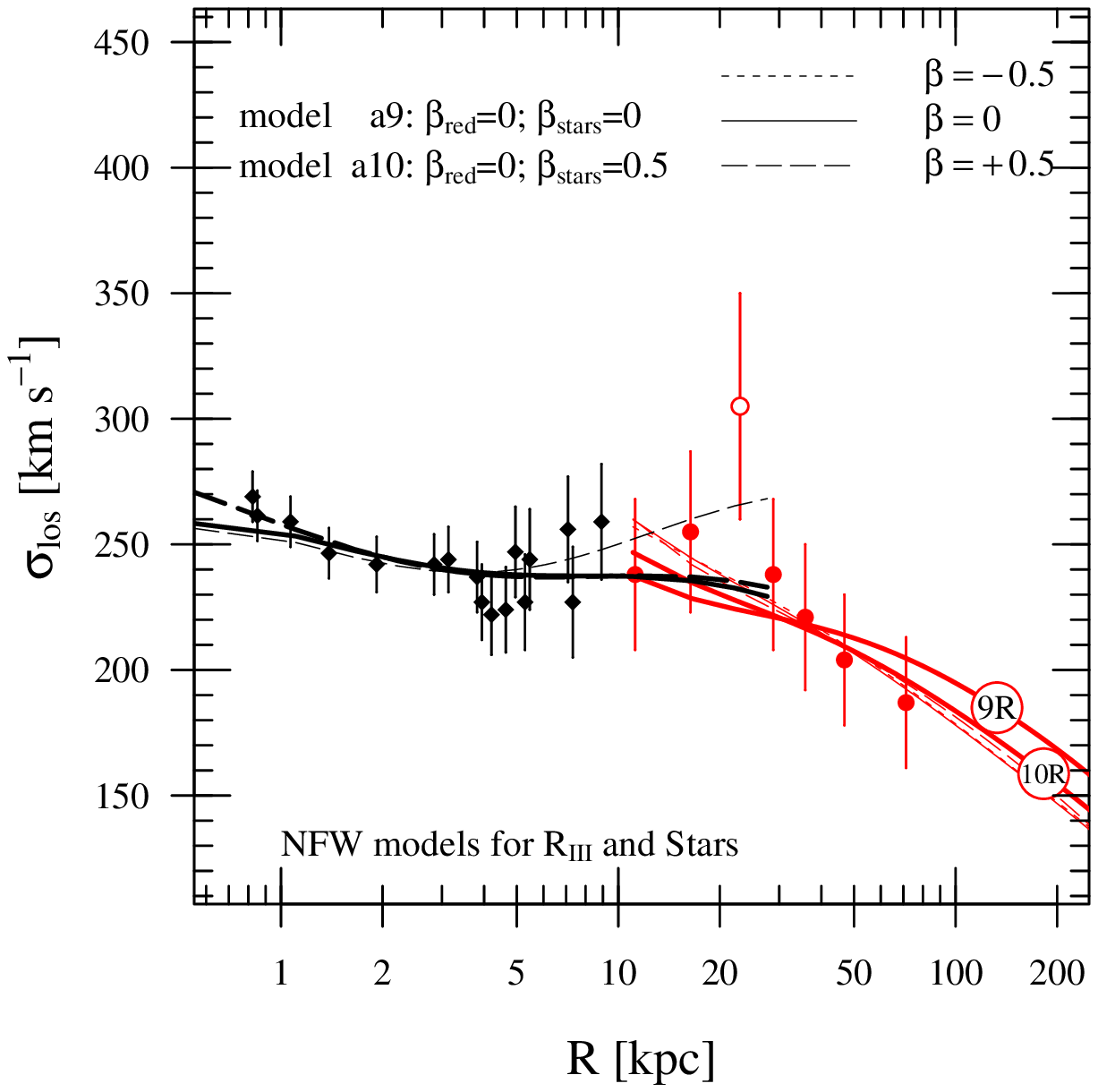}}

\\
\end{tabular}
\caption{Best fit Jeans models (NFW halo). \emph{Left panel:}
NGC\,1399 GC data and modelled velocity dispersion profiles.  Crosses
and dots show the red (\textsc{Riii}) and blue (\textsc{Biii})
velocity dispersion profiles (for a moving bin of 35 GCs,
cf.~Fig.~\ref{fig:dispersion}, lower panels), respectively.  The
\emph{thin} curves are the best fit models obtained for the red
(models {\it{a1--a3}}) and the blue GCs (models {\it{a4--a6}})
separately.  The \emph{thick} lines (with labels) show the dispersions
for the combined model for red and blue GCs (models \emph{a7} and
\emph{a8}).  \emph{Middle panel:} The same for the `extended' samples
including the \cite{bergond07} measurements (samples \textsc{Rv} and
\textsc{Bv}).  \emph{Right panel:} Red GCs and stars.  Diamonds show
the stellar velocity dispersion \citep{saglia00}, the red GCs
(\textsc{Riii}) are shown as dots. The thin lines are the models for
the individual tracer populations, i.e.~models \emph{c1} and \emph{c2}
for the stars and models \emph{a1--a3} for the red GCs. The thick
curves are the joint solutions, i.e.~models \emph{a9} and \emph{a10}.
In all panels, solid lines are isotropic models, and the models with
$\beta=-0.5$ and $\beta=+0.5$ are shown as dashed and long--dashed
lines, respectively.  The parameters of the halos shown here are
listed in Table\,\ref{tab:models}.}
\label{fig:nfwmod}
\label{fig:combi}
\end{figure*}

One immediately notices that the contours for the GCs (shown in the
top and middle row of Fig.~\ref{fig:contours}) are quite elongated,
with a negative correlation between the virial mass and concentration.
Below we discuss the results obtained for the different tracer
populations.

\subsection{NFW--models for the metal--rich population:}
The top row of Fig.~\ref{fig:contours} shows the NFW model parameters
for the metal--rich (red) GCs.  For a given value of $\beta$, the
halos derived for the sample \textsc{Riii} (i.e.~metal--rich GCs
fainter than $m_R=21.1$, after removal of GCs in the vicinity of
NGC\,1404 and the outlier--rejection as described in
Sect.~\ref{sect:interlopers}) are somewhat less massive and less
concentrated than the halos found for the `extended' sample
\textsc{Rv}. The reason is that the latter includes bright GCs for
which we found a larger velocity dispersion
(e.g.~Fig.~\ref{fig:dispersion}, lower left panel). Note however
that the models for \textsc{Riii} and \textsc{Rv} agree within their
respective 90\% contours. The results also agree well with the halo
parameters presented in Paper\,I (shown as a circle in
Fig.~\ref{fig:contours}).

\par The modelled dispersion profiles are
compared to the observational data in the left and middle panel of
Fig.~\ref{fig:nfwmod}. For both \textsc{Riii} and \textsc{Rv}, the
agreement between models and data is very good, and the best--fit
models (shown as thin lines) for the different values of the
anisotropy parameter $\beta$ are indistinguishable. The model
parameters listed in Table\,\ref{tab:models} illustrate the
mass--anisotropy degeneracy, where the virial mass increases with
$\beta$. \par

\subsection{NFW models for the metal--poor GCs:}
The shape of the velocity dispersion profiles of the metal--poor GCs
is not as smooth as that of the metal--rich GC population. The `jumps'
in the profiles may be caused by the presence of interlopers.  Since
the dispersion values of adjacent bins show quite strong variations,
the $\chi^2$ values derived when modelling the blue GCs are
substantially larger than those found for the red GCs. \par The
parameters derived for the sample \textsc{Biii} and the `extended'
sample \textsc{Bv} differ significantly: As can be seen from the
middle row in Fig.~\ref{fig:contours}, the solutions for the extended
sample (solid contours) have a smaller virial mass and a larger
concentration. 
\par Compared to the models for the red GCs, the dark halos derived
from the dispersion profile of the blue GCs are up to an order of
magnitude \emph{more massive} and about four times \emph{less
concentrated} which, as will be detailed below, considerably
complicates the task of finding  a common solution.

\subsection{Finding joint solutions}
Given that the different tracer populations (i.e. metal--rich and
metal--poor GCs, and stars) move in the same potential, it is necessary 
to find a joint solution. For a set of tracer populations (labelled
$a$ and $b$), we determine the combined parameters by minimising the
sum $\chi^2=\chi_a^2+\chi_b^2$ in the
$(r_{\rm{dark}},\varrho_{\rm{dark}}$)--parameter space. Since the
tracer populations may have different orbital anisotropies, this
procedure is performed for different combinations of $\beta$.
\subsubsection{Combining red and blue GCs}
First, we assume that $\beta=0$ for both blue and red GCs.  The
dispersion profiles for the combinations \textsc{Biii} $+$
\textsc{Riii} (model\,{\it{a7}}) and the `extended' samples
\textsc{Bv} $+$ \textsc{Rv} (model {\it{b7}}) are compared to the
observations in the left and middle panel of Fig.~\ref{fig:nfwmod},
respectively: The thick solid lines (labelled 7B and 7R for blue and
red GCs, respectively) do not agree with the data in the sense that
the models fail to reproduce the large difference one observes
regarding the dispersions of red and blue GCs.\par To find a better
agreement between model and data, we combine the most massive halo
found for the red GCs (i.e.~$\beta_\mathrm{red}=+0.5$) with the least
massive halo compatible with the blue GCs
($\beta_\mathrm{blue}=-0.5$).

The modelled dispersions for blue and red GCs
are shown as thick dashed and long--dashed lines (labelled 8B and 8R)
in the left and middle panel of Fig.~\ref{fig:nfwmod}. Again, the
agreement between data and model is poor. \par

The bottom right panel of Fig.~\ref{fig:solcombine} shows the contours
for this combination: The joint solution (shown as a diamond) for the
blue `extended' sample (\textsc{Bv},$\beta=-0.5$) and the red sample
(\textsc{Rv}, $\beta=+0.5$) lies outside the $68\%$ contour levels of
the individual tracer populations.

\subsubsection{Red GCs and the stellar velocity dispersion profile}
\label{sect:saglia}
The stellar velocity dispersion profile given by Saglia et
al.~(\citeyear{saglia00}) is used to put constraints on the mass
profile.  To be consistent, we use the same stellar mass profile as
for the calculations of the GC dispersion profiles (see
Sect.~\ref{sect:luminous}). Since this profile is an approximation
valid for $R\geq 500\,\rm{pc}$, we only consider the Saglia et
al.~data points outside $0\farcm 15$. The spatial density of the stars
is given by Eqns.~\ref{eq:surfb} and \ref{eq:mylum}.\par The detailed
modelling by \cite{saglia00} revealed that the stars have a radially
variable anisotropy parameter $\beta(R)$, which is positive and almost
reaches $\beta=+0.5$ near $\sim 10\arcsec$ (see their Fig.~5). We
therefore model the stars for the constant anisotropies $\beta=0$ and
$\beta=+0.5$ (models {\it{c1}} and {\it{c2}}, respectively). The model
with the radial bias gives a slightly better fit to the data.  The
stellar data cover the central $\sim10$\,kpc, i.e.~a region with
$R\leq r_{s}$. Thus, the scale radius and hence the total mass of the
dark halo cannot be constrained by these data.  However, as can be
seen from the bottom left panel in Fig.~\ref{fig:contours} (solid
contours) the concentration parameter $c_{vir}$ is quite tightly
constrained.  The contours for the red GCs (sample \textsc{Riii},
$\beta_\mathrm{red}=0$) shown in the same panel (dashed contours) have
a substantial overlap with the parameter space allowed by the stars.
The contours for the joint solution (model {\it{a10}}) are shown in
the bottom middle panel of Fig.~\ref{fig:contours}.\par The
corresponding models are compared to the observed velocity dispersion
profiles in the right panel of Fig.~\ref{fig:nfwmod}. The thick line
labelled `10R' shows the joint solution for the red GCs and the thick
long--dashed line shows the corresponding model for the stars.  For
stars and red GCs alike the agreement between the combined model
(a10) and the data is excellent, and we adopt this solution (which
also agrees very well with the result from Paper\,I) as our preferred
mass model for NGC\,1399. The NFW halo has a scale parameter of
$r_s=34\,\rm{kpc}$, a density $\varrho_s=0.0088\,M_\odot\,\rm{pc}^{-3}$,
and the virial radius is $R_\textrm{vir}=510\,\rm{kpc}$.

\label{sect:burkert}
We now assume that the cumulative mass of the dark halo is described
by the expression given in Eq.~\ref{eq:massburkert}. This cored halo,
introduced by Burkert\,(\citeyear{burkert95}),  has two free
parameters as well, namely the central density $\varrho_0$ and the scale
radius $r_0$.  The solutions of the Jeans equation are found in the
same manner as outlined in Sect.~\ref{sect:modcomp}, with the radii
being fixed to values $r_0 \in \{1,2,3,,\ldots 100 \}\,\rm{kpc}$. \\
The best--fit parameters are given in the last columns of
Table\,\ref{tab:models}.\par
\begin{figure*}
\includegraphics[width=0.49\textwidth]{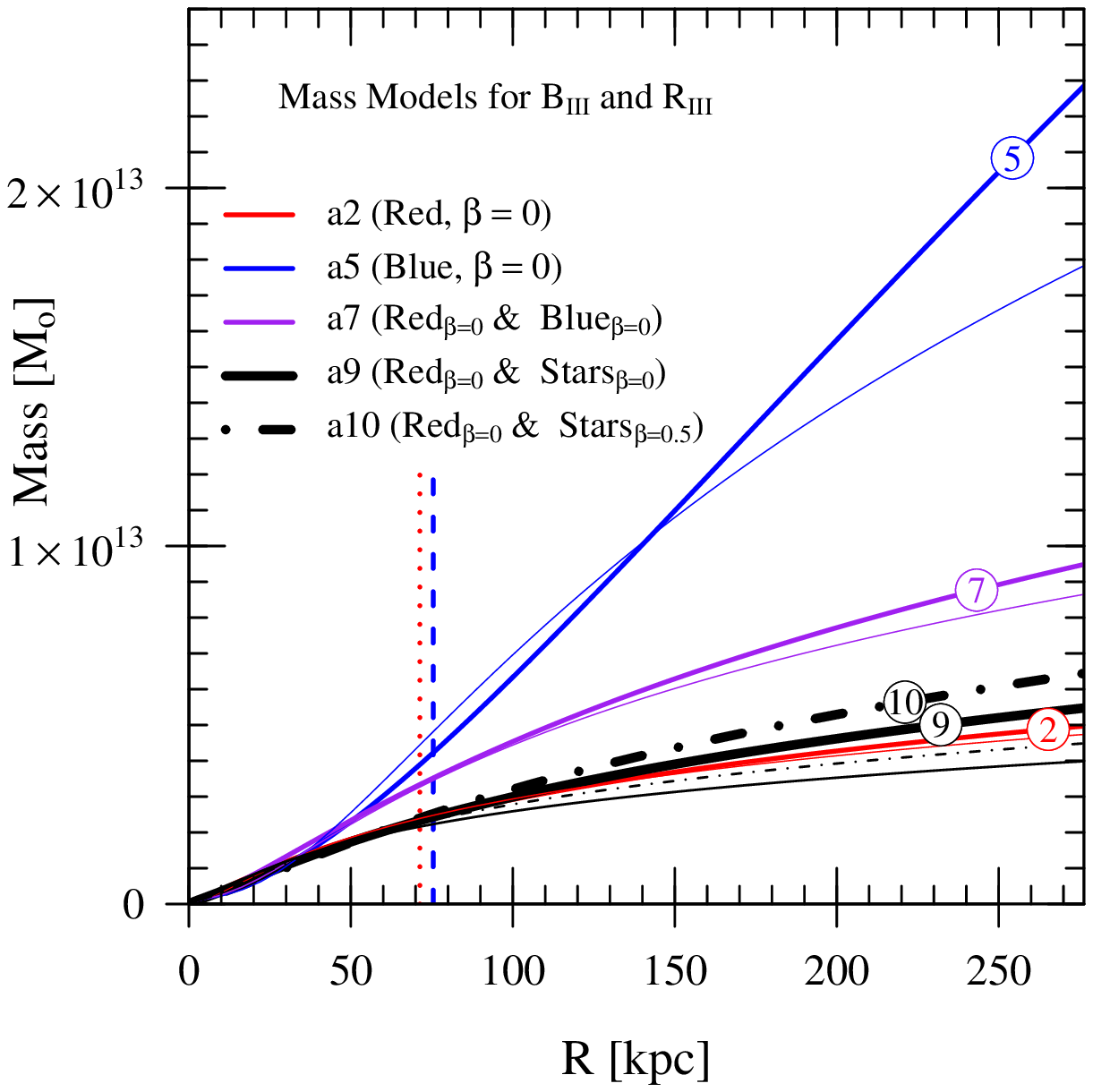}
\includegraphics[width=0.49\textwidth]{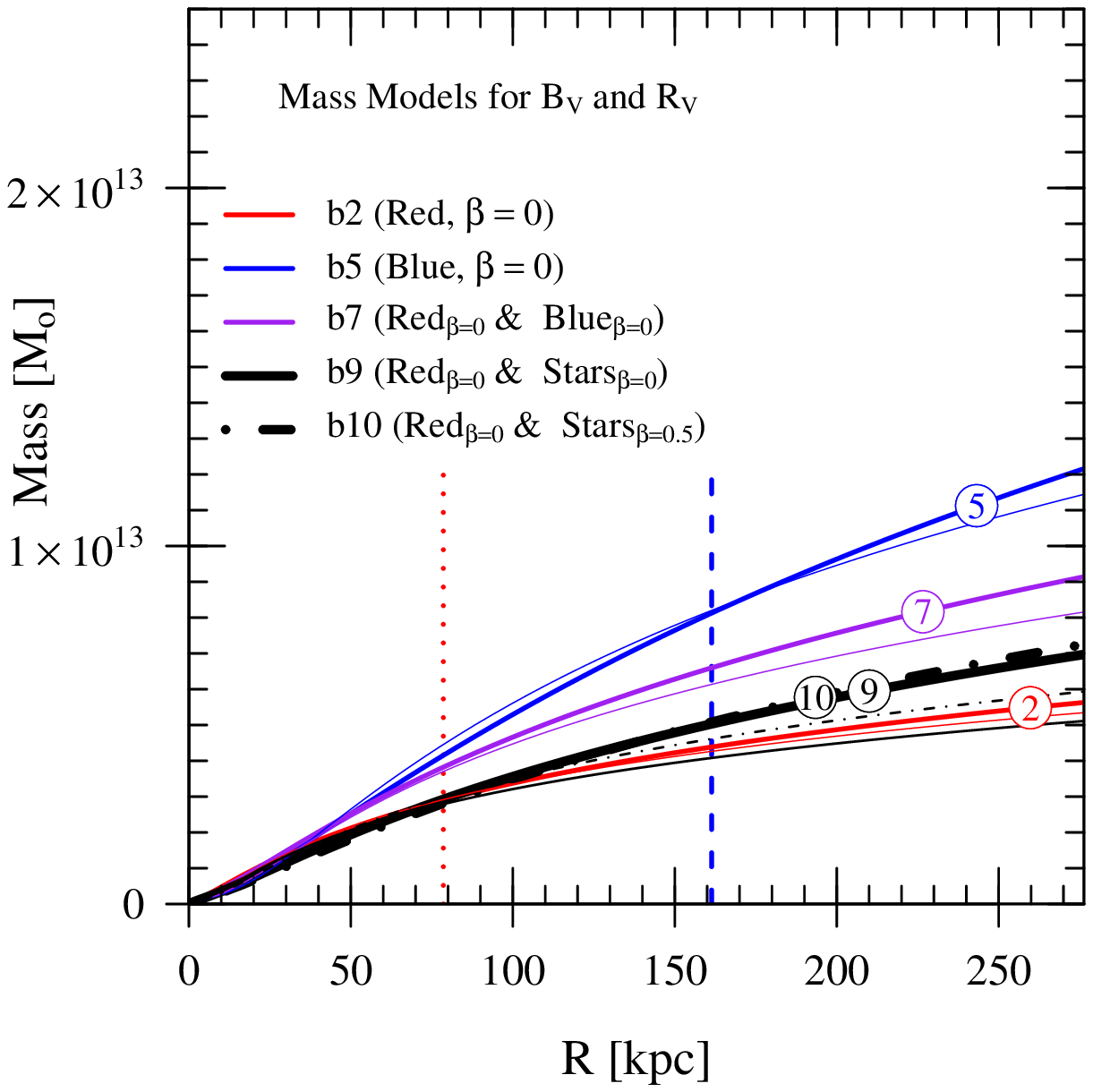}
\caption{ \emph{Left panel:} Mass profiles derived for the GC
dispersion profiles \textsc{Riii} and \textsc{Biii},
{i.e.}. The thick lines with labels referring to the
identifiers in Table\,\ref{tab:models} show the mass assuming an
NFW--type dark halo. The cored Burkert halos are shown as thin curves
with the same line style. For a given tracer population, the Burkert
halo is less massive than the best--fit NFW halo. {\emph{Right
panel:} The same for the extended data sets \textsc{Rv} and
\textsc{Bv} (i.e.~including velocities from B+07). In both panels, the
dotted and dashed vertical lines indicate the radial distance of the
outermost velocity dispersion data point for the red and blue GCs,
respectively (cf.~Table\,\ref{tab:dispersion}). } }
\label{fig:burkertcompare}
\end{figure*}

 Again, the solutions found for the red
GCs have substantially smaller scale--radii than the halos describing
the blue GCs. For a given tracer population, the cored halo models
provide a fit of similar quality, and we compare the derived mass
profiles in Fig.~\ref{fig:burkertcompare}. We find that for a given
dispersion profile the Burkert halos are less massive than the
corresponding NFW halos. Within the central $\sim80\,\rm{kpc}$, which
is about the radial extent of the data sets \textsc{Riii} and
\textsc{Biii}, however, the corresponding mass profiles are
almost indistinguishable.

\section{Discussion}
\label{sect:discussion}

\subsection{Population aspects}
\begin{figure}
\centering
\resizebox{\hsize}{!}
{\includegraphics[width=0.48\textwidth]{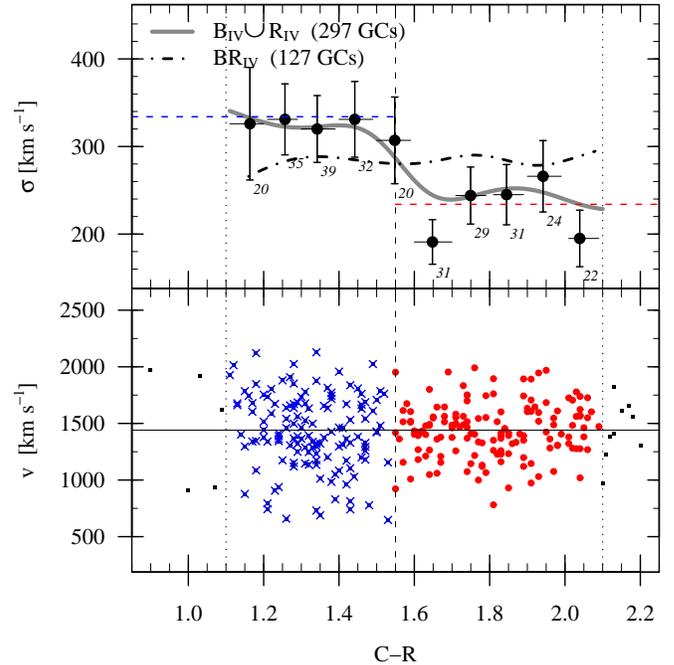}}
\caption{\emph{Upper panel}: Velocity dispersion as a function of
$C\!-\!R$--colour.  The data points show the velocity dispersion for
the Class\,A GCs which are fainter than $m_R=21.1$ (i.e.~the union of samples
\textsc{Biv} and \textsc{Riv}) for bins of $0.1\,\rm{mag}$ width. The
labels show the number of GCs in a given bin. For bins containing less
than 10 GCs, no dispersion was calculated. The dashed horizontal lines
are the velocity dispersions of the samples \textsc{Biv} and
\textsc{Riv} as given in Table\,\ref{tab:losvd}. The solid curve shows
the dispersion calculated using a Gaussian kernel with
$\sigma\!=\!0.1\,\rm{mag}$. For comparison, the dot--dashed curve
shows the same for the bright Class\,A GCs (sample \textsc{BRiv}).
\emph{Lower panel}: Velocities vs.~colour. Crosses and dots show blue
and red GCs (from the samples \textsc{Biv} and \textsc{Riv},
respectively) used in the upper panel.  Small squares indicate GCs in
the too sparsely populated colour bins.  In both panels, the dashed
line at $C\!-\!R=1.55$ indicates the colour used in this work to
divide blue from red GCs.  The vertical dotted lines indicate the
colour range for which the dispersions are calculated.}
\label{fig:sigmacr}
\end{figure}

{The strongest evidence pointing towards the existence of
substructure within the GCSs of elliptical galaxies is the bimodal
distribution of GC colours.}  The works by
e.g.~\cite{kundu01a,kundu01b,larsen01} and \cite{pengACS06} confirmed
that colour bimodality is a ubiquitous feature in the GCS of giant
ellipticals.  {As opposed to the Milky Way, where the GC
metallicity can be directly measured from spectra of individual GC
stars and the presence of kinematically distinct subpopulations has
been firmly established (e.g.~\citealt{zinn85,cote99}), there are but
very few spectroscopic metallicity measurements for GCs surrounding
giant elliptical galaxies (e.g.\citealt{cohen98,cohen03} for M87 and M49,
respectively).}  {Thus, the interpretation of the \emph{colour
bimodality} of elliptical galaxies in terms of a bimodal
\emph{metallicity} distribution, as proposed by e.g.~\cite{az92}, has
not gone unchallenged:}
The fundamental question in this respect is whether the description
as two separate populations is true or whether we rather face a
continuum of properties where the bimodal appearance is just a
morphological feature in the colour distribution, caused by a
non--linear colour--metallicity relation
(e.g.~\citealt{richtler06,yoon06}).\par

{That in NGC\,1399 blue and red GCs show a distinct dynamical
behaviour was already shown in Paper\,I. To gain a more quantitative
understanding of the kinematical differences between blue and red GCs,
we plot in Fig.~\ref{fig:sigmacr} the velocity dispersion
as function of $C\!-\!R$--colour. Our sample shows a jump in the
velocity dispersion at $C\!-\!R \simeq 1.55$ rather than a velocity
dispersion which smoothly changes with colour. {{In the
case of elliptical galaxies,}} this has not yet been {shown} before
and it is clear evidence that we indeed face two different
populations.
}

It is clear
however  that a simple colour cut does not cleanly separate
these two populations, but that there is some contamination of the
blue population by members of the red population and vice
versa. Intrinsically, these populations are expected to differ more
strongly in their kinematical and structural properties than our
observationally motivated blue and red populations.

Motivated by this strong evidence for two dynamically distinct
subpopulations in the NGC\,1399 GCS, we now address the question
concerning the nature and origin of the blue (metal--poor) and red
(metal--rich) GCs in NGC\,1399.


In the following, we discuss the idea that a major part of the GCS,
namely the blue GCs, has its origin predominantly in the accretion of
material, supposedly in the form of dwarf galaxies, during the
assembly of NGC\,1399 and the entire Fornax cluster. 
This {{concept}} is not at all new
\citep{cote98,hilker99III} but our large sample enables us to
distinguish the two subpopulations better than it was previously
possible.

The comparison with the Galactic GCS is instructive. The picture of
the Milky Way's halo has undergone a dramatic change from an
homogeneous stellar distribution to a strongly sub--structured halo
(see the review article by \citealt{helmi08}). The Sagittarius dwarf
donated six GCs to the Galactic system (e.g.
\citealt{bellazzini03}). The extragalactic origin of $\omega$ Centauri
as a former nucleus of a dwarf galaxy is hardly doubted any more
\citep{hilker00,hilker04,bedin04,villanova07}, and many of the
Galactic halo GCs with anomalous {horizontal branches} (HBs)
are suspected to have their origin outside the Milky Way
\citep{lee07,catelan09}.

Given the many tidal streams detected  (and yet to be detected),
the hypothesis that a large part of the Galactic halo has fallen in
has strong arguments in its favour.  If this applies to a relatively
isolated spiral galaxy, we would expect the effects of accretion to be
much more important in a giant elliptical in the centre of an
assembling galaxy cluster.


\subsubsection{Red GCs}
The metal--rich GCs resemble the stellar field population of NGC\,1399
under various aspects: The density profiles (shown in
Fig.~\ref{fig:numdens}) and the velocity dispersions
(Fig.~\ref{fig:kappa}, upper panel) are indistinguishable in the
radially overlapping domain.  {Further}, the field population
and the red GCs show similar anisotropies in terms of $h_4$ and
$\kappa$, respectively.  This strongly points towards a common
formation history. {This is further supported by the spatial
distribution of red GCs in elongated galaxies, which closely follows
the galaxy light, (e.g..~NGC\,1380 \citealt{kissler97};
NGC\,1052 \citealt{forbesN1052})}.

{Currently, the mechanisms leading to the assembly of the giant
ellipticals which now reside at the centres of galaxy clusters are not
fully understood and a matter of debate
(e.g.~\citealt{collins09,colavitti09} and references therein).  The
very old stellar ages derived for giant ellipticals
(e.g.~\citealt{trager00}) point towards their formation at high
redshift followed by passive evolution with ~no major episodes of star
formation (or GC formation, see below).  Such an early assembly is,
for instance, an inherent feature of the  `revised monolithic collapse
scenario' by \cite{pipino08}.}

According to the available observational evidence, GC formation is
enhanced in regions of high star formation rate
\citep{larsen00,weidner04,schweizer06}. In the case of NGC\,1399, the
first epoch of massive cluster formation was supposedly during the
first major merger or the collapse phase assembling the main body of
NGC\,1399. As can be seen from on--going mergers
(e.g.~\citealt{whitmoreschweizer95,wang04}), the starburst commences
well before the components are actually merged (for theoretical models
see e.g.~\citealt{mihoshernquist96} and \citealt{dimatteo08}). Thus,
also in the merger scenario, one expects the formation of metal--rich
GCs at a quite early stage. The processes leading to equilibrium,
whether it is phase mixing or violent relaxation, then act equally on
the stellar field population and the metal--rich GCs, leaving the same
imprints on both populations.  {However, the same processes
which lead to the observed similarities between the field stars and
the metal--rich GCs also erased all dynamical traces from the epoch
when the GCs were formed -- thus rendering the distinction between a
monolithic collapse scenario and a series of (early) gas--rich mergers
impossible. In their very deep images of NGC\,1399 \cite{tal09} did
not find any of the classical merger signatures (i.e.~shells,
isophotal twists, tidal tails, plumes and fans) -- which further
points towards an early assembly of NGC\,1399 and its predominantly
old \citep{kissler98,forbes01,kundu05,hempel07} metal--rich GCs.}

\subsubsection{Blue GCs}
The blue GCs are distinguished from their red counterparts by a
shallower density profile and a higher velocity dispersion. It had
been concluded in Paper\,I that the difference in the density profile
would be enough to explain the difference in the velocity dispersion
profile. Our present treatment, however, differs from the one in
Paper\,I. Firstly, we do not impose such a strict velocity
cut. Secondly, we use an updated density profile which is steeper for
large radii. Because of these factors, finding a halo which
simultaneously satisfies the constraints from the red and the blue
GCs is more difficult.

Moreover, the fluctuations of the velocity dispersion on small radial
scales are larger than one would expect from Poissonian noise
alone. The velocity uncertainties of the blue GCs are admittedly
larger but this feature persists also after imposing a quality
selection.

{ In Paper\,I, we identified a set of predominantly blue GCs
with extreme velocities and showed that their apogalactic distances
must be about $500$--$700$\,kpc, i.e.~significantly larger than the
traceable extent of the NGC\,1399 GCS, or the core of the Fornax
cluster which both are of the order $\sim 250\,\rm{kpc}$
(\citealt{bassino06} and \citealt{fergusonfornax}, respectively)}.

{To explain the complex kinematical properties of the
metal--poor GCs, we propose a scenario in which the majority of the
blue GCs surrounding NGC\,1399 belong to this galaxy's very extended
GCS as suggested by \cite{schuberth08}.  In addition, there is
a population of `vagrant' GCs which formerly belonged to (i) dwarf
galaxies (which were disrupted in the encounter with NGC\,1399) or
(ii) one of the more massive early--type galaxies in the Fornax
cluster core: While the large relative velocities of the galaxies near
the cluster centre make mergers very unlikely (compared to group
environments where the velocity dispersion is smaller), the fly--by of
a galaxy might lead to tidal stripping of GCs
\citep{forbes97,bekki03}. Since, in general, the metal--poor GCs have
a shallower number density distribution than the metal--rich GCs, the
former are more likely to get stripped. The observed kinematic
properties of these stripped GCs would depend strongly on the
orientation of the `donor' galaxy's orbit with respect to the plane of
the sky and, of course, the impact parameter and the effectiveness of
the stripping. }

{
Assuming that the stripped GCs would form a stream which roughly
follows the donor's orbit, GCs stripped in a galaxy--galaxy encounter
taking place in the plane of the sky would be next to undetectable
since their velocities would be very close to the systemic
velocity.}

{
 For a galaxy such as NGC\,1404, on the other hand, which has
a large relative velocity (and there are signs that its orbit is very
inclined, cf.~\citealt{1404infall}), one would expect the stripped GCs
to have extreme velocities.}

{In this picture, the observed
velocity field of the blue GCs is the superposition of a dynamically
old population and streams of GCs whose phase--space coordinates are
still correlated with the orbits of the donor galaxies. Depending on
the geometry, the observed line--of sight velocity dispersion derived
from this composite distribution can easily exceed the intrinsic
dispersion of the dynamically old GCs belonging to NGC\,1399 itself.}

{ However, as pointed out in Sect.~\ref{sect:1404inter}, a
robust test of this scenario requires a more complete spatial coverage
than provided by our current data set in conjunction with numerical
simulations.}  {Assuming that the contamination by stripped GCs
is indeed the cause for the jagged appearance of the line--of--sight
velocity dispersion profile of the NGC\,1399 blue GCs
(cf.~Fig.~\ref{fig:dispersion}, right panels), one would expect the
dispersion profiles of blue GCs of ellipticals residing in lower
density environments to be more regular. NGC\,4636 is such a galaxy:
It is situated in the very outskirts of the Virgo cluster (its nearest
giant neighbour has a projected distance of $\sim 175\,\rm{kpc}$) and
\cite{tal09} found no signs for a (recent) merger. Aside from two blue
GCs with extreme velocities, the quality--selected velocity dispersion
profile of the metal--poor GCs declines and has a smooth appearance
\citep{schuberth06,schuberth4636}. }

\subsection{Dynamical aspects}
In the following, we discuss the stellar--dynamical quantities and
compare the derived mass profiles to the results from X--ray studies
and cosmological N--body simulations.

\subsubsection{Rotation}
So far, no clear picture regarding the rotational properties of the
GCSs of giant ellipticals has emerged. For instance, M87 shows
significant rotation for both the metal--rich and the metal--poor GCs,
albeit with different axes of rotation \citep{cote01}. In
NGC\,4472, only the metal--poor GCs show a strong rotation signal
\citep{cote03}. {{For NGC\,4636, no rotation is detected
for the metal--poor GCs, while there are indications for a rotation of
metal--rich GCs \citep{schuberth06}}} ({for an overview
and discussion of the rotational behaviour of the GCSs of several
elliptical galaxies see \citealt{romanowsky08}}).  

\par In the case of NGC\,1399, we find no rotation signal for the red
(metal--rich) GCs. This is in agreement with the findings for the
field population for which \cite{saglia00} quote an upper limit of
$\varv_{\mathrm{rot}}\simeq30\,\rm{km\,s}^{-1}$.  { For the
blue GCs, however, the situation is more complicated: Although the
($\sim\!95\%$ CL) rotation signature found for the entire blue sample
(B\textsc{i}, for the full radial range) vanishes when culling the
more uncertain (Class\,B) velocity measurements from the analysis
(sample B\textsc{iv}), the rotation signal for the radial range
$4\arcmin < R \leq8\arcmin$ ($22 \lesssim R \lesssim 44\,\rm{kpc}$,
i.e.~the outer subsamples) appears to be robust with respect to
interloper removal and quality selection. The values (after interloper
removal; $A=110\pm 53\textrm{km\,s}^{-1}, \Theta_0=130\pm53^\circ$)
are consistent with the rotation signature reported in Paper\,I.}
\par { Unfortunately,  for the outermost parts
($R\gtrsim50\,\rm{kpc}$) of the NGC\,1399 GCS where, according to
numerical simulations \citep{bekki05}, the rotation amplitude of the
metal--poor GCs is expected to be largest, the data suffer from a very
incomplete angular coverage, thus precluding any statement regarding
the rotation of the GCS at very large galactocentric distances.}
\subsubsection{Comparison to cosmological simulations}
\label{sect:bullock}
In Fig.~\ref{fig:nfwmod} we compare our solutions to the results
presented in the study of \cite{bullock01} (cf.~their
Fig.\,4). These authors analysed a sample of $\sim\!5000$ simulated
halos with virial masses in the range
$10^{11}$--$10^{14}\,M_{{\odot}}$.  The concentration parameter
$c_{\rm{vir}} \equiv R_{\rm{vir}}/r_{\rm{dark}}$ decreases with growing
halo mass (solid line). The scatter in this relation, however, is
quite large, as the dashed lines encompassing 68 per cent of the
simulated halos show. 

We find that our best--fit combined halos with
$r_{\rm{dark}}\simeq30\,\rm{kpc}$ have somewhat higher concentrations
than simulated halos of a similar mass. Yet one has to bear in mind
that the numerical experiments were carried out using dark matter
particles only. It is conceivable that the presence of baryons and
dissipative effects might act to increase the concentration of real
halos. Thus we conclude that our best--fit halos do not stand in
stark contrast to the halos found in cosmological N--body
simulations.

\subsection{Comparison with X--ray mass profiles}
NGC\,1399 and the Fornax cluster have, due to their proximity and
brightness, been extensively observed by several X--ray satellite
missions, and the data support the existence of a massive dark halo
(e.g.~\citealt{jones97,ikebe96}, and \citealt{paolillo02}). Recently,
\cite{churazov08} used Chandra data to demonstrate that, for the
central 4\farcm5 ($\simeq 25\,\rm{kpc}$), the gravitational potentials
derived from stellar kinematics (cf.~Saglia et
al.~\citeyear{saglia00}) and X--rays agree very well. Regarding larger
scales, the Fornax cluster is known to be morphologically quite
complex: For example, \cite{ikebe96}, using ASCA data, reported the
discovery of substructure in the X--ray maps which they interpreted in
terms of a galaxy--sized dark halo embedded in a cluster--wide
halo. \par Using deep high--resolution ROSAT data, \cite{paolillo02}
confirmed this finding and identified three different structures: The
innermost component is centred on NGC\,1399 and dominates the inner
$50\arcsec{}(\simeq\!4.6\,\rm{kpc})$.  Then follows a second, almost
spherical `galactic halo' component, the centre of which lies
1\arcmin{} SW of NGC\,1399, its radial extent is of the order
$400\arcsec{}(\simeq35\,\rm{kpc})$. The third and most extended
`cluster' component is more elongated and its centre lies 5\farcm6{}
northeast of NGC\,1399. \par In terms of the total mass profile, the
existence of different components leads to `shoulder--like' features
near the interface regions (i.e.~where the gas--densities of two
components become equal). This behaviour is clearly seen for the range
of mass--profiles derived in both studies, see Fig.~17 of
\cite{paolillo02} for a comparison.

\cite{makishima01} who analysed
a sample of 20 galaxy clusters (including Fornax and Virgo) propose
that such hierarchical dark matter profiles, parametrised in terms of
a central excess mass superimposed on a cored King--type profile, are
in general linked to the presence of a cD galaxy. Cuspy profiles, on
the other hand, are found in the absence of a cD galaxy
(e.g.~Abell\,1060 which has two central giant ellipticals).

\par In the case of NGC\,1399, the transition from the central
component to the cluster--wide component takes place at about 60\,kpc
($\simeq11\arcmin$), a region probed by our GC sample.

The question arises, accordingly, which consequences such a `nested'
mass distribution would have on the GC dynamics.  Would one be able to
detect signs of such a halo--in--halo structure in the velocity
dispersion profile of the GCs?  \par
\begin{figure}
\centering
\resizebox{\hsize}{!}
{\includegraphics[width=0.48\textwidth]{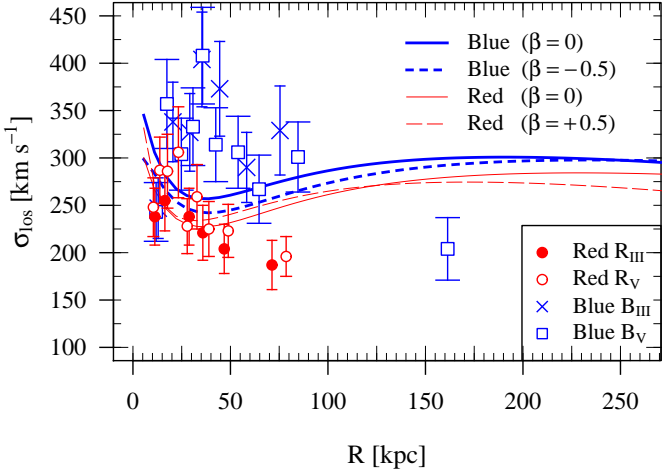}}
\caption{Comparison to the `nested halo' mass profile by
\cite{ikebe96}. Here, we plot the dispersion profiles expected for
their Model\,1, see text for details.  The thick curves are the
dispersion profiles for the blue GCs, where the solid and dashed lines
correspond to $\beta=0$ and $\beta=-0.5$, respectively. The thin lines
show the models for the red GCs. The solid line is $\beta=0$, and the
long--dashed line shows the model for $\beta=+0.5$.  The data for the
red GC samples \textsc{Riii} and \textsc{Rv} are shown as dots and
circles, respectively. The corresponding blue samples \textsc{Biii}
and \textsc{Bv} are shown as crosses and unfilled squares,
respectively.  }
\label{fig:ikebedisp}
\end{figure}

\begin{figure}
\centering
\resizebox{\hsize}{!}
{\includegraphics[width=0.48\textwidth]{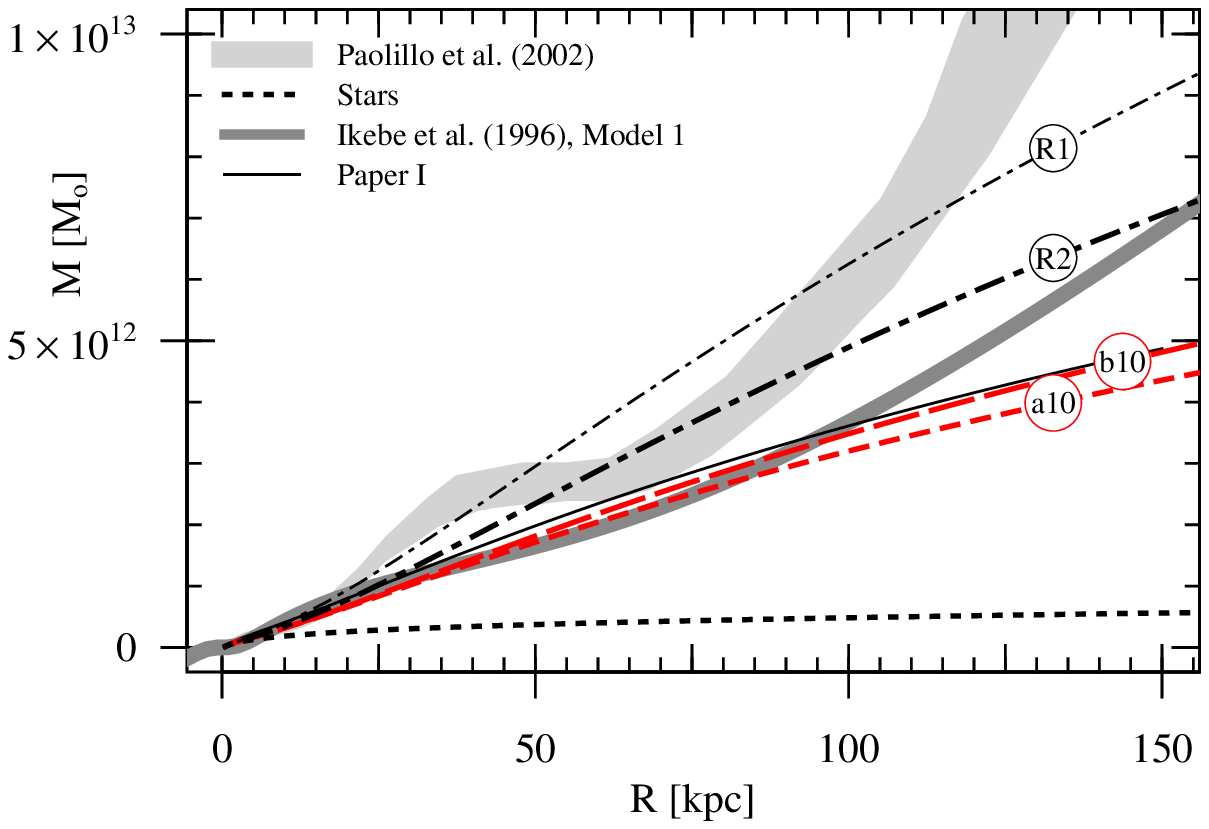}}
\caption{ Comparison to X--ray measurements.  The grey area shows the
range of the mass models presented by \cite{paolillo02}. The thick
solid line shows the mass derived by \cite{ikebe96} (their Model\,1).
The thin solid line indicates the mass profile derived in Paper\,I.
The dash--dotted curves are the two halos presented in
\cite{richtler08}: The more massive one (labelled R1) has the
parameters $r_s=50\,\rm{kpc}$ and
$\varrho_s=0.0085\,\rm{M_\odot\,pc}^{-3}$. The halo derived for the
`safe' sample in \cite{richtler08} (R2) has $r_s=50\,\rm{kpc}$ and
$\varrho_s=0.0065\,\rm{M_\odot\,pc}^{-3}$.  The dashed and
long--dashed curves are the best--fit NFW mass profiles {\it{a10}} and
{\it{b10}} derived in this work (cf.~Table\,\ref{tab:models}). The
stellar mass is shown as a short--dashed line.}
\label{fig:fp}
\end{figure}

 We chose the isothermal profile
of Ikebe et al.~(their Model\,1) and, using the standard
$\beta$--model
for the X--ray gas density profiles
\begin{equation}
\varrho_{\textrm{gas}}(r) = \varrho_{\textrm{gas,0}} \left( 1+ \frac{r^2}{r_{c}^2}\right)^{-\frac{3}{2}\beta_X},
\label{eq:betamod}
\end{equation}
calculate  the total gravitating mass using 
\begin{equation}
M \left( < r \right) = \frac{-kT}{G\mu \rm{m}_{\rm{H}}} \left(
\frac{\rm{d} \ln \varrho}{\rm{d} \ln r}
+
\frac{\rm{d} \ln T}{\rm{d} \ln r}
\right)\,r  \; 
\label{eq:xmass}
\end{equation}
\citep{FLG80}. Here, $\beta_X$ is the
power--law exponent,

$m_H$ is the mass of the hydrogen atom, and $\mu=0.6$ the mean
molecular weight (assuming a fully ionised plasma with primordial
element abundances).  The total density is the sum of two components
with the parameters given in Table\,1 of Ikebe et al.:
$\varrho_{1,0}=2.3\times 10^{-2}\,\rm{cm}^{-3}$,
$\varrho_{2,0}=8.2\times 10^{-4}\,\rm{cm}^{-3}$,
$r_{c,1}=4.8\,\rm{kpc}$ $r_{c,2}=127\,\rm{kpc}$, $\beta_{X,1}=0.51$,
and $\beta_{X,2}=0.60$.  Since we assume that the gas is isothermal
($T=1.2\times 10^7\,\rm{K}$), the second term in Eq.~\ref{eq:xmass}
vanishes.  Figure\,\ref{fig:ikebedisp} shows the velocity dispersion
profiles expected for the blue and red GC subpopulations (assuming
constant anisotropy parameters of $-0.5,0,+0.5$) for the mass profile
given in Eq.~\ref{eq:xmass} and plotted in Fig.~\ref{fig:fp}. Indeed,
the modelled velocity dispersion profiles show a dip near
$\sim40\,\rm{kpc}$, but the feature is quite shallow and appears to be
at variance with the findings for the red GCs. \par The mass profiles
derived by \cite{paolillo02} would lead to more pronounced features in
the predicted velocity dispersion profiles. Note however that a
simple isothermal profile based on the gas density distribution
ignoring the offsets of the different components (shown as thick line
in their Fig.~15) leads to a non--monotonic (i.e.~unphysical)
density--profile at the transition from the galaxy to the cluster
component.

Figure\,\ref{fig:fp} compares the mass profiles derived in this study
(shown as solid and dot--dashed lines) with the X--ray mass profiles
presented by \cite{paolillo02} and \cite{ikebe96}.  {{Within the
central 100\,Mpc, where the vast majority of our dynamical probes is
found, the preferred mass profiles derived in this study agree well
with the \cite{ikebe96} mass model, although we do not find any signs for the transition from galaxy to cluster halo in our kinematic data.}}

\section{Conclusions}
\label{sect:conclusions}
Using the largest data set of globular cluster (GC) radial velocities
available to date, we revisit and extend the investigation of the
kinematical and dynamical properties of the NGC\,1399 GC system with
respect to Richtler et al.~(\citeyear{richtler04}, Paper\,I). We
measure about 700 GC radial velocities out to approximately
$14.5\arcmin\simeq 80\,\rm{kpc}$ (in Paper\,I we reached 40 kpc). To
this sample we add 56 GC velocities from Bergond et al.
(\citeyear{bergond07}, B+07) which go as far out as 200 kpc.\\ 
Our main
findings are the following: There is no significant rotation signal
among the red (metal--rich) subpopulation.  We find rotation around
the minor axis for the blue (metal--poor) clusters in a radial
interval of $4\arcmin{} < R < 8\arcmin{}$ (i.e.~$22\,\rm{kpc}\lesssim
R\lesssim44\,\rm{kpc}$), however weak.  
The blue and red clusters form
two {{\emph{kinematically distinct}} subpopulations rather than showing a continuum of
kinematical properties.

The red clusters correspond under various kinematical aspects to the
stellar field populations. 

Their velocity dispersion declines
outwards and their velocity distribution suggests {orbital isotropy}.

The jump to the higher dispersion of the blue clusters occurs at a
colour ($C\!-\!R = 1.55$), which is also suggested by the morphology
of the colour distribution.  The blue GCs, however, show a more
complex behaviour. Their velocity dispersion profile is not as smooth
as that of the red GCs. There exist very high/low individual
velocities, which suggest very large apogalactic distances, as already
found in Paper\,I. These objects seem to belong to an intergalactic
cluster population, which {may be made up of GCs stemming from
disrupted dwarfs and GCs stripped off neighbouring galaxies during
close encounters}.

We performed a Jeans analysis in order to constrain the mass profile
(stellar plus dark).  We found it difficult to find a dark halo which
simultaneously accounts for the red and the blue clusters, which again
argues for a different dynamical history.

The dark NFW--halo which was found to represent the kinematics of the
red GCs ($\beta_{\mathrm{GCs}}=0$) and the stellar velocity dispersion
profile presented by \cite{saglia00} ($\beta_{\mathrm{stars}}=+0.5$),
our Model\,{\it{a10}}, has the parameters
$\varrho_{\mathrm{dark}}=0.0088\,M_\odot\,\rm{pc}^{-3}$ and
$r_{\mathrm{dark}}=34\,\rm{kpc}$. The virial mass is
$M_{\textrm{vir}}=8.0\times10^{12} M_\odot$. \\
Including the velocities from B+07, (Model\,{\it{b10}}) yields a
slightly more massive halo
($\varrho_{\mathrm{dark}}=0.0077\,M_\odot\,\rm{pc}^{-3}$ and
$r_{\mathrm{dark}}=38\,\rm{kpc}$, and a virial mass
$M_{\textrm{vir}}=9.5\times10^{12} M_\odot$).

These NFW halos, which were found to represent the observations, are
marginally less massive than the one from Paper\,I.  The total mass
profile fits agree reasonably well with the X--ray based mass profiles
out to 80\,kpc. However, our model halos when extrapolated stay
significantly below the X--ray masses. Moreover, we do not see the
transition from the galaxy halo to the cluster halo claimed to be
present in the X--ray data.

We argue that these findings are consistent with a scenario where the
red GCs are formed together with the bulk of the field population, most
probably in {early} multiple mergers with many progenitors
involved. {This is consistent with the NGC\,1399 GCs being
predominantly old \citep{kissler98,forbes01,kundu05,hempel07}, with ages very
similar to the stellar age of NGC\,1399 ($11.5\pm2.4$\,Gyr,
\citealt{trager00}) } \par A large part of the blue GC population has
been acquired by accretion processes, most plausibly through dwarf
galaxies during the assembly of the Fornax cluster. Moreover, there
should be a significant population of intra-cluster GCs.

\begin{acknowledgements}
We thank the anonymous referee for helpful comments.
We also thank Mike Fellhauer for fruitful discussions.  TR and AJR
acknowledge support from the Chilean Center for Astrophysics, FONDAP
No.~15010003.  AJR was further supported by the National Science
Foundation grants AST--0507729 and AST--0808099.  This research has
made use of the NASA/IPAC Extragalactic Database (NED) which is
operated by the Jet Propulsion Laboratory, California Institute of
Technology, under contract with the National Aeronautics and Space
Administration.
\end{acknowledgements}

\bibliographystyle{aa}

\bibliography{copyofdissrefs} 
\appendix
\section{Radial velocity dispersion profiles}
\centering
\begin{table*}
\caption{Velocity dispersion profiles for fixed annular bins
(cf.~Fig.~\ref{fig:dispersion}, middle panels). The bins start at
$R=1\farcm0, 3\farcm5, 5\farcm5, 7\farcm5, 9\farcm5, 12\farcm5,
15\farcm5$, and $30\farcm0$; at a distance of $19\,\rm{Mpc}$, 1\farcm0
corresponds to 5.2\,kpc. The left and right table show the values
obtained for the blue and red GCs, respectively. The Roman numerals
refer to the sample definition in
Sect.~\ref{sect:sampledefine}. Column (1) gives the bin number, the
mean radius of the GCs in a bin is given in Col.~(2). The velocity
dispersion (in units of $\rm{km\,s}^{-1}$) and its uncertainty are
given in Cols.~(3) and (4). Col.~(5) is the number of GCs in a given
bin, and Col.~(6) their mean velocity.}
\begin{tabular}{lr}

\begin{tabular}{llllll} \hline \hline
\multicolumn{6}{c}{Blue (metal--poor) GCs}\\
\hline
Bin&$\bar{R}$ &	$\sigma_{\mathrm{los}}$&	$\Delta\sigma_{\mathrm{los}}$&	$N_{\mathrm{gc}}$&	$\bar{\varv}$ \\
(1) & (2) & (3) & (4) & (5) & (6) \\
\hline
\multicolumn{6}{l}{Sample \textsc{Bi}}\\
\hline
1&2.59&	288&	30&	50&	1398 \\
2&4.59&	312&	35&	42&	1440 \\
3&6.43&	396&	36&	61&	1459 \\
4&8.52&	363&	46&	33&	1488 \\
5&11.06&	351&	39&	42&	1447 \\
6&13.77&	369&	58&	22&	1332 \\
7&17.02&	--&	--&	6&	2101 \\

\hline
\multicolumn{6}{l}{Sample \textsc{Bii}}\\
\hline

1&2.59&	288&	30&	50&	1398 \\
2&4.59&	312&	35&	42&	1440 \\
3&6.39&	393&	37&	58&	1387 \\
4&8.53&	350&	46&	30&	1444 \\
5&11.06&	325&	38&	38&	1412 \\
6&13.77&	369&	58&	22&	1332 \\
7&17.02&	--&	--&	6&	2101 \\

\hline
\multicolumn{6}{l}{Sample \textsc{Biii}}\\
\hline

1&2.59&	288&	30&	50&	1398 \\
2&4.59&	312&	35&	42&	1440 \\
3&6.39&	393&	37&	58&	1387 \\
4&8.53&	350&	46&	30&	1443 \\
5&11.02&	296&	36&	37&	1395 \\ 
6&13.76&	338&	58&	21&	1320 \\
7&17.56&	--&	--&	2&	1471 \\

\hline
\multicolumn{6}{l}{Sample \textsc{Biv}}\\
\hline

1& 2.62&	234&	33&	27&	1411 \\
2 & 4.48&	317&	47&	24&	1445 \\
3& 6.37&	371&	43&	38&	1478 \\ 
4& 8.52&	375&	62&	19&	1466 \\
5& 11.01&	351&	59&	20&	1363 \\
6 & 13.87&	338&	88&	11&	1343 \\
7 & --&	--&	--&	0&	-- \\

\hline
\multicolumn{6}{l}{Sample \textsc{Bv}}\\
\hline
1 & 2.56&	303&	28&	64&	1398 \\ 
2 & 4.56&	317&	32&	53&	1494 \\
3 & 6.40&	369&	31&	73&	1386 \\
4 & 8.51&	325&	40&	35&	1486 \\
5 & 11.10&	271&	27&	53&	1429 \\
6 & 13.75&	322&	45&	28&	1380 \\ 
7 & 21.66&	254&	40&	21&	1424 \\
8 & 34.75&	136&	31&	10&	1466 \\
\hline 
\hline
\end{tabular}
&
\begin{tabular}{llllll} \hline \hline
\multicolumn{6}{c}{Red (metal--rich) GCs} \\ \hline
Bin&$\bar{R}$ &	$\sigma_{\mathrm{los}}$&	$\Delta\sigma_{\mathrm{los}}$&	$N_{\mathrm{gc}}$&	$\bar{\varv}$ \\
(1) & (2) & (3) & (4) & (5) & (6) \\

\hline
\multicolumn{6}{l}{Sample \textsc{Ri}}\\
\hline
1&2.47&	244&	22&	68&	1446 \\
2&4.52&	278&	27&	64&	1461 \\
3&6.40&	244&	26&	47&	1450  \\
4&8.42&	260&	35&	36&	1556  \\
5&11.07&	282&	41&	27&	1425  \\
6&13.69&	--&	--&	7&	1222  \\
7&16.10&	--&	--&	7&	1463  \\
\hline

\multicolumn{6}{l}{Sample \textsc{Rii}}\\
\hline
1 &2.47&	244	&22&	68&	1446 \\
2&4.52&	278	&27&	64&	1461 \\
3&6.40&	244	&26&	47&	1450 \\
4&8.41&	216	&30&	28&	1504 \\
5&11.09&	281	&41&	26&	1422 \\
6&13.69&	--	&--	&7&	1222 \\
7&16.10&	--	&--&	7&	1463 \\ 
\hline
\multicolumn{6}{l}{Sample \textsc{Riii}}\\
\hline

1& 2.47&	244&	22&	68&	1446 \\
2& 4.52&	278&	27&	64&	1461 \\
3& 6.40&	225&	25&	46&	1461 \\
4& 8.41&	216&	30&	28&	1504 \\
5& 11.13&	188&	30&	23&	1418\\ 
6& 13.69&	--&	--&	7&	1222 \\
7 & 16.10&	--&	--&	7&	1463 \\
\hline
\multicolumn{6}{l}{Sample \textsc{Riv}}\\
\hline

1& 2.45&	260&	30&	42&	1398 \\
2& 4.57&	256&	29&	43&	1467 \\
3 & 6.43&	209&	27&	32&	1428 \\
4& 8.38&	206&	32&	22&	1485 \\
5& 10.93&	167&	33&	14&	1392 \\
6 & 13.18&	--&	--&	4&	1208 \\
7 & 15.57&	--&	--&	1&	1485 \\

\hline
\multicolumn{6}{l}{Sample \textsc{Rv}}\\
\hline

1 & 2.44&	280&	21&	97&	1405 \\
2 & 4.55&	265&	22&	80&	1454 \\
 3& 6.44&	245&	23&	63&	1472 \\
4& 8.38&	220&	29&	32&	1488 \\
5 &11.04&	213&	29&	29&	1418 \\
6& 13.80&	190&	43&	14&	1366 \\
7&19.33&	107&	23&	13&	1463 \\
8 &--&	--&	--&	--&  -- \\

\hline 
\hline

\end{tabular}

 \\
\end{tabular}
\label{tab:fixbin}
\end{table*}

\begin{table*}
\caption{Velocity dispersion profiles for the subsets of our data used
for the Jeans modelling.  The dispersion profiles are obtained using a
moving bin of 35 GCs. The bins do not overlap, and the minimum number
of GCs per bin was set to 20, hence the large number of GCs in the
last bin of sample \textsc{Rv}.  For each bin, $\bar{R}$ is the mean
radial distance from NGC\,1399, and $R_{\mathrm{min}}$ and
$R_{\mathrm{max}}$ give the radial range covered by the GCs of the bin
(at a distance of 19\,Mpc, 1\farcm0{} corresponds to
$5.52\,\rm{kpc}$). $N_{\mathrm{GC}}$ is the number of GCs in a given
bin. The line--of--sight velocity dispersion $\sigma$ and its
uncertainty $\Delta\sigma$ are calculated using the estimator by
\cite{pm93}. The asterisk marks the data point which was omitted when
modelling the sample\,\textsc{Riii}.  }
\resizebox{0.99\textwidth}{!}{
\begin{tabular}{rr@{.}llllr@{.}lr@{.}lr||r@{.}llllr@{.}lr@{.}lr} \hline \hline

$\rm{N}^{o}$&\multicolumn{2}{c}{$\bar{R}_{\mathrm{blue}}$} & $\sigma_{\mathrm{blue}}$ & $\Delta\sigma_\mathrm{blue}$ &$N_{\mathrm{GC}}$ & \multicolumn{2}{c}{$R_{\mathrm{min}}$} & \multicolumn{2}{c}{$R_{\mathrm{max}}$} 
&\multicolumn{1}{r||}{$\bar{\varv}$}
& \multicolumn{2}{c}{$\bar{R}_{\mathrm{red}}$} & $\sigma_{red}$ & $\Delta\sigma_\mathrm{red}$ &$N_{\mathrm{GC}}$ & \multicolumn{2}{c}{$R_{\mathrm{min}}$} & \multicolumn{2}{c}{$R_{\mathrm{max}}$\rule{0ex}{2.5ex}} &\multicolumn{1}{c}{$\bar{\varv}$}\\ 

&\multicolumn{2}{c}{$[${\tiny{arcmin}}$]$} & $[\rm{km\,s}^{-1}]$ &  $[\rm{km\,s}^{-1}]$  &$$ & \multicolumn{2}{c}{$[${\tiny{arcmin}}$]$} & \multicolumn{2}{c}{$[${\tiny{arcmin}}$]$}& & \multicolumn{2}{c}{$[${\tiny{arcmin}}$]$} &  $[\rm{km\,s}^{-1}]$  &  $[\rm{km\,s}^{-1}]$  &$$ & \multicolumn{2}{c}{$[${\tiny{arcmin}}$]$} & \multicolumn{2}{c}{$[${\tiny{arcmin}}$]$ \rule{0ex}{2.5ex}} &$[\rm{km\,s}^{-1}]$\\ \hline 

\multicolumn{11}{l||}{Blue GCs (sample\,\textsc{Biii})}&
\multicolumn{10}{l}{Red GCs (sample\,\textsc{Riii})}\rule{0ex}{2.5ex}

\\\hline

1& 2&34&  247& 32& 35& 1&72&   2&83& 1406&2&03&  238& 29& 35&  1&12&  2&43  &1450  \\
2& 3&71&  338& 42& 35& 2&90&   4&63& 1400&2&96&  255& 32& 35&  2&45&  3&53 &1427  \\
3& 5&27&  327& 40& 35& 4&63&   5&82& 1465&4&$15$&  305& 44& 35&  3&53&  4&$69$ $(^\star)$ &1562  \\
4& 6&40&  404& 49& 35& 5&82&   7&08& 1455&5&21&  238& 30& 35&  4&74&  5&82 &1436 \\
5& 8&05&  373& 46& 35& 7&08&   9&14& 1397&6&48&  221& 29& 35&  5&84&  7&31 &1496  \\
6& 10&55& 290& 36& 35& 9&24&  11&66& 1414&8&47&  204& 26& 35&  7&40&  9&95 &1451 \\
7& 13&65& 329& 46& 30& 11&70& 18&24& 1364&12&90& 187& 25& 33& 10&10& 16&98 &1418 \\ \hline \hline

\multicolumn{11}{l||}{Blue GCs (sample\,\textsc{Bv})}&
\multicolumn{10}{l}{Red GCs (sample\,\textsc{Rv})}\rule{0ex}{2.5ex}
\\\hline 

1 & 2&15& 243& 30& 35& 1&44& 2&73 &1439& 1&89& 248& 30& 35& 1&12& 2&19\rule{0ex}{2.ex}  &1456 \\
2 & 3&15& 357& 47& 35& 2&75& 3&75 &1346& 2&49& 287& 35& 35& 2&20& 2&82  &1451  \\
3 & 4&49& 304& 37& 35& 3&76& 4&94 &1470& 3&20& 286& 38& 35& 2&83& 3&74  &1363  \\
4 & 5&57& 333& 41& 35& 4&98& 5&95 &1473& 4&22& 306& 47& 35& 3&76& 4&66  &1577  \\
5 & 6&46& 408& 50& 35& 6&03& 6&93 &1386& 5&04& 228& 28& 35& 4&67& 5&45  &1422  \\
6 & 7&68& 314& 39& 35& 6&96& 8&71 &1419& 5&94& 259& 34& 35& 5&46& 6&44  &1502  \\
7 & 9&76& 306& 37& 35& 8&80& 10&93 &1452& 7&04& 225& 28& 35& 6&54& 7&63  &1404  \\
8 & 11&70& 267& 35& 35& 10&94& 12&84 &1375& 8&84& 223& 28& 35& 7&69& 10&68  &1447  \\
9 & 15&32& 301& 37& 35& 12&87& 20&88 &1419& 14&21& 196& 21& 49& 10&69& 29&19  &1449  \\
10 & 29&20& 204& 32& 21& 21&16& 41&66  &1416  \\

\hline \hline
\end{tabular}
}

\label{tab:dispersion}
\end{table*}

\end{document}